\documentclass[aps,onecolumn,floats,prd,nofootinbib,superscriptaddress,longbibliography]{revtex4-2}

\usepackage[utf8]{inputenc}
\usepackage{graphicx}
\usepackage{dcolumn}
\usepackage{bm}
\usepackage{hyperref}
\usepackage{lipsum}
\usepackage{xcolor}
\usepackage{calc}
\usepackage{comment}
\usepackage{float}
\usepackage{diagbox}
\usepackage{soul}
\usepackage{ulem}
\usepackage{amsmath}
\usepackage{multirow}
\usepackage{adjustbox}
\usepackage{subfigure}

\makeatletter
\newcommand\footnoteref[1]{\protected@xdef\@thefnmark{\ref{#1}}\@footnotemark}
\makeatother

\begin{document}

\title{Confronting Cold New Early Dark Energy and its Equation of State with Updated CMB, Supernovae, and BAO Data}

\author{Aleksandr Chatrchyan}\email{aleksandr.chatrchyan@su.se}
\author{Florian Niedermann}\email{florian.niedermann@su.se}
\affiliation{
Nordita, KTH Royal Institute of Technology and Stockholm University\\
Hannes Alfv\'ens v\"ag 12, SE-106 91 Stockholm, Sweden}
\author{Vivian Poulin}\email{vivian.poulin@umontpellier.fr}
\affiliation{Laboratoire Univers \& Particules de Montpellier (LUPM), CNRS \& Universit\'e de Montpellier (UMR-5299), Place Eug\`ene Bataillon, F-34095 Montpellier Cedex 05, France}
\author{Martin S. Sloth}\email{sloth@sdu.dk}
\affiliation{Universe-Origins, University of Southern Denmark, Campusvej 55, 5230 Odense M, Denmark}

\begin{abstract}
Cold New Early Dark Energy (Cold NEDE) addresses the Hubble tension through a triggered vacuum phase transition in the dark sector. In this paper, we constrain a phenomenological fluid model using recent cosmic microwave background likelihoods based on Planck NPIPE data alongside baryonic acoustic oscillations (BAO) and supernovae data from Pantheon+. Exploiting the enhanced constraining power of the datasets, we introduce and study an extended version of the NEDE fluid model in which the equation of state parameter $w_\mathrm{NEDE}$, characterizing the post-phase transition fluid, is allowed to evolve with non-vanishing derivatives ${d}w_\mathrm{NEDE}/d\ln a$ and ${d^2}w_\mathrm{NEDE}/{d}(\ln a)^2$. Our results indicate that data is compatible with a rather simple time dependence that could arise from a mixture of radiation and a stiff fluid. With the updated datasets, the base and extended models still show a significant reduction of the DMAP tension from $6.3 \sigma$ in $\Lambda$CDM down to $3.5\sigma$ with a small simultaneous reduction of the $S_8$ tension, slightly improving over recent findings for the axion-like early dark energy model. Finally, we also provide a first test of the model against new BAO data from the Dark Energy Spectroscopic Instrument (DESI) survey. Replacing the previous BAO constraints in our analysis with the new ones, the tension is further reduced to $2.6 \sigma$, reaffirming the Cold NEDE model as a promising solution to the Hubble tension.
\end{abstract}

\maketitle

\section{Introduction}

The standard model of cosmology, the $\Lambda$CDM model, has undergone rigorous testing  from a variety of observations, and foremost using high-precision measurements of the Cosmic Microwave Background (CMB) anisotropies and Baryonic Acoustic Oscillation (BAO) data. 
These measurements are widely regarded as well-understood and unlikely to be affected by any unknown systematic effects. 
Therefore, it would appear highly surprising
if any of the precisely measured cosmological parameters, assuming the $\Lambda$CDM model, could deviate substantially
from their true values, and significantly distort our understanding of the Universe's evolution. Nevertheless, the direct measurement of the Hubble parameter $H_0$ by SH$_0$ES~\cite{Riess:2021jrx}, using Type Ia supernovae (SN1a) calibrated with Cepheids, yields $H_0 = 73.04 \pm 1.04$ km s${}^{-1}$Mpc${}^{-1}$ exhibiting a $5\sigma$ tension with the $\Lambda$CDM model prediction of $H_0= 67.36 \pm 0.54$ km s${}^{-1}$Mpc${}^{-1}$ using Planck 2018 data \cite{Planck:2018vyg}.
$\Lambda$CDM is a simple six parameter model, and if the tension is not due to a yet unknown systematic error, it indicates that we need additional parameters encompassing new degrees of freedom in order to bring it in agreement with SH$_0$ES, potentially related to the fundamental nature of dark matter and dark energy.

Other ongoing and future measurements will soon test the local determination of $H_0$ by SH$_0$ES with the Hubble Space Telescope (HST). Most noteworthy are the recent measurements made with the James Webb Space Telescope (JWST) from the CCHP \cite{Lee:2024qzr,Freedman:2024eph} and the SH$_0$ES team \cite{Anand:2024nim,Li:2024yoe,Li:2024pjo}. The CCHP analyses relies on three different methods to calibrate the SN1a based on Cepheids, the Tip of the Red Giant Branch (TRGB), and the J-Region Asymptotic Giant Branch, leading to a lower value of $H_0 = 69.96 \pm 1.05$ (stat)  $\pm 1.12$ (sys) km s${}^{-1}$Mpc${}^{-1}$. Though it appears in less significant tension with $\Lambda$CDM using CMB and BAO data, it is important to note that it relies on a smaller sample of SN1a calibrators with only one anchor, and it is in agreement with the spread in $H_0$ values established with the HST by the SH$_0$ES team \cite{Riess:2024vfa}. Moreover, alternative TRGB measurements also show significant disagreements with Planck (e.g.~\cite{Scolnic:2023mrv}). Regardless, if it is confirmed from direct measurements that the value of $H_0$ predicted by $\Lambda$CDM using CMB and BAO data is significantly higher than the $\Lambda$CDM model prediction, a method for reconciling the direct measurements with the CMB and BAO measurements is needed. Since the BAO measurements provide an almost model-independent measurement of the Hubble constant times the sound horizon, $H_0\cdot r_s$, any attempt to raise $H_0$ will require a smaller sound horizon $r_s$, in order to stay compatible with BAO. This points towards an energy injection altering the expansion rate shortly before recombination. This is the principle exploited in Early Dark Energy (EDE) type models~\cite{Kamionkowski:2022pkx,Poulin:2023lkg,McDonough:2023qcu} where the EDE component becomes important and decays just before recombination. In the \textit{AxiEDE} model~\cite{Karwal:2016vyq,Poulin:2018cxd,Smith:2019ihp}, the energy injection comes from a slow-rolling scalar field, which becomes heavy just before recombination and decays. However, this approach requires, for phenomenological reasons, a rather contrived potential for the scalar field. On the other hand, it is well-known that several phase transitions have taken place during the course of the cosmological evolution, and it is, therefore, a straightforward possibility, building on well-tested principles in quantum field theory and particle physics, that the injected energy comes from the latent heat of a fast cosmological phase transition between big bang nucleosynthesis and recombination, a possibility known as New Early Dark Energy (NEDE)~\cite{Niedermann:2023ssr}. The faster decay and the microphysics triggering the phase transition make NEDE models phenomenologically distinct from EDE-type models. So far mainly, two types of NEDE models have been explored in depth in the literature: the Cold NEDE \cite{Niedermann:2019olb,Niedermann:2020dwg,Cruz:2023lmn} and the Hot NEDE models \cite{Niedermann:2021vgd,Niedermann:2021ijp,Garny:2024ums}. For other implementations of EDE, see the review \cite{Poulin:2023lkg}. Here, we will focus on the Cold NEDE model.

It has indeed been shown that Cold NEDE fits the CMB and BAO slightly better than $\Lambda$CDM: Using only CMB (Planck 2018), BAO and supernovae (Pantheon) data, without including a local prior on $H_0$, Cold NEDE was found to be preferred at the 2$\sigma$ level while bringing the tension with SH$_0$ES below 2$\sigma$~\cite{Cruz:2022oqk,Cruz:2023cxy} and, simultaneously, reducing the $S_8$ tension also below 2$\sigma$~\cite{Cruz:2023lmn}. When including the prior on $H_0$ from SH$_0$ES, the preference of Cold NEDE over $\Lambda$CDM is more than 5$\sigma$.

Since Cold NEDE, like other EDE-type models, fit the CMB with a much lower sound horizon and larger $H_0$, it is really a significantly different fit to the CMB than $\Lambda$CDM. In (N)EDE models the lower sound horizon and the increased expansion rate due to the presence of the (N)EDE component is accompanied by a shortening of the CMB damping scale and a faster decay of the Weyl potential due to new acoustic oscillations in the (N)EDE fluid, which have to be compensated by a bluer primordial spectral index $n_s$ and a higher amplitude of primordial perturbation $A_s$, as well as a larger fraction of Cold Dark Matter (CDM) $\omega_{cdm}$, respectively. The different predictions of the cosmological parameters and of the details of the CMB TT, TE and EE power spectrum as well as of the matter power spectrum depending on the details of the specific (N)EDE model, leads to the expectation that more precise CMB and BAO data will on their own be able to single out the right model beyond $\Lambda$CDM. Here, we will take one more step in this direction by testing Cold NEDE against $\Lambda$CDM with the improved CMB data based on the NPIPE maps \cite{Planck:2020olo}, and compare with an earlier similar test of AxiEDE also using NPIPE data~\cite{Efstathiou:2023fbn}. On top of a better handle of a number of systematic effects, the new NPIPE maps have a larger signal-to-noise at small scales thanks to improvements in the processing of time ordered data \cite{Planck:2020olo}, allowing to use a larger sky fraction (about $\sim 80\%$) and resulting in $\sim 10\%$ higher precision on $\Lambda$CDM parameters~\cite{Rosenberg:2022sdy,Tristram:2023haj}. In addition, the Pantheon dataset used in previous studies of the Cold NEDE model is updated with Pantheon+~\cite{Brout:2022vxf}. We also provide a first test of the model against new BAO data from the Spectroscopic Dark Energy Instrument (DESI)~\cite{DESI:2024mwx}. To that end, we perform an analysis where we replace our baseline BAO datasets with the new constraints. Interestingly, DESI data has previously shown some hints of deviations from $\Lambda$CDM that go in the direction of {\it relaxing} constraints on EDE models \cite{Qu:2024lpx,Poulin:2024ken}.

To harness the constraining power of the updated datasets, we have also improved the phenomenological treatment of the NEDE fluid after the phase transition in the Cold NEDE model in two ways. First, in previous work, the NEDE fluid after the phase transition was treated as a single fluid with a moderately stiff but constant equation of state, $1/3 \leq w_\mathrm{NEDE} \leq 1$. While a good approximation, it is not necessarily a precise description of the fluid, which can consist of several components with different equations of state or a fluid of decaying particles.  To be specific, after the phase transition, part of the latent heat is naturally converted into gravitational waves and other forms of radiation, while part of the latent heat can be carried away by vector anisotropic stress, kinetic energy, and particles with repulsive interactions, which all will behave like fluid components with a stiff equation of state. It is therefore natural to expect the NEDE fluid after the phase transition to be a mixture of two fluids, one with the same equation of state as that of radiation, $w= 1/3$, and one with the same equation of state as a stiff fluid $w=1$. In a realistic microscopic particle physics implementation of the model, there will typically also be channels for particle decay, leading to an additional redshift dependence of the effective equation of state of the NEDE fluid. These scenarios can be encompassed into a more general treatment in terms of a fluid with a redshift-dependent equation of state $w_\mathrm{NEDE}(a)$.
Second,  we test the trigger mechanism inherent to Cold NEDE. Indeed in this model it is predicted, as a theoretical consistency requirement, that the ratio of the Hubble rate at the time of the phase transition and mass of the trigger field satisfies the relation $H_*/m\approx 0.2$. If Cold NEDE is not the correct microscopic model solving the Hubble tension, there would be no reason that data would prefer this ratio of $H_*/m$. However, if Cold NEDE is the correct microscopic model, we would expect data to single out the predicted ratio. It is therefore instructive to let $H_*/m$ be a free parameter and test this prediction of the model.
Last but not least, additional data will also enable us to better compare Cold NEDE against other models. We will therefore also contrast our findings with respect to earlier result of fitting AxiEDE with NPIPE data \cite{Efstathiou:2023fbn}, in order to gauge whether improved data can help disentangle between EDE models.

The paper is organized as follows. In Sec.~\ref{sec:model}, we review the Cold NEDE model and introduce the extended model with the improved NEDE fluid treatment. Then in Sec.~\ref{sec:data_analysis}, we discuss the data analysis and the results, and then in Sec.~\ref{sec:concl}, we conclude. In Appendix~\ref{app:profile}, we provide likelihood profiles to further support our statistical findings.

\section{The NEDE model}\label{sec:model}

We start by reviewing the phenomenological NEDE model in Sec.~\ref{sec:pheno_model}. It is valid on cosmological scales and enables testing against cosmological datasets. It models a first-order phase transition as a sudden change in the equation of state parameter. In Sec.~\ref{sec:extension}, we discuss a natural extension to this model, followed  by a review of a microscopic realization in terms of two interacting scalar fields in Sec.~\ref{sec:micro}.

\subsection{Minimal Phenomenological Setup} \label{sec:pheno_model}
The phenomenological NEDE model in its simplest form consists of an ideal fluid component $\rho_\mathrm{NEDE}(\tau,\mathbf{x})$ to which we refer as the NEDE fluid, and a trigger field $\phi(\tau,\mathbf{x})$ with ultra-light mass $m$. In short, the trigger determines the time $\tau_*$ at which the fluid's equation of state parameter changes abruptly from
\begin{align}\label{eq:w_NEDE}
& w_\mathrm{NEDE} = -1 \, \quad \text{for} \quad \tau < \tau_* 
\end{align}
to
\begin{align}
1/3 \leq & w_\mathrm{NEDE}(\tau) \leq 1   \quad \text{for} \quad \tau \geq \tau_* \,. 
\end{align}

In other words, an early dark energy component is converted to a quickly decaying  fluid. 
As we will argue later in Sec~\ref{sec:micro}, such a behavior can be achieved in a first-order phase transition where the dynamics of $\phi$ removes the potential barrier seen by another field $\psi$ stuck in the false minimum of its potential. This happens when the Hubble friction that initially traps $\phi$ close to its initial value is released. Decomposing $\rho_\mathrm{NEDE}(\tau, \mathbf{x}) = \bar{\rho}_\mathrm{NEDE}(\tau) + \delta \rho_\mathrm{NEDE}(\tau, \mathbf{x})$, the background energy density evolves as
\begin{align}
\label{eq:bg_sol}
\bar{\rho}_\mathrm{NEDE}(\tau) =\bar{\rho}_\mathrm{NEDE}(\tau_*) \exp\left[-3 \int_{\tau_*}^\tau \mathrm{d}\tilde \tau \left[1 + w_\mathrm{NEDE}(\tilde \tau) \right] \mathcal{H}(\tilde \tau)  \right]\,,
\end{align}
where $ \bar{\rho}_\mathrm{NEDE}(\tau_*)$ corresponds to the amount of early dark energy before the phase transition and $\mathcal{H} =a H$ is the conformal Hubble parameter. Its fraction at the time of the decay is
$f_\mathrm{NEDE} = \bar{\rho}_\mathrm{NEDE}(\tau_*)/  \bar{\rho}_\mathrm{tot} (\tau_*)$.  

While the above parametrization would be enough to describe a consistent background evolution, we will see that we also need to keep track of the trigger field's dynamics to describe the perturbation sector consistently.  To be explicit, decomposing $ \phi(\tau,\mathbf{x})=\bar{\phi}(\tau) + \delta \phi(\tau,\mathbf{x})$, the background and perturbation equations in synchronous gauge read
\begin{subequations}
\label{eq:KG}
\begin{align}
\bar{\phi}^{\prime\prime} + 2 \mathcal{H} \bar{\phi}^{\prime} + m^2 \bar{\phi}&=0\,,\\
\delta \phi^{\prime\prime}  + 2 H \delta \phi^\prime + (k^2/a^2 +  m^2) \delta \phi & = -\frac{1}{2} h^\prime \bar{\phi}^\prime  \,, \label{eq:pert_phi}
\end{align}
\end{subequations}
where we switched to momentum space, a prime denotes differentiation with respect to conformal time $\tau$, and $h$ denotes the spatial trace of the metric perturbation.  For $H \gg m$, the background equation describes an overdamped harmonic oscillator with $\bar{\phi} \simeq \bar{\phi}_\mathrm{ini}$. As $H$ drops below the trigger mass $m$, the field thaws and starts to roll down to the bottom of its potential where it oscillates with frequency $m$. The transition in $w_\mathrm{NEDE}$ is triggered during the thawing phase when a threshold value $\phi_*$ is reached. This happens before the field reaches the minimum of its potential for the first time, corresponding to $H(z_*) \sim m$. The precise value is determined by the microscopic model. For example, in the case of Cold NEDE it is~\cite{Niedermann:2020dwg} $H(z_*)/m \simeq 0.2$. This condition implicitly fixes the transition redshift $z_*$ (and hence also $\phi_*$). After it has served its main purpose by triggering the transition in $w_\mathrm{NEDE}$, due to its coherent oscillations, $\phi$ acts like a fuzzy dark matter component~\cite{Preskill:1982cy,Abbott:1982af,Dine:1982ah} (see also \cite{Turner:1983he,Brandenberger:1984jq}), where its abundance today is controlled by $\phi_\mathrm{ini}$~\cite{Cruz:2023lmn},
\begin{align}\label{eq:Omega_phi}
\Omega_\phi \approx 0.4 \times \left( \frac{1+z_*}{5000}\right) \left(\frac{\phi_\mathrm{ini}}{M_\mathrm{Pl}} \right)^2 \left( 1- f_\mathrm{NEDE}\right)\,.
\end{align}
The model's parameter space therefore admits two natural regimes: one, where $\Omega_\phi \ll 0.01 $ and thus the trigger contribution to dark matter is negligible, and another, where $0.01 \lesssim \Omega_\phi$ and thus the trigger makes a sizable contribution to dark matter. In particular, in the latter regime the trigger field's pressure perturbations act against gravitational collapse and can therefore suppress the matter power spectrum on scales~\cite{Hu:2000ke,Hwang:2009js} $k > k_\mathrm{J,eq}\simeq  0.16 \; {\rm Mpc}^{-1} \left[m/(10^{-27} {\rm eV})\right]^{1/2}$. As shown in~\cite{Cruz:2023lmn}, this allows the model to address the $H_0$ and $S_8$ tension simultaneously.   

Now we come to a crucial point for the model's phenomenology, which also sets it apart from other EDE-type models. In short, the NEDE perturbations $\delta \rho_\mathrm{NEDE}(\tau, \mathbf{x})$ after the phase transition inherit the trigger field's adiabatic perturbations $\delta \phi(\tau,\mathbf{x})$. This can be understood as follows. Before the phase transition, NEDE behaves like a cosmological constant and thus has vanishing adiabatic perturbations, i.e.\ $\delta \rho_\mathrm{NEDE}(\tau< \tau_*) = 0$.\footnote{Adiabatic perturbations can be understood as arising from a space-dependent time shift, which will leave a constant invariant.}. However, this is no longer true after the phase transition when the fluid has an equation of state $w_\mathrm{NEDE} > 1/3$. Therefore, we need to find a physical way of initializing these perturbations both on sub and super-horizon scales. This is achieved through the trigger field. Due to its perturbations $\delta \phi$, the threshold value $\phi_*$ is reached earlier or later at different positions in space. This gives rise to a (slightly) space-dependent transition time $\tau_*(\mathbf{x})= \bar{\tau}_*+ \delta \tau_*(\mathbf{x})$. It is defined implicitly through $\phi(\tau_*(\mathbf{x}), \mathbf{x}) = \phi_*$, implementing the requirement that the phase transition occurs on a surface of constant $\phi$. From this one can derive that (see~\cite{Cruz:2022oqk} for more details) $\delta \tau_*(\mathbf{x}) = - \delta \phi(\bar{\tau}_*,\mathbf{x})/\bar{\phi}^\prime(\bar{\tau}_*) $. The initial perturbation can then be obtained by performing a simple shift (taking into account that $\rho_\mathrm{NEDE}$ in a given region started to dilute earlier or later for $\delta \tau_* < 0$ or $\delta \tau_* > 0$, respectively),
\begin{align}
\bar{\rho}_\mathrm{NEDE} (\tau_*) \to \bar{\rho}_\mathrm{NEDE} (\tau_* - \delta \tau_*(\mathbf{x})) 
&= \bar{\rho}_\mathrm{NEDE} (\tau_*) -\bar{\rho}^\prime_\mathrm{NEDE} (\tau_*)\delta \tau_*(\mathbf{x}) \nonumber\\
&=\bar{\rho}_\mathrm{NEDE} (\tau_*) \left[ 1 - 3 \mathcal{H}_* (1 + w_\mathrm{NEDE}^*)  \frac{\delta \phi(\bar{\tau}_*,\mathbf{x})}{\bar{\phi}^\prime(\bar{\tau}_*)} \right]\,,
\end{align}
where in the second line we used the conservation equation $\bar{\rho}_\mathrm{NEDE}^\prime + 3 \mathcal{H} (1 + w_\mathrm{NEDE})\bar{\rho}_\mathrm{NEDE}=0$.
From this, we read off the initial density contrast, $\delta_\mathrm{NEDE} = \delta \rho_\mathrm{NEDE}/\bar{\rho}_\mathrm{NEDE}$, after the phase transition as [switching to momentum space $\phi(\tau,\mathbf{x}) \to \phi(\tau,\mathbf{k})$]
\begin{subequations}\label{eq:match}
\begin{align}
\delta_\mathrm{NEDE}(\tau_*,\mathbf{k}) = - 3 \mathcal{H}_*(1 + w_\mathrm{NEDE}^*)  \frac{\delta \phi(\bar{\tau}_*,\mathbf{k})}{\bar{\phi}^\prime(\bar{\tau}_*)}\,,
\end{align}
which reproduces the result in the literature~\cite{Niedermann:2020dwg}.
A more involved calculation relying on a covariant matching across the transition surface~\cite{Israel:1966rt,Deruelle:1995kd} also determines the initial velocity contrast as~\cite{Niedermann:2020dwg}
\begin{align}
\theta_\mathrm{NEDE}(\tau_*,\mathbf{k}) = k^2 \frac{\delta \phi(\bar{\tau}_*,\mathbf{k})}{\bar{\phi}^\prime(\bar{\tau}_*)}\,,
\end{align}
\end{subequations}
where both equations assume an instantaneous transition occurring on time scales $\ll 1/k$. The matching conditions are valid both on sub and super-horizon scales. In particular, they are controlled by the ratio $\delta \phi / \bar{\phi}^\prime$, which is independent of $\phi_\mathrm{ini}$ (and hence also $\Omega_\phi$) for adiabatic perturbations~\cite{Niedermann:2020dwg, Cruz:2023lmn}. Of course, beyond these adiabatic perturbations, a phase transition will also lead to isocurvature perturbations arising from the stochastic nature of a first-order phase transition. As was pointed out in~\cite{Niedermann:2020dwg} (and later also confirmed in \cite{Elor:2023xbz}), these contributions can be avoided on phenomenologically relevant scales if the trigger acts sufficiently fast.   The perturbations are then evolved with the perturbed continuity and Euler equation
\begin{subequations}
\label{eq:perts}
\begin{align}
\delta'_\mathrm{NEDE} &= -(1 + w_\mathrm{NEDE})\left( \theta_\mathrm{NEDE} + \frac{h'}{2}\right) - 3(c_s^2 - w_\mathrm{NEDE}){\cal H} \delta_\mathrm{NEDE}\,, 
\\
\theta'_\mathrm{NEDE} &= -(1-3c_s^2) {\cal H}\theta_\mathrm{NEDE} + \frac{c_s^2 k^2}{1 + w_\mathrm{NEDE}}\delta_\mathrm{NEDE}\,,
\end{align}
\end{subequations}
where we assumed a barotropic fluid with sound speed
\begin{align}
c_s^2 = w_\mathrm{NEDE}(\tau) - \frac{1}{3} \frac{ w_\mathrm{NEDE}^\prime(\tau)}{1 + w_\mathrm{NEDE}(\tau)} \frac{1}{\mathcal{H}}\,.
\end{align}
This is a crucial difference compared to AxiEDE, which deep inside its dissipating phase can be described in terms of a fluid with a $k$-dependent sound speed~\cite{Poulin:2018dzj} (whereas $c_s$ here is $k$-independent). Moreover, in the case of the phenomenological NEDE model, the fluid perturbations are initialized through the $k$-dependent matching equations in \eqref{eq:match}. In particular, the choice of the trigger plays an important role in determining the subsequent dynamics. This was recently demonstrated in the case of Hot NEDE, where the ultralight field was replaced by a dark sector temperature~\cite{Garny:2024ums}, which led to a decisively different phenomenology. For Cold NEDE, the parameter combination $H(z_*)/m$, which is predicted by the microscopic model, controls how much the trigger field has thawed by the time of the phase transition. As a result, it determines $\bar{\phi}^\prime$ on the right hand side of the matching equations \ref{eq:match} and thus controls the initial size of the perturbations $\delta_\mathrm{NEDE}(\tau_*, \mathbf{k})$ in the dissipating dark sector fluid. As we will see in Sec.~\ref{sec:thawing}, this leads to observable consequences. In any event, this trigger sector in the NEDE model does not have a counterpart in AxiEDE.

In summary, the background evolution, as implemented in the Boltzmann code \texttt{TriggerCLASS}, is described in terms of \eqref{eq:bg_sol} and \eqref{eq:KG}. In our minimal implementation, we will assume that the equation of state after the phase transition is constant, {i.e.} $w_\mathrm{NEDE}(\tau \geq \tau_*) =w_\mathrm{NEDE}^* =\rm const $. This assumption will be relaxed when considering the extended model in Sec.~\ref{sec:evolving_w}. The solution is uniquely determined by fixing $f_\mathrm{NEDE}$, $\phi_\mathrm{ini}$, $w_\mathrm{NEDE}^*$, and $m$. In practical terms, the Boltzmann code takes $ \Omega_\phi$ and $z_*$ as the input parameters and determines $m$ (for fixed $H(z_*)/m$) and $\phi_\mathrm{ini}$ trough a shooting algorithm. For simplicity, in what follows we omit the ``NEDE'' subscript from $w$ and $\rho$ in our formulas.

\subsection{Time-Dependent Equation of State}\label{sec:extension}

The benefit of the phenomenological NEDE model is that it provides a rather general framework to capture the physics of different field theoretic models that exhibit a triggered vacuum phase transition. A crucial parameter is the equation of state parameter $w$ that changes abruptly in the phase transition from $-1$ to a value $>1/3$. Here, previous works relied on the simplifying assumption that the equation of state after the phase transition can be modeled as a constant. The reasoning was simple: even if $w(\tau> \tau_*)$ is time-dependent, since the NEDE fluid becomes quickly subdominant for $w$ large enough, data is not very sensitive to its subsequent evolution. For example, $\rho/\rho_\mathrm{tot} \propto 1/a $ for $w \approx 2/3$. However, this argument requires the time-dependence to be sufficiently weak and it therefore makes sense to relax that assumption. In particular, we will test if the effective time-evolution of a simple, tightly coupled two-component fluid is compatible with data. 

A general time-dependent equation of state can be parametrized in terms of its Taylor expansion,
\begin{align}
w(a) = w^* + \frac{dw}{d\ln a}\Big|_{a_*} \ln\Bigl(\frac{a}{a_*} \Bigr) + \frac{1}{2}\frac{d^2w}{d(\ln a)^2}\Big|_{a_*} \Bigl[ \ln\Bigl(\frac{a}{a_*} \Bigr) \Bigr]^2 +...\,,
\label{eq_Taylor}
\end{align}
where $w^* =w(a_*) $ and we found it more convenient to parametrize the time-dependence in terms of the scale factor $a$ (rather than $\tau$).
Usually, the NEDE fluid decays efficiently after the transition, and thus quickly becomes negligible in the energy budget of the universe. It is thus reasonable to assume that the qualitative features of a given NEDE model are captured already by the lowest-order terms in the Taylor expansion, before the higher-order terms start to become relevant. Here, we will include two higher-order terms and constrain their coefficients with current datasets. This allows us to test whether the $w = \rm const$ assumption is justified in light of newer datasets and whether relaxing it can give us a clue about the underlying microphysics. Let us also add a remark on the regime of applicability of the truncation in \eqref{eq_Taylor}.  
To be more precise, the time interval for which $w(a)$ can still be described by the truncated series is determined by how large higher-order coefficients are. In particular, for the second derivative, we impose a general convergence criteria expressed as
\begin{align}\label{eq:convergence}
\frac{1}{2}\left|\frac{d^2w}{d(\ln a)^2}\right|_{a_*} \left[\ln\Bigl(\frac{a}{a_*} \Bigr)\right]^2 \lesssim  \max \left\{w^*, \left|\frac{dw}{d\ln a} \Big|_{a_*} \ln\Bigl(\frac{a}{a_*} \Bigr)  \right|  \right\}.
\end{align}
Thus, for a given microscopic model one should ensure that the above condition holds for values of $a$ for which the NEDE fluid is still expected to be relevant, i.e.,~it constitutes a sizable component of the energy budget. On the other hand,  Eq.~\eqref{eq:convergence} may be violated at late times provided the NEDE fluid is negligible (and remains so). An additional self-consistency requirement, which can be checked on the level of the data analysis, is that the values of the higher-order coefficients of the Taylor expansion are less constrained when fitted to the data. We will check this a posteriori in the case of the first and second derivatives.

Moreover, we always impose a lower bound on the value of the equation of state parameter, $w(a)>0$, at any time between the transition and today. This is done to avoid exponentially growing modes in the NEDE perturbations, which otherwise arise for $w<0$ (these could cause the numerical integration to fail and would in any case lead to an unacceptable fit to data). This implies that if we were to include only the first derivative in the Taylor expansion~(\ref{eq_Taylor}), it would be allowed to take only values in the range
\begin{align}
\label{eq:bound1}
\frac{dw}{d\ln a}\Big|_* > - \frac{w(a_{*})}{\ln[z_{*}]}.
\end{align}
Such a linear approximation turns out to be quite restrictive and removes an interesting parameter region where the first derivative is violating \eqref{eq:bound1} initially for $\tau \gtrsim \tau_*$ but is then stabilized by higher order terms later on. Therefore, including a second derivative allows more negative values of the first derivative initially while at the same time not running into a negative equation of state at late times, when we expect the NEDE fluid to be irrelevant. This is why we keep the first two derivatives in our analysis.

As before, the bound $w(a)>0$ implies a lower bound on the first derivative for given values of the remaining coefficients (or, alternatively, a lower bound on the second derivative). More specifically, if $\frac{d^2w}{d\ln a^2} |_{*}<0$, it is sufficient to  require that $w(a)$ is positive today as it can only decrease monotonically. We thus require that $w(a_0)>0$ or, equivalently,
\begin{align}
w_* + \frac{dw}{d\ln a}\Big|_* \ln\Bigl(z_{*} \Bigr) + \frac{1}{2}\frac{d^2w}{d(\ln a)^2}\Big|_* \Bigl[ \ln\Bigl(z_{*} \Bigr) \Bigr]^2>0,
\label{eq_w_today_positive}
\end{align}
from which the lower bound follows. The same is required if $\frac{d^2w}{d\ln a^2} |_{*}>0$, however, then $w(a)$ can be non-monotonic and one additionally needs that, in the case when $w(a)$ has a local minimum between the transition time and today, i.e., $dw(a)/d(\ln a)= 0$, the equation of state parameter is positive at that minimum. This condition translates into 
\begin{align}
\frac{dw}{d\ln a}\Big|_* > - \sqrt{ 2 w_* \frac{d^2w}{d\ln a^2} \Big|_*}.
\end{align}
This sets a stronger lower bound on $\frac{dw}{d\ln a}\Big|_*$ compared to the one from (\ref{eq_w_today_positive}) for values of the second derivative in the range $\frac{d^2w}{d(\ln a)^2}\Big|_* > \frac{2 w_{*}}{\ln (z_{*})^2}$. For the data analysis in section~\ref{sec:extensions}, we introduce an auxiliary variable $y$ defined via
\begin{align}\label{eq:y}
\frac{dw}{d\ln a}\Big|_*  = y + 
\begin{cases} 
- \frac{w_*}{\ln z_{*} } - \frac{1}{2} \frac{d^2w}{d\ln a^2}\Big|_* \ln z_{*},  &   \frac{d^2w}{d\ln a^2}\Big|_*  \leq \frac{2w_*}{  (\ln z_{*})^2  } \\
- \sqrt{ 2 w_{*} \frac{d^2w}{d\ln a^2}\Big|_*  }, & \frac{d^2w}{d\ln a^2}\Big|_*  > \frac{2w_*}{  (\ln z_{*})^2  }\\ 
\end{cases}
\end{align}
With this definition, it is easy to show that $w(a)>0$ (for $a<a_0\equiv1$) is equivalent to simply demanding $y>0$.

We now turn to two simple but physical scenarios, which we will use as benchmark when analyzing the data constraints.

\subsubsection{Two-Fluid Model}\label{sec:evolving_w}

An interesting yet simple example of a system with a nontrivial time-dependence of the equation of state is the mixture of two fluids, each characterized by their equation of state parameters $w_1$ and $w_2$. In a simple case, these two components could be a stiff fluid, $w_1=1$, and radiation, $w_2=1/3$. The only free parameter characterizing this system is the ratio of the energy densities of the two components at a certain time. We denote this ratio at the time of the phase transition by $r_f$. In other words, the total equation of state at that time is given by 
\begin{align}
w_*= \frac{P}{\rho} = w_1 r_f + w_2 (1-r_f).
\label{eq:r}
\end{align}
The energy densities of the two fluids dilute differently with the Hubble expansion, according to $d\rho_i/d \ln a = -3(1+w_i)\rho_i$ and, therefore, $w(a)$ changes with time. One can show that the derivative of the equation of state parameter satisfies
\begin{align}
\frac{dw}{d\ln a}\Big|_{a=a_*} = 3(w_*- w_1) ( w_* - w_2).
\label{eq_mixture_dw}
\end{align}
As can be expected, the derivative of the equation of state parameter is always negative. One can also compute the second derivative of the equation of state,
\begin{align}
\frac{d^2w}{d(\ln a)^2}\Big|_{a=a_*} = 9 (w_*-w_1) (w_*-w_2) (2w_*-w_1 - w_2).
\label{eq_mixture_d2w}
\end{align}
The sign of this term can be either positive (if $w_*<(w_1+w_2)/2$) or negative (in the other case). 

Let us now demonstrate that the criterion \eqref{eq:convergence} can be satisfied in the special case where we start with a mixture of a stiff fluid ($w_1=1$) and radiation ($w_2 = 1/3$). Substituting \eqref{eq_mixture_dw} and \eqref{eq_mixture_d2w}, we derive the (sufficient) condition
\begin{align}
\left[\ln\left( \frac{a}{a_*} \right)\right]^2 \lesssim \frac{1}{9} \left|\frac{w_*}{(1-w_*)(w_*-1/3)(w_*-2/3)}\right|\;.
\end{align}
The right-hand side is minimal for $w_* \simeq 0.45$, which yields the sufficient condition $a/a_* \lesssim 7$. If the phase transition happens at redshift $z_* \lesssim 5000$, we have therefore already entered deep into matter domination (where NEDE is clearly subdominant as it falls off even faster than radiation) before the approximation breaks down. As a result, we expect our phenomenological model in \eqref{eq_Taylor} to capture the main features of the two-fluid case, although a more detailed analysis where $w(a)$ is treated exactly would provide an important test.

\subsubsection{Decaying Fluid Model}
Another interesting possibility is to have a mixture of two fluids, where one is converted into the other with time. Microscopically, this could describe the decay of a particle with rate $\Gamma_{1\to 2}$.  The equations of motion for the energy density components can be written as (we adopt a fluid formalism similar to the one in~\cite{Bringmann:2018jpr})
\begin{align}
\begin{cases} 
\frac{d\rho_1}{d \ln a} &= -3(1+w_1)\rho_1(a)  - \frac{\Gamma_{1\to 2}(a)}{H(a)} \rho_1(a), \\
\frac{d\rho_2}{d \ln a} &= -3(1+w_2)\rho_2(a)  + \frac{\Gamma_{1\to 2}(a)}{H(a)} \rho_1(a)\,, \\ 
\end{cases}
\label{eq_eom_Q}
\end{align}
which differs from the previous case by the inclusion of the $\propto \rho_1 \Gamma_{1\to 2} $ terms, describing the decay of the first fluid. To be relatively general, we define $Q(a) = \Gamma_{1\to 2}(a)/H(a)$ and assume the time-dependence to be of the form 
\begin{align}
Q(a) = Q_* \Bigl( \frac{a}{a_{*}}  \Bigr)^{q_{*}}.
\end{align}
With this parametrization, (\ref{eq_eom_Q}) can be solved, yielding the following relations
\begin{subequations}
\label{eq_Q_dependence}
\begin{align}
\frac{dw}{d\ln a}\Big|_{a=a_*} &= 3 \Bigl( w_*- w_1-\frac{Q_*}{3} \Bigr) ( w_* - w_2),\\
\frac{d^2w}{d(\ln a)^2}\Big|_{a=a_*} &= 9\Bigl( w_*-w_1 - \frac{Q_*}{3} \Bigr) (w_*-w_2) \Bigl( 2w_*-w_1 - w_2 - \frac{Q_*}{3} \Bigr) - q_* Q_* (w_* - w_2),
\end{align}
\end{subequations}
between $w_{*}$ and the first two derivatives at $a_{*}$. Note that $q_*$ does not enter the formula for the first derivative. In other words, as we are primarily interested in constraining $(dw/{d\ln a})|_*$, we do not need to make an assumption about the time-dependence of $Q$.

In the two examples mentioned above, the description of the system as a fluid with a time-dependent equation of state parameter $w(a)$ holds at background level. The fluid perturbations of the two components may, however, evolve independently in the most general case. To make an effective fluid approach applicable, we will, therefore, assume that both fluids are tightly coupled, i.e., their velocity perturbations are the same, $\theta_1 = \theta_2$. In that case, one can indeed describe the system as a single fluid with the time-dependent equation of state parameter $w(a)$ and perturbations that are evolved according to~\eqref{eq:perts}.

\subsection{Example of a Microscopic Model}
\label{sec:micro}

Here, we briefly review the Cold NEDE model as a concrete example of a field theoretic model (for a more extensive discussion see also \cite{Niedermann:2020dwg,Cruz:2023lmn} and the review~\cite{Niedermann:2023ssr}). In particular, we are interested in how its microscopic parameters are related to the phenomenological parameters introduced above.  

In a nutshell, the idea is to consider a field $\psi$ that is trapped in a false minimum of its potential. Initially it is separated from the true minimum by a high potential barrier. The barrier height is controlled by another field $\phi$ with ultralight mass $m$ through a very small coupling term $\tilde \lambda \phi^2 \psi^2$. As the trigger field starts rolling when the Hubble friction is released for $H \lesssim m$, it effectively shrinks the potential barrier and enables the field $\psi$ to tunnel to its true minimum. This idea can be realized in terms of the two-field potential (this is an eV-scale adaption of a model realizing a first-order inflation scenario~\cite{Adams:1990ds,Linde:1990gz})
\begin{align}\label{coldNEDE_pot}
V(\psi,\phi) =\frac{\lambda}{4}\psi^4+\frac{\beta}{2} M^2 \psi^2 -\frac{1}{3}\alpha M \psi^3
+ \frac{1}{2}m^2\phi^2 +\frac{1}{2}\tilde\lambda \phi^2\psi^2  \ldots\,,
\end{align}
where the parameters $\beta$, and $\alpha$ are assumed to be of order unity, $\lambda < 1$, and the ellipses stand for higher order operators in an effective field theory (EFT) approach. The mechanism relies on a hierarchy of scales: The tunneling field is described by eV-scale physics, i.e., $M \sim \mathrm{eV}$, while the trigger field has a mass of order $m\sim 10^{-27} \mathrm{eV}$ in order for its thawing to occur around matter-radiation equality. Moreover, the coupling between both fields needs to be hierarchically small to keep loop corrections to the light mass scale $m$ under control, explicitly~\cite{Niedermann:2020dwg} $\tilde \lambda \leq \mathcal{O}(1) \times 10^3 m^2/ (\beta M^2) \sim 10^{-51}$. If the trigger field is an ultra-light axion, then this hierarchy of scales follows naturally from the Planck scales and the eV scale. Assuming that the scale $\Lambda$, which controls non-perturbative instanton corrections to the axion potential, is related to the mass of the NEDE boson, i.e. $M\sim \Lambda \sim$ eV, then with a Planckian decay constant $f\sim M_\mathrm{Pl}$, the mass of the trigger field is naturally $m= \Lambda^2/f=10^{-27}\mathrm{eV}$, and thus, the smallness of the trigger mass can be understood as arising from the breaking of the shift symmetry. Similarly, the smallness of the coupling parameter can be obtained in a generalized two-axion setting where also $\psi$ is an axion with decay constant $\tilde f $ and non-perturbative instanton scale $\tilde \Lambda $~~\cite{Cruz:2023lmn}. In that case, one finds $\tilde \lambda = \tilde \Lambda^4/ (2 \tilde f^2 f^2) \sim (\tilde \Lambda/ \Lambda)^4 (m/\tilde f)^2 $, which indeed is parametrically suppressed and naturally satisfies the upper bound on $\tilde \lambda$ for $\tilde f^2 \gtrsim 10^{-3} \beta \tilde \Lambda^4 / M^2$.

For $\lambda < 2 \alpha^2/(9 \beta)$, the model exhibits a false vacuum at $\psi = 0$ and a true one at 
\begin{align}\label{eq:true_minimum}
( \psi, \phi)_\text{True} =  \left(\frac{M}{2 \lambda}\,\left[\alpha + \sqrt{\alpha^2-4 \lambda \beta }\right] ,0 \right)\,.
\end{align}
We initially assume $\psi$ to be trapped in its false vacuum and the trigger field to be frozen close to its initial field value, i.e., $\phi \approx \phi_\mathrm{ini}$. Then the trigger's dynamics is indeed captured by \eqref{eq:KG} introduced before. Within this scenario, we identify early dark energy, introduced in \eqref{eq:bg_sol}, with the latent heat $\bar{\rho}_\mathrm{NEDE}(\tau_*) =V(0,\bar{\phi}(\tau_*)) - V(\psi_\mathrm{True},0)$ that is released when the field tunnels to the true minimum. We assume that it makes up a sizeable fraction $f_\mathrm{NEDE}$ of the total energy density. In terms of the EFT parameters, we obtain
\begin{align}\label{def:f_NEDE}
f_\mathrm{NEDE} = \frac{\bar{\rho}_\mathrm{NEDE} (\tau_*)}{\rho_\mathrm{tot}(\tau_*)} = \frac{c_\delta}{36} \frac{\alpha^4}{\lambda^3} \frac{M^4}{M_\mathrm{Pl}^2 H_*^2}\,.
\end{align}
Here, we introduced the function
\begin{align}
c_\delta = \frac{1}{216} \left( 3 + \sqrt{9 -4 \delta}\right)^2 \left( 3 - 2 \delta + \sqrt{9 -4 \delta}\right)\,,
\end{align}
where $\delta =   9 \lambda \beta/ \alpha^2$. The probability of the field to tunnel to the true minimum can be approximated as $\Gamma \sim M^4 \exp{\left(-S_E\right)}$, where the Euclidian action $S_E(\delta_\mathrm{eff})$ depends on the trigger through
\begin{align}
\delta_\text{eff}(\tau) =  \left( \delta +  9\frac{ \lambda \tilde{\lambda}}{\alpha^2} \frac{\bar{\phi}^2(\tau)}{M^2}\right) \,.
\end{align}
The phase transition takes place when $\Gamma \lesssim H^4$, corresponding to the sudden nucleation of vacuum bubbles that convert space from the false to the true vacuum as they expand. This percolation phase starts when the Euclidian action drops below a critical value (assuming a phase transition shortly before matter-radiation equality, the threshold is $S_E \lesssim 250 $). For our purposes, we only need to know that $S_E \gg 250 $ for $\delta_\mathrm{eff} \gtrsim 2$, corresponding to a sizable barrier separating both vacua, and $S_E \to 0$ for $\delta_\mathrm{eff} \to 0$, corresponding to a disappearing barrier. As a result, for $\delta$ sufficiently small and $\phi_\mathrm{ini}$ sufficiently large, we will pass from a regime with $\Gamma \gg H^4$ when $\bar{\phi} \approx \phi_\mathrm{ini}$ (blocked percolation) to a regime with $\Gamma \ll H^4$ when $\bar{\phi} \to 0$ (strong percolation). Importantly, as $\Gamma$ scans many orders of magnitude, this can happen on a very short time scale $1/\bar{\beta} \ll 1/H_*$ (which prevents the creation of unacceptably large anisotropies in the CMB as discussed in \cite{Niedermann:2020dwg} and more recently in~\cite{Elor:2023xbz}). Here, we introduced the parameter $\bar{\beta} \simeq \frac{d\Gamma}{a d \tau} / \Gamma  \simeq\  \frac{d S_E}{a d \tau} $, which corresponds to the average bubble separation after nucleation and hence the typical bubble size at collision time (using that bubbles expand at nearly the speed of light in vacuum).

Assuming that the phase transition still takes place during radiation domination, the dictionary between the microscopic and phenomenological parameters reads
\begin{subequations}
\label{eq:dict}
\begin{align}
m &\simeq 1.7 \times 10^{-27} \, \text{eV} \, (1-f_\text{NEDE})^{-1/2}\left( \frac{g_*}{3.4} \right)^{1/2} \left( \frac{z_*}{5000} \right)^2 \, \left(\frac{0.2}{H_*/m}\right) \,,\label{eq:dict2}\\
M^4 &\simeq (0.4 \, \text{eV})^4 \, \frac{1}{c_\delta}\, \left(\frac{\lambda^3\alpha^{-4}}{0.01} \right) \left( \frac{f_\text{NEDE} / (1-f_\text{NEDE})}{0.1} \right) \left( \frac{g_*}{3.4} \right) \left( \frac{z_*}{5000} \right)^4  \,,\label{eq:dict1}\\
\left(\frac{\phi_\mathrm{ini}}{M_\mathrm{Pl}} \right)^2 & \simeq 0.13 \times \left( \frac{1+z_*}{5000}\right) ^{-1} \left(\frac{\Omega_\phi}{0.005}\right)  \left( 1- f_\mathrm{NEDE}\right)^{-1}\,,\label{eq:dict3}\\
H_* \bar{\beta}^{-1} & \approx  10^{-3}\, \left(\frac{\delta}{0.1}\right)^{1/2}\,\left(\frac{f_\text{NEDE}}{0.1}\right)^{1/2}\, \left(\frac{{0.1}}{\delta^*_\text{eff}-\delta}\right)^{1/2} \left(\frac{M_{\mathrm{Pl}}/\phi_\text{ini}}{10^4} \right)\, \left(\frac{H_*/m}{0.2} \right)\, \left( \frac{\lambda}{0.01}\right)^{3/2}\,.\label{eq:dict4}
\end{align}
\end{subequations}
Here, $g_*$ denotes the number of relativistic degrees of freedom in the visible sector at the time of the phase transition. For obtaining    \eqref{eq:dict2}, we used 
\begin{align}\label{eq:H_star}
H_* = \frac{\pi}{\sqrt{90}} \frac{\sqrt{g_\mathrm{rel,vis}^*}}{\sqrt{1 -f_\mathrm{NEDE}}} \frac{T^2_\mathrm{vis*}}{M_\mathrm{Pl}}\,
\end{align}
together with $T_\mathrm{vis*} \simeq 2.3 \times  10^{-4} (1+z_\ast) \,\mathrm{eV}$. The relation \eqref{eq:dict1} follows similarly from \eqref{def:f_NEDE}, and \eqref{eq:dict3} is a direct rewriting of \eqref{eq:Omega_phi}. Finally, the result for $\bar{\beta}$ in \eqref{eq:dict4} we cite from \cite{Niedermann:2020dwg}. Overall, this confirms the before-mentioned mass hierarchy $m \ll M$ and shows that a Planckian value for $\phi_\mathrm{ini}$ gives rise to a sizable amount of fuzzy dark matter (this assumes that the shift in the trigger mass during the phase transition can be neglected, which is valid for the values of $\tilde \lambda$ considered, here~\cite{Cruz:2023lmn}).

This discussion applies to the phase transition itself. Describing the system \textit{after} the phase transition, on the other hand, requires further input. On small scales $ \ell_0 \lesssim 1/\bar{\beta} (\ll 1/H_*)$, the colliding bubbles will give rise to an inhomogeneous field condensate. Instead, on large scales $\ell_0 \gg 1/\bar{\beta}$, which can be probed by cosmological observables, this state will average to an ideal fluid with time-dependent equation of state $w(\tau)$. While the generic expectation is that $w(\tau) \to 1/3$ asymptotically, the precise functional form of $w(\tau > \tau_*)$ will depend in a highly non-trivial way on the shape of the potential and the available decay channels for $\psi$.  While at this stage, we lack a unique prediction of $w(\tau)$, we outline different possible scenarios leading to $w(\tau > \tau_*) > 1/3$ as preferred by data:

\begin{enumerate}
\item \textbf{Small-scale shear:}
The small-scale anisotropic stress after the phase transition sources shear metric perturbations $\sigma_{\mu\nu}$ that can be decomposed into a transverse-traceless tensor (i.e., gravitational waves), transverse vector, and a scalar component. While the averaged energy density associated with the tensor component (alongside the sourcing fluid) behaves like radiation ($w=1/3$), the vector and scalar component is known to behave like a stiff fluid ($w=1$) on large scales~\cite{Xue:2011nw}. Analog to the sourcing of gravitational waves, the amount of vector and scalar shear will for example depend on the duration $\bar{\beta}$ and the strength $\alpha = f_\mathrm{NEDE}/(1-f_\mathrm{NEDE})$ of the phase transition. While we leave a more quantitative exploration, which requires an averaging procedure over small scales, to future work, at background level, this scenario motivates the two-fluid extension discussed in Sec.~\ref{sec:evolving_w}.

\item \textbf{Quick decay of tunneling field:} While the previous possibility requires a rather efficient conversion of the anisotropic stress into vector shear, we could instead assume that the $\psi$ particles decay, populating a different sector that features a stiff and a radiation component. Following Zel'dovich's old proposal~\cite{Zeldovich:1961sbr}, a high density $n_\chi$ of particles $\chi$ that repel each other, leads to a quickly decaying energy density with $\rho_\chi \propto n_\chi^2 \propto 1/a^6$ \cite{NiedermannandSlothtoappear}. 

\item \textbf{Slow dissipation of kination phase:}  Another possibility is to change the shape of the potential in \eqref{coldNEDE_pot} such that the field tunnels into an extended monotonic section (rather than a global minimum), where it picks up kinetic energy, leading to a phase of kination with $w=1$. In the presence of a marginally efficient decay channel with $\Gamma \sim H_*$, the field condensate could populate a relativistic species on time scales $\sim 1/H_*$. In terms of the phenomenological model discussed in Sec.~\ref{sec:evolving_w}, this would correspond to a scenario with $Q_*$ of order unity.  
\end{enumerate}

All three scenarios share the characteristic of introducing a nontrivial time-dependence, where $w(\tau) > 1/3$ right after the phase transition and $w(\tau) \to 1/3$ asymptotically. In particular, we would in all cases have $d w/ d \ln a <0$. The aim of this work is therefore to establish if these microscopic scenarios could be compatible with data by constraining our extended fluid model, and if the new data has the strength to discriminate from a constant equation of state as used as an approximation in previous treatments.\footnote{It should be noted that our extended fluid model would capture the background evolution of these scenarios while the perturbation sector might differ depending on the details of the model building. In that sense, our analysis provides an indication as to what types of scenarios can be viable and should be explored further both phenomenologically and in model building.} 

\section{Data Analysis}\label{sec:data_analysis}

The aim of this section is two-fold. First, we investigate how the updated CMB and supernovae datasets impact the base model's ability to address the Hubble tension. Second, we will constrain the time-evolution of the equation of state parameter introduced in Sec.~\ref{sec:evolving_w}, leveraging the enhanced constraining power of the updated datasets. Alternative extensions, including the trigger threshold parameter $H_*/m$, are briefly discussed towards the end.

\subsection{Implementation and Datasets}

The phenomenological NEDE model, implemented in the Boltzmann code \texttt{TriggerCLASS}\footnote{We use \texttt{TriggerCLASS} version 6.1, which is publicly available: \href{https://github.com/NEDE-Cosmo/TriggerCLASS.git}{https://github.com/NEDE-Cosmo/TriggerCLASS.git} and modifies \texttt{CLASS}~\cite{Blas:2011rf}.}, is interfaced with the MCMC sampler \texttt{MontePython}~\cite{Audren:2012wb,Brinckmann:2018cvx}. We scan the six $\Lambda$CDM parameters ($\omega_b$, $\omega_\mathrm{cdm}$, $h$, $\log_{10}(A_s)$, $n_s$, $\tau_{\rm reio}$) using the Metropolis-Hastings algorithm with standard prior ranges. The neutrino sector is modeled with two massless and one massive neutrino with mass $m_3=0.06 \mathrm{eV}$ and temperature $T_3=0.716 \mathrm{eV}$, amounting to $N_\mathrm{eff}=3.046$. In accordance with the discussion in Sec.~\ref{sec:model}, we sample the four NEDE parameters\footnote{We checked for our base model runs with NPIPE data that relaxing the upper prior boundary did not lead to any relevant changes in the parameter inference.} $f_\mathrm{NEDE} \in [0,0.3]$, $\log_{10}(z_*) \in [2.8,4]$, $ 3 w \in [1,3]$, and $\Omega_{\phi} \in [0,0.1]$  when testing the NEDE model in Sec.~\ref{sec:NEDE_NPIPE}. As explained in more detail in Sec.~\ref{sec:extensions}, in the case of the extended model, we add additional parameters allowing for a non-trivial time evolution of the equation of state parameter after the transition [see also \eqref{eq_Taylor}].

We use the following datasets for our MCMC analysis:

\begin{itemize}
\item \bf CMB\rm: Planck NPIPE data~\cite{Planck:2020olo} implemented through the \texttt{CamSpec}~\cite{Rosenberg:2022sdy,Efstathiou:2019mdh} likelihood code. This dataset updates the previous Planck 2018  high-$\ell$ dataset~\cite{Planck:2018vyg}. For comparison we also run an analysis using Planck 2018 high-$\ell$ data relying on the \texttt{Plik} likelihood code and another one using Planck NPIPE data implemented through the \texttt{HiLLiPoP} likelihood code~\cite{Tristram:2023haj}. In addition, all runs are complemented with the Planck 2018 lensing, and low-$\ell$ TT+EE CMB anisotropy measurements~\cite{Planck:2018vyg,Planck:2018lbu}.

\item \bf BAO\rm: Baryonic acoustic oscillation (BAO) measurements from the 6dF Galaxy Survey~\cite{Beutler:2011hx} (at $z=0.106$), SDSS DR7~\cite{Ross:2014qpa} (at $z=0.15$), and CMASS and LOWZ galaxy samples of BOSS DR12~\cite{BOSS:2016wmc} (at $z=0.38$, $0.51$, $0.61$). The latter includes a constraint on the growth rate $f \sigma_8$ from redshift space distortions.

\item \bf Pantheon+\rm: Catalog of 1550 Type Ia supernovae (SNe), constraining the shape of the expansion history over the redshift range $0.001 < z < 2.26$~\cite{Brout:2022vxf}.

\item \bf SH$_0$ES\rm: Constraint on the SNe Type Ia absolute magnitude implemented as a Gaussian prior, $M_b = -19.253 \pm 0.027$~\cite{Riess:2021jrx}.

\end{itemize}

\begin{figure}[tbp]
\centering
\includegraphics[width=\columnwidth*2/3]{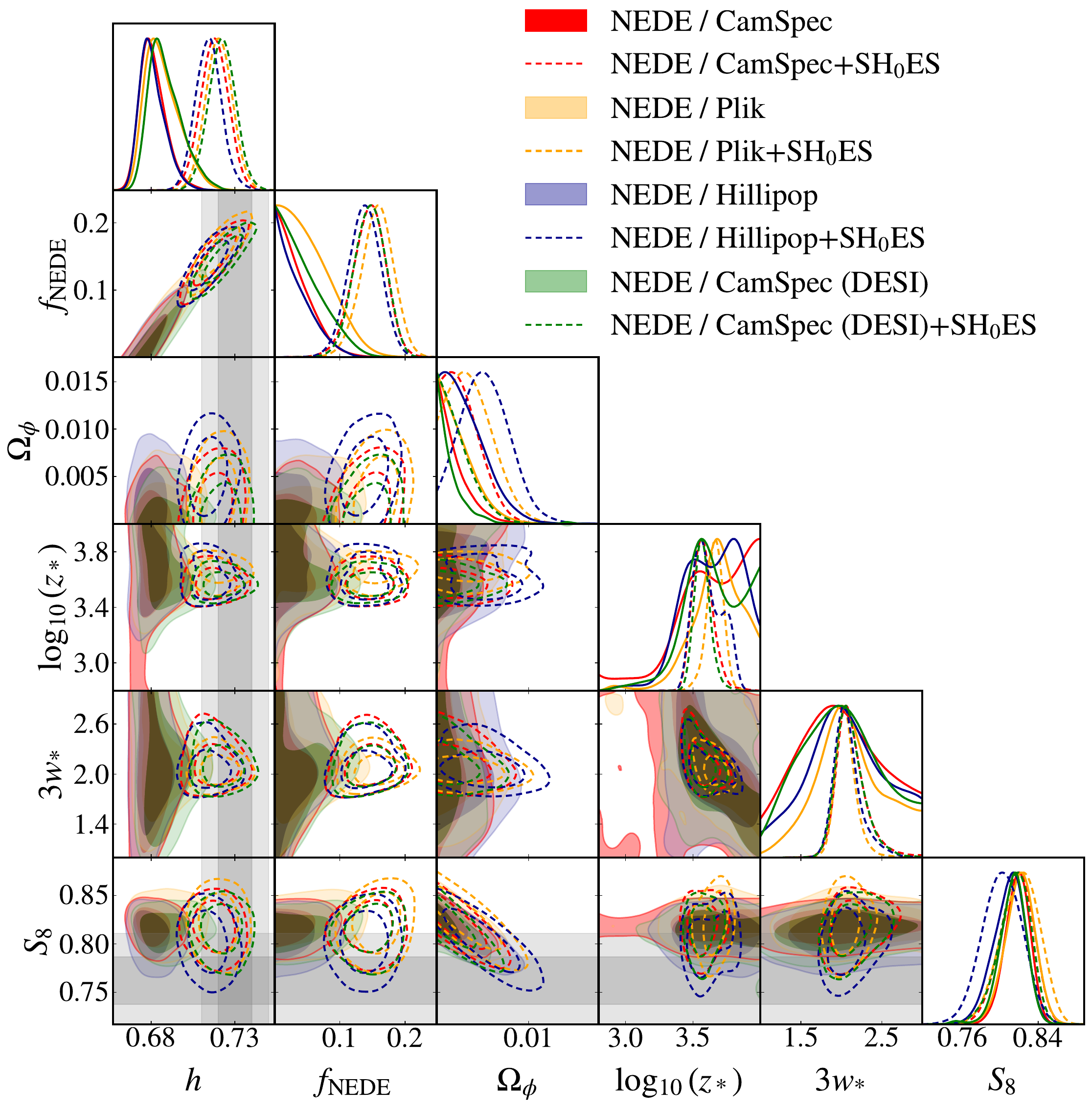}
\caption{Comparison of 2D posterior distributions reconstructed from either the \texttt{CamSpec}/NPIPE, \texttt{HiLLiPoP}/NPIPE, or \texttt{Plik}/Planck 2018 likelihoods (alongside our baseline datasets) in the base NEDE model. The green contours are obtained from an analysis where our baseline BAO data has been exchanged with DESI data. Here and henceforth, the horizontal and vertical gray bands show the $1\sigma$ and $2\sigma$ constraints on $S_8$ from~\cite{Joudaki:2019pmv} and on $H_0$ from~\cite{Riess:2021jrx}. Compared to Planck 2018, both NPIPE likelihoods lead to tighter constraints on $f_\mathrm{NEDE}$ and a later decay (closer to matter-radiation equality). The model's ability to address the $S_8$ tension is preserved for \texttt{CamSpec}/NPIPE and even enhanced for \texttt{HiLLiPoP}/NPIPE.}
\label{fig:NEDE_hillipop-vs-CamSpec}
\end{figure}

We will refer to BAO and Pantheon+ as our baseline dataset combination.
In this work, we also explore the impact of the BAO measurements from the recently released Dark Energy Spectroscopic Instrument (DESI) survey~\cite{DESI:2024mwx}. More specifically, for the runs labeled ``DESI'' we replace the BAO baseline dataset from above with DESI BAO data. To that end, we employ the same likelihood that has recently been used for constraining the axiEDE model in~\cite{Poulin:2024ken} and was build from the data in Tab.~1 of~\cite{DESI:2024mwx}.

All our MCMC analyses are done with and without including the SH$_0$ES prior on $M_b$. This in turn allows us to compute the \textit{difference of the maximum a posteriori} (DMAP) tension measure, defined as~\cite{Raveri:2018wln}
\begin{align}
Q_\mathrm{DMAP}(H_0) = \sqrt{ \chi^2_\mathrm{w\backslash\, H_0} -\chi^2_\mathrm{w\backslash o\, H_0} }\,.
\end{align} 
Here, $\chi^2$ represents the best-fit chi-squared with ($\mathrm{w\backslash\, H_0}$) and without ($\mathrm{w\backslash o\, H_0}$) including the SH$_0$ES measurement in the analysis. The DMAP tension is a quantitative estimate of the inconsistency between the SH$_0$ES measurement and the other datasets consisting of CMB, BAO, and Pantheon+ as described above. For Gaussian posteriors this measure is equivalent to the usual tension measure computed from the mean values and their confidence intervals. In particular if the two dataset combinations are consistent, their maximum likelihoods should be close, resulting in a small $Q_{\mathrm{DMAP}}$. In~\cite{Schoneberg:2021qvd} the DMAP measure was used to compare different models in their attempt to resolve the Hubble tension. The benefit of this method is that it is insensitive to prior volume effects, which are known to make the Gaussian tension less reliable for EDE-type models~\cite{Herold:2022iib,Cruz:2023cxy}.
We use the simulated-annealing optimizer \texttt{Procoli}~\cite{Karwal:2024qpt} to find the global maximum likelihoods value reported in our data tables. All chains satisfy a Gelman-Rubin convergence level of $|R-1| \lesssim 0.05$~\cite{Gelman:1992zz}. When plotting marginalized posteriors, we show the $68\%$ C.L. and $95\%$ C.L. contours. 
Finally, in order to assess the overall fit quality of a given model, we employ the Akaike Information Criterium (see also~\cite{Schoneberg:2021qvd}). It assigns a penalty of $2$ to the usual $\chi^2$ measure for each free parameter the model adds to the six $\Lambda$CDM parameters. More concretely, 
$
\Delta {\rm AIC} = \chi^2_{\rm min}({\rm NEDE}) - \chi^2(\Lambda{\rm CDM}) + 2( N_{p, \mathrm{NEDE}}  - N_{p, \Lambda{\rm CDM}}),
$
where $N_{p, \Lambda{\rm CDM}} = 6$ and $N_{p, \mathrm{NEDE}}$ is the number of free parameters in a given NEDE model ($N_{p, \mathrm{NEDE}} = 10$ for the base model).

\begin{figure}[tbp]
\centering
\includegraphics[width=\columnwidth*2/3]{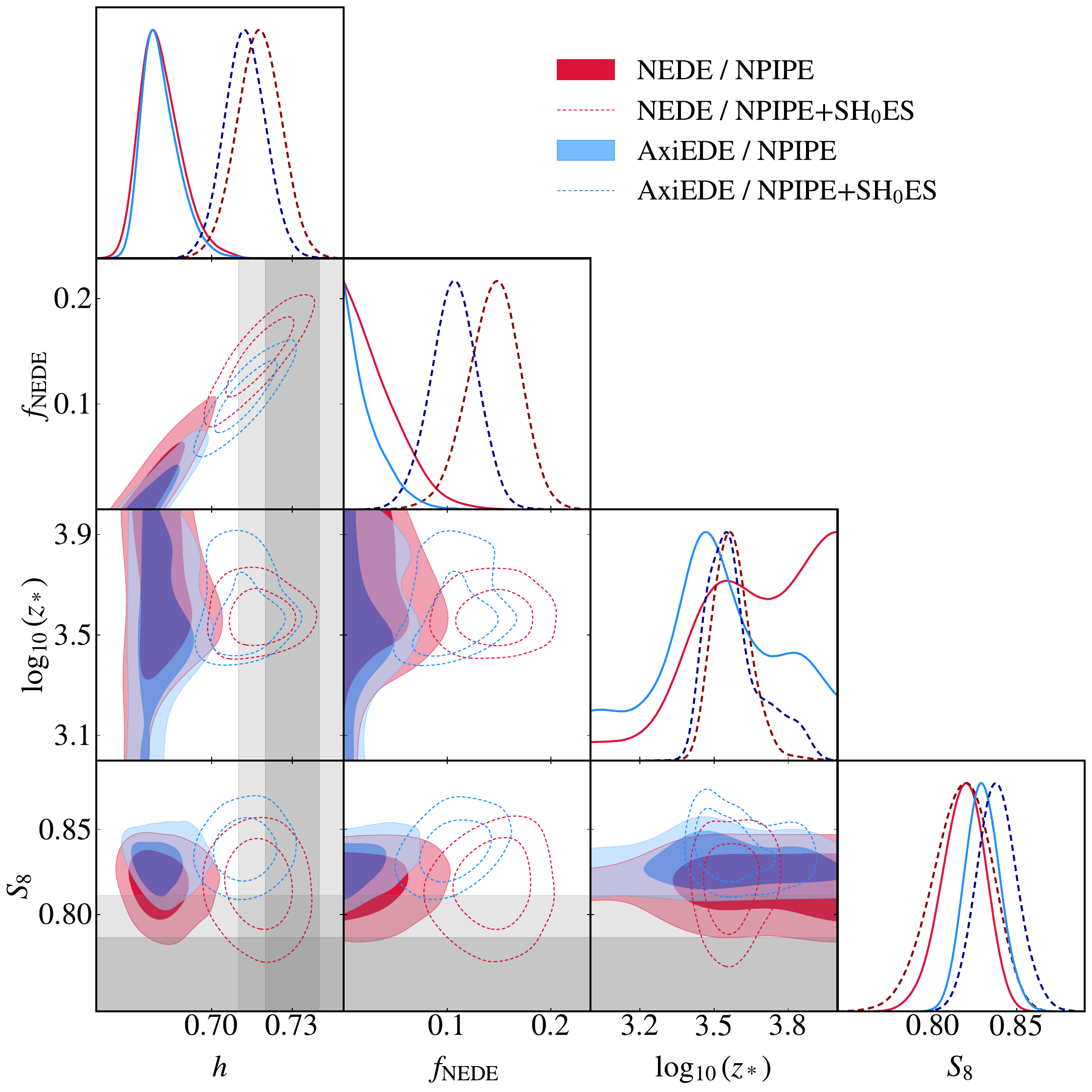}
\caption{Comparison of 2D posterior distributions in the NEDE and AxiEDE model obtained from our baseline datasets alongside the \texttt{CamSpec}/NPIPE likelihood. 
Both models respond to NPIPE similarly, with NEDE admitting overall larger values of $f_\mathrm{NEDE}$ and hence $H_0$. 
}
\label{fig:NEDE_vs_axEDE}
\end{figure}

\subsection{Confronting NEDE with NPIPE (and Pantheon+)}
\label{sec:NEDE_NPIPE}

\begin{table}[tbp]
\centering

\begin{tabular}{|c|c|c|c|c|}
\hline
& \multicolumn{2}{c|}{\texttt{CamSpec}/NPIPE} & \multicolumn{2}{c|}{\texttt{HiLLiPoP}/NPIPE }\\
\hline
SH$_0$ES prior? & no & yes & no & yes \\ 
\hline
\hline
$h$ 
& $0.6811(0.6786)^{+0.0058}_{-0.0090}$ 
& $0.7177(0.7208)\pm 0.0083$ 
& $0.6808(0.6818)^{+0.0054}_{-0.0090}$ 
& $0.7143(0.7179)\pm 0.0078$ 
\\
$f_{\rm NEDE}$
& $ < 0.0829(0.0139)$
& $0.145(0.151)^{+0.027}_{-0.023}$ 
& $ < 0.0852(0.0352)$ 
& $0.138(0.142)\pm 0.025$ 
\\
$\log_{10}(z_*)$
& unconstrained $ (3.999)$ 
& $3.574(3.579)^{+0.062}_{-0.086}$ 
& $3.64(3.82)^{+0.25}_{-0.19}$
& $3.61(3.56)\pm 0.11$ 
\\
$\Omega_{\phi}$
& $< 0.0053(0.0002)$ 
& $ < 0.0066(0.0027)$ 
& $< 0.0077 (0.0029)$ 
& $0.0053(0.0059)^{+0.0024}_{-0.0029}$ 
\\
$3w_*$
& unconstrained $(1.134)$  
& $2.13(2.04)^{+0.12}_{-0.21}$ 
& unconstrained $(1.79)$
& $2.08(2.05)^{+0.11}_{-0.19}$ 
\\
$10^{2}\omega{}_{b }$
& $2.228(2.218)^{+0.016}_{-0.019}$ 
& $2.259(2.25)^{+0.019}_{-0.022}$ 
& $2.236(2.24)^{+0.014}_{-0.021}$ 
& $2.278(2.269)^{+0.022}_{-0.027}$ 
\\
$\omega{}_{\rm cdm }$
& $0.1217(0.1212)^{+0.0015}_{-0.0028}$ 
& $0.1310(0.1317)\pm 0.0029$ 
& $0.1211(0.1214)^{+0.0017}_{-0.0028}$ 
& $0.1295(0.1298)\pm 0.0028$ 
\\
$n_{s }$
& $0.9682(0.9696)^{+0.0050}_{-0.0060}$ 
& $0.9853(0.9861)\pm 0.0058$ 
& $0.9678(0.9706)^{+0.0046}_{-0.0062}$ 
& $0.9855(0.9865)\pm 0.0059$ 
\\
$10^9 A_s$
& $2.110(2.103)^{+0.027}_{-0.033}$ 
& $2.150(2.147)^{+0.029}_{-0.032}$ 
& $2.112(2.114)\pm 0.028$ 
& $2.157(2.165)\pm 0.029$ 
\\
$\tau{}_{\rm reio }$
& $0.0561(0.0543)^{+0.0063}_{-0.0072}$ 
& $0.0571(0.0562)^{+0.0064}_{-0.0076}$ 
& $0.0588(0.0589)\pm 0.0062$ 
& $0.0614(0.0621)\pm 0.0063$ 
\\
$\Omega_m$
& $0.3122(0.3114)\pm 0.0053$ 
& $0.3010(0.2995)\pm 0.0051$ 
& $0.3129(0.3123)\pm 0.0057$ 
& $0.3038(0.3018)\pm 0.0055$ 
\\
$S_8$
& $0.820(0.83)^{+0.014}_{-0.012}$ 
& $0.819(0.819)^{+0.019}_{-0.016}$ 
& $0.811(0.817)^{+0.019}_{-0.014}$ 
& $0.803(0.795)\pm 0.021$ 
\\

\hline
$\Delta \chi^2({\rm NEDE}-\Lambda$CDM) & $-1.44$ & $-28.54$ &  $-2.4$ & $-30.4$ \\
\hline
$Q_{\rm DMAP}$ & \multicolumn{2}{c|}{3.5$\sigma$} & \multicolumn{2}{c|}{3.7$\sigma$} \\
\hline
\end{tabular}
\caption{Confidence intervals and $1\sigma$ error bars (best-fit in parenthesis) for the base NEDE model reconstructed from the \texttt{CamSpec}/NPIPE  and \texttt{HiLLiPoP}/NPIPE likelihoods (alongside our baseline datasets) with and without the SH$_0$ES prior on $M_b$.}  
\label{tab:npipe}
\end{table}

\begin{table}[!b]
\centering
\begin{tabular}{|c|c|c|c|c|}
\hline
& \multicolumn{2}{c|}{\texttt{Plik}/Planck 2018} & \multicolumn{2}{c|}{\texttt{CamSpec}/NPIPE (DESI)}\\
\hline
SH$_0$ES prior? & no & yes & no & yes \\ 
\hline
\hline
$h$
	 & $0.6867(0.6913)^{+0.0071}_{-0.013}$ 
	 & $0.7205(0.7216)\pm 0.0079$ 
	 & $0.6886(0.6839)^{+0.0070}_{-0.012}$ 
	 & $0.7223(0.7225)\pm 0.0080$ 
	 \\
$f_{\rm NEDE}$
	 &  $ <0.119(0.068)$ 
	 & $0.157(0.162)^{+0.026}_{-0.023}$ 
	 &  $ <0.109(0.019)$ 
	 & $0.146(0.145)\pm 0.024$ 
	 \\
$\log_{10}(z_*)$
	 & $3.67(3.72)^{+0.22}_{-0.13}$ 
	 & $3.662(3.678)^{+0.065}_{-0.055}$ 
	 & unconstrained $(4.0)$ 
	 & $3.570(3.585)^{+0.051}_{-0.065}$ 
	 \\
$\Omega_{\phi}$
	 &  $ <0.00629(0.0003)$ 
	 & $0.0038(0.0032)^{+0.0014}_{-0.0033}$ 
	 &  $ <0.00497(0.00008)$ 
	 &  $ <0.00589(0.00011)$ 
	 \\
$3w_{*}$
	 & $2.10(1.98)^{+0.38}_{-0.48}$ 
	 & $2.058(2.04)^{+0.098}_{-0.14}$ 
	 & $1.99(1.22)^{+0.44}_{-0.58}$ 
	 & $2.11(2.09)^{+0.13}_{-0.20}$ 
	 \\
$10^{-2}\omega{}_{b }$
	 & $2.261(2.266)^{+0.019}_{-0.026}$ 
	 & $2.307(2.311)\pm 0.024$ 
	 & $2.234(2.229)^{+0.016}_{-0.020}$ 
	 & $2.261(2.254)^{+0.019}_{-0.021}$ 
	 \\
$\omega{}_{\rm cdm }$
	 & $0.1234(0.1257)^{+0.0022}_{-0.0040}$ 
	 & $0.1325(0.1331)\pm 0.0030$ 
	 & $0.1217(0.1209)^{+0.0019}_{-0.0035}$ 
	 & $0.1304(0.1309)\pm 0.0027$ 
	 \\
$n_{s }$
	 & $0.9718(0.9769)^{+0.0057}_{-0.0080}$ 
	 & $0.9909(0.9934)\pm 0.0056$ 
	 & $0.9718(0.9738)^{+0.0056}_{-0.0071}$ 
	 & $0.9878(0.9899)\pm 0.0055$ 
	 \\
$10^9 A_s$
	 & $2.126(2.123)^{+0.029}_{-0.035}$ 
	 & $2.163(2.157)^{+0.029}_{-0.033}$ 
	 & $2.124(2.113)^{+0.031}_{-0.035}$ 
	 & $2.160(2.151)^{+0.030}_{-0.034}$ 
	 \\
$\tau{}_{\rm reio }$
	 & $0.0569(0.0548)^{+0.0067}_{-0.0076}$ 
	 & $0.0580(0.0565)^{+0.0068}_{-0.0077}$ 
	 & $0.0592(0.0577)\pm 0.0075$ 
	 & $0.0596(0.0578)^{+0.0066}_{-0.0079}$ 
	 \\
$\Omega_m$
	 & $0.3122(0.3107)\pm 0.0058$ 
	 & $0.3034(0.3032)\pm 0.0053$ 
	 & $0.3056(0.3063)\pm 0.0053$ 
	 & $0.2956(0.2941)\pm 0.0047$ 
	 \\
$S_8$
	 & $0.822(0.84)^{+0.017}_{-0.014}$ 
	 & $0.823(0.828)^{+0.022}_{-0.019}$ 
	 & $0.815(0.825)^{+0.015}_{-0.012}$ 
	 & $0.816(0.834)^{+0.019}_{-0.014}$ 
	 \\

\hline
$\Delta \chi^2({\rm NEDE}-\Lambda$CDM) & $ -1.94 $ & $-33.34$ & $ -2.5 $ & $-33.42$ \\

\hline
$Q_{\rm DMAP}$ & \multicolumn{2}{c|}{2.8$\sigma$} & \multicolumn{2}{c|}{2.6$\sigma$}\\
\hline
\end{tabular}
\caption{Confidence intervals and $1\sigma$ error bars (best-fit in parenthesis) for the base NEDE model reconstructed from the \texttt{Plik}/Planck 2018 and \texttt{CamSpec}/NPIPE likelihood together with our baseline datasets. Both analyses where performed with and without the SH$_0$ES prior on $M_b$. In the case of NPIPE, the BAO data is taken from the DESI survey resulting in the smalles residual tension.}
\label{tab:plik}
\end{table}

\begin{table*}[tbp]

\def\arraystretch{1.2}
\scalebox{1}{
\begin{tabular}{|l|c|c|c|c|c|c|c|c|}
\hline
 & \multicolumn{2}{c|}{\texttt{CamSpec}/NPIPE}  & \multicolumn{2}{c|}{\texttt{HiLLiPoP}/NPIPE} & \multicolumn{2}{c|}{ \texttt{Plik}/Planck 2018} & \multicolumn{2}{c|}{ \texttt{CamSpec}/NPIPE (DESI)} \\

\hline
SH$_0$ES prior? & no & yes & no & yes & no & yes & no & yes \\ 
\hline
\hline
{\textit{Planck}}~high$-\ell$ TTTEEE & 11237.56 & 11245.36& 30494.32&30501.97& 2344.87 & 2350.34 & 11238.2 & 11244.44\\
{\textit{Planck}}~low$-\ell$ TT  & 22.61 & 21.39& 21.95& 21.04&  21.98 & 20.73 & 22.02 & 20.98 \\
{\textit{Planck}}~low$-\ell$ EE & 396.07 & 396.41 &  $-$& $-$ & 396.09 & 396.28 & 396.73 & 396.61 \\
{\textit{Lollipop} EB} & $-$ & $-$ & 125.66& 126.78& $-$ &  $-$ & $-$ & $-$ \\
{\textit{Planck}}~lensing & 9.27 &9.84 &9.13 & 9.42& 9.1 & 9.85 & 9.54 & 9.83 \\
BOSS BAO low$-z$ &1.13 & 1.94& 1.09& 1.78&1.11 & 1.53 & $-$ & $-$ \\ 
BOSS BAO/$f\sigma_8$ DR12& 6.82& 6.04 & 5.91& 6.00& 7.19 & 6.22 & $-$ & $-$\\
DESI BAO & $-$ & $-$ & $-$ & $-$ & $-$ & $-$ & 16.13 & 12.92 \\
Pantheon+ & 1411.06& 1413.16& 1411.08& 1412.87 & 1411.18 & 1412.4 & 1411.81 & 1414.36 \\
SH$_0$ES & $-$ & 2.66& $-$ & 3.91& $-$ & 2.27 & $-$ & 2.34 \\
\hline
total $\chi^2_{\rm min}$ &13084.52 &13096.83 & 32070.27 & 32083.69 &  4191.52 & 4199.21 & 13094.44 & 13101.48\\
$\Delta \chi^2_{\rm min}({\rm NEDE}-\Lambda{\rm CDM})$ &  $-1.44$ & $-28.54$ & $-2.4$ & $-30.04$ & $-1.94$ & $-33.34$ & -2.5 & -33.42\\
\hline
$Q_{\rm DMAP}(\Lambda \mathrm{CDM})$&\multicolumn{2}{c|}{6.3$\sigma$} &\multicolumn{2}{c|}{6.4$\sigma$} & \multicolumn{2}{c|}{$6.3\sigma$}  & \multicolumn{2}{c|}{$6.2\sigma$}   \\
\hline    
$Q_{\rm DMAP}({\rm NEDE})$&\multicolumn{2}{c|}{3.5$\sigma$} &\multicolumn{2}{c|}{3.7$\sigma$} & \multicolumn{2}{c|}{$2.8\sigma$ } & \multicolumn{2}{c|}{$2.6\sigma$ }\\
\hline
AIC& 6.56&-20.54&5.6&-22.04&6.06&-25.34 & 5.5 & -25.42\\
\hline
\end{tabular}
}
\caption{Best-fit $\chi^2$ per experiment (and total) for the base NEDE model when fitted to our baseline datasets supplemented  either by the \texttt{CamSpec}/NPIPE, \texttt{HiLLiPoP}/NPIPE or \texttt{Plik}/Planck 2018 likelihood. In the fourth column with \texttt{CamSpec}/NPIPE, the BAO data is taken from the DESI survey. We compare the fits with and without the SH$_0$ES prior on $M_b$. We report the $\Delta \chi^2_{\rm min}\equiv\chi^2_{\rm min}({\rm NEDE})-\chi^2_{\rm min}(\Lambda{\rm CDM})$, the tension metric $Q_{\rm DMAP}$ and the value of $\Delta {\rm AIC}$ (for reference, $Q_\mathrm{DMAP}$ values for $\Lambda$CDM are also shown.)
}
\label{tab:chi2_Planck_EDE}

\end{table*}
Previous analyses of NEDE relied on Planck 2018 data implemented through the likelihood code \texttt{Plik} and the Pantheon dataset. This consistently reduced the Hubble tension below $2 \sigma$, as was confirmed by independent works (see for example~\cite{Niedermann:2020dwg,Schoneberg:2021qvd,Cruz:2023cxy}). Recently, these results were strengthened by including ground-based CMB data~\cite{Cruz:2022oqk} and relaxing the assumption $\Omega_\phi \ll 1$~\cite{Cruz:2023lmn}, corresponding to a small trigger contribution to dark matter.

In Tab.~\ref{tab:npipe} alongside Fig.~\ref{fig:NEDE_hillipop-vs-CamSpec}, we report and compare our findings using NPIPE data. As mentioned before, all analyses include also (for the first time) Pantheon+. We use two different likelihood codes for NPIPE, \texttt{CamSpec} and \texttt{HiLLiPoP}, to double-check the consistency of our findings and to test to what extent they are sensitive to the likelihood code. 

We find that Cold NEDE reduces the $Q_\mathrm{DMAP}$ tension from $6.3 \sigma$ within $\Lambda$CDM to $2.8 \sigma$ (\texttt{Plik}/Planck 2018),  $3.5 \sigma$ (\texttt{CamSpec}/NPIPE), and $3.7 \sigma$ (\texttt{HiLLiPoP}/NPIPE).  When comparing this with the previous works mentioned above, the residual tension with the newer datasets is larger by up to $1.5 \sigma$. However, this comparison has to be taken with a grain of salt: The previous analyses, which relied on Pantheon and \texttt{Plik}/Planck 2018, also gave rise to a weaker $\Lambda$CDM tension. For example, \cite{Cruz:2023lmn} (\texttt{Plik}/Planck 2018+BAO+Pantheon) used a prior on $H_0$ that corresponded to a $\Lambda$CDM $Q_\mathrm{DMAP}$ tension of $4.7 \sigma$, which was reduced to $1.4 \sigma$ within Cold NEDE. In other words, the level of relative improvement is similar to what we find here for \texttt{Plik}/Planck 2018 (in both cases, the relaxation of the tension is larger than $3\sigma$). Exchanging \texttt{Plik}/Planck 2018 with \texttt{CamSpec}/NPIPE diminishes this improvement slightly to around $2.5 \sigma$. In all cases, we find a significant $\chi^2$ improvement of at least 28 units. While compared to previous analyses, the base model remains similarly efficient at reducing the Hubble tension, it cannot fully resolve it anymore for our baseline datasets (provided we accept that the Hubble tension has passed the $Q_\mathrm{DMAP}$ threshold of $6 \sigma$). 
The observation that NPIPE data slightly disfavors the NEDE model compared to \texttt{Plik}/Planck 2018 is also supported by the contours in Fig.~\ref{fig:NEDE_hillipop-vs-CamSpec}. Comparing the orange (\texttt{Plik}/Planck 2018) with the red (\texttt{CamSpec}/NPIPE) contours, it is apparent that the evidence for NEDE is diminished. To be specific, we find $f_\mathrm{NEDE} < 8.3 \% $ ($95\%$ C.L.) for NPIPE and $f_\mathrm{NEDE} < 11.9 \% $ ($95\%$ C.L.) for Planck 2018.\footnote{We stress that this should be understood as a relative comparison as both values are known to be affected by prior volume effects that become sizeable in the $f_\mathrm{NEDE} \to 0$ limit~\cite{Niedermann:2020dwg,Murgia:2020ryi}. A profile likelihood for $f_\mathrm{NEDE}$ can be expected to support a non-vanishing fraction of $f_\mathrm{NEDE}$~\cite{Herold:2022iib,Cruz:2023cxy}. Moreover, it should be noted that we are comparing $f_\mathrm{NEDE}$ at the time of decay, which is not exactly the same in Plik and NPIPE fits. Specifically, the later decay found for NPIPE might in part explain the difference.}  This result is reproduced by the blue contours (\texttt{HiLLiPoP}/NPIPE). Overall, we find good agreement between the \texttt{CamSpec} and \texttt{HiLLiPoP} likelihood code.\footnote{In addition, we also investigated the internal consistency of the \texttt{CamSpec} likelihood within NEDE by combining our baseline datasets with only the NPIPE TT and the TEEE data. We find consistency between these reduced NPIPE runs and the full one. In particular, we find that polarization data alone has limited constraining power, leaving, for example, $w_\mathrm{NEDE}$ and $f_\mathrm{NEDE}$ rather unconstrained. 
} 
The only exception, albeit not significant, concerns the constraint on trigger dark matter $\Omega_\phi$, which is $\Omega_\phi < 0.8 \%$ ($95\%$ C.L.) for \texttt{HiLLiPoP} and strengthens to $\Omega_\phi < 0.5 \%$ for \texttt{CamSpec}. In that context, it is important to note that in both cases the model preserves its ability to reduce the value of $S_8$, thereby keeping the Gaussian $S_8$ tension below $2.1 \sigma$ for all analyses performed here (using the weak lensing constraint in \cite{Joudaki:2019pmv} as a reference value). Moreover, we find in agreement with the previous analysis in \cite{Cruz:2023lmn} that including the SH$_0$ES prior further reduces the tension to below $2\sigma$. This trend is particularly pronounced in the case of \texttt{HiLLiPoP}.

Another difference between Planck 2018 and NPIPE data arises for the decay time. When including the SH$_0$ES prior on $H_0$, the NPIPE analysis yields $z_* \approx  {3600} $, which is a relatively low value close to radiation-matter equality when compared to the Planck 2018 analysis, which yields $z_*\approx {4600}$. Again, this difference is not very significant but as we will argue later in Sec.~\ref{sec:evolving_w} it has implications for the extended model. On the other hand, without the prior on $H_0$, the decay redshift $z_*$ is less constrained (as is expected for a lower fraction $f_\mathrm{NEDE}$).

Determining the origin of the difference between both datasets is nontrivial. To obtain more insight into this, in App.~\ref{app:profile} we construct likelihood curves for both datasets and compare them. Interestingly we observe a bi-modality feature in the NPIPE data, which is absent for Planck 2018.

For the first time, we also tested the Cold NEDE model against the recently released DESI BAO dataset. To that end, we performed a combined analysis with \texttt{CamSpec}/NPIPE and Pantheon+ (green contour in Fig.~\ref{fig:NEDE_hillipop-vs-CamSpec}). Notably, this leads with $2.6 \sigma$ to the smallest residual tension in this work. A similar tendency of DESI data to relax constraints on $H_0$ has been observed recently for the AxiEDE model~\cite{Qu:2024lpx,Poulin:2024ken}. This is further confirmed through the likelihood profiles provided in App.~\ref{app:profile}. Notably, when combining with SH$_0$ES, we find a $~6\sigma$ evidence for a non-vanishing fraction of NEDE, a significance level that had not been obtained in previous analyses.

In Fig.~\ref{fig:NEDE_vs_axEDE}, we compare Cold NEDE (in red) and AxiEDE (in blue). We find that both models are similarly affected by NPIPE data, although the constraints on $f_\mathrm{NEDE}$ (and hence $H_0$) are somewhat weaker, leading to a marginally smaller residual tension in the case of Cold NEDE (in~\cite{Efstathiou:2023fbn} a residual $3.7 \sigma$ tension was reported for \texttt{CamSpec}). This raises the expectation that newer CMB datasets will be increasingly capable of distinguishing between the two models (see also Refs.~\cite{Poulin:2021bjr,Poulin:2023lkg} for other comparisons). As mentioned before, while both models share similarities on background level, differences arise predominantly on perturbation level. To be specific, the perturbations in AxiEDE correspond to perturbations of a homogeneous scalar field background and are described by the perturbed Klein-Gordon equation, whereas perturbations in Cold NEDE, are perturbations of a barotropic fluid with an amplitude that, on sub-horizon-scales, is controlled by the momentum-dependent perturbations of a subdominant scalar field evaluated at $\tau_*$.

Moreover, the trigger in Cold NEDE acts like a fuzzy dark matter component at late times. This in turn reduces the value of $S_8$ and, as mentioned before, this effect correlates positively with $f_\mathrm{NEDE}$). A similar mechanism is not present in AxiEDE where the value of $S_8$ tends to be larger, a difference that can be seen in the lowest row of Fig.~\ref{fig:NEDE_vs_axEDE}.

\subsection{Constraining a Time-Dependent Equation of State}
\label{sec:extensions}

\begin{figure}[tbp]
\centering
\subfigure{
\adjustbox{valign=c}{\includegraphics[width=0.5\textwidth]{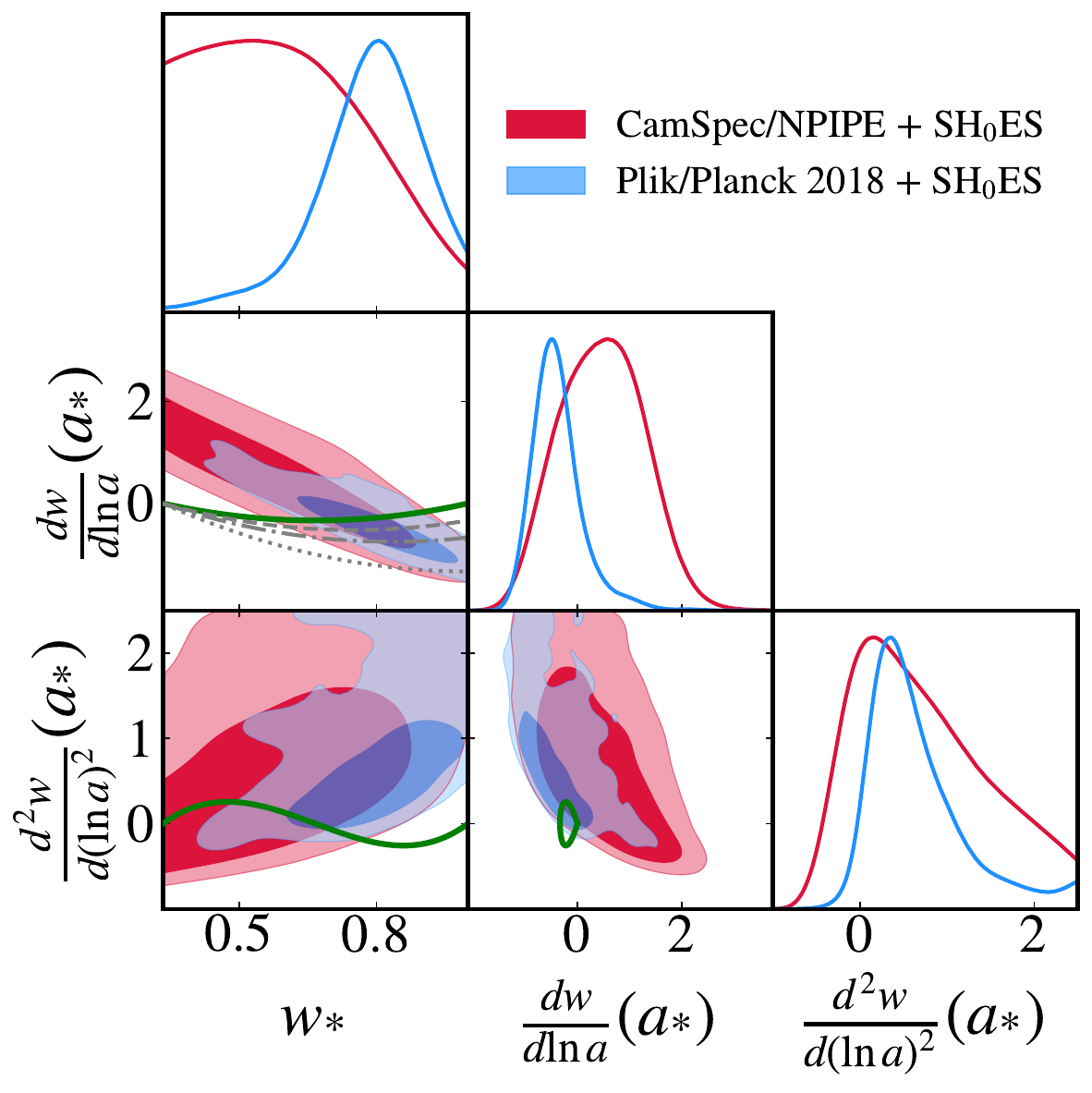}}
\label{fig:w_w1_w2_1}
}
\hfill
\subfigure{
\adjustbox{valign=c}{\includegraphics[width=0.333\textwidth]{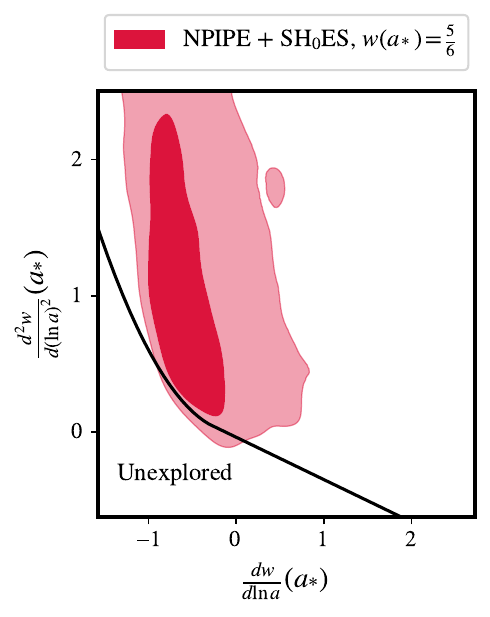}}
\label{fig:w_w1_w2_2}
}
\caption{\textbf{Left:} Comparison of the 2D posterior distributions obtained for the extended NEDE model with a time-dependent equation of state using either the \texttt{CamSpec}/NPIPE (red) or the \texttt{Plik}/Planck 2018 (blue) likelihood (alongside our baseline datasets and the SH$_0$ES prior on $M_b$). The green curve shows the prediction for the two-fluid mixture in \eqref{eq_mixture_dw} and \eqref{eq_mixture_d2w} (stiff fluid and radiation). The gray lines in the $w$ vs $dw/d\ln a$ plane show the case when the stiff fluid decays into radiation, where the dashed, dash-dotted and dotted lines correspond to $Q_{*} = 1/2$, $Q_{*} = 1$ and  $Q_{*} = 2$, respectively. \textbf{Right:} The distribution for the first two derivatives of the equation of state parameter restricted to $w_{\mathrm{NEDE}}(a_*) = 5/6$, demonstrating partial insensitivity to the value of the second derivative (in the case of \texttt{CamSpec}/NPIPE). The black line corresponds to $y=0$. In the region below it, $w$ would turn negative at a later time and, hence, that parameter region is left unexplored.}
\label{fig:w_w1_w2}
\end{figure}

Here, we summarize the results of the MCMC analysis for the extended NEDE model with a time-dependent equation of state parameter. Before doing so, we mention a technical subtlety that arises when implementing the Taylor ansatz \eqref{eq_Taylor} in \texttt{TriggerCLASS}. As it was explained in section~\ref{sec:extension}, a lower bound should be imposed on the value of the equation of state parameter, specifically $w(a)>0$, at any time between the transition and today. Implementing this bound directly as prior ranges for the derivatives is complicated, since these ranges are then no longer constant. On the other hand, in terms of the auxiliary variable $y$, defined in~(\ref{eq:y}), the requirement $w(a)>0$ (for $a<a_0\equiv1$) is equivalent to $y>0$. We, accordingly, perform a parameter fit to the extended model using the auxiliary variable $y$ instead of the first derivative. To be specific, we use boundaries $[0, 3] $ and $[-1, 2.5]$ for $y$ and $d^2w_{\mathrm{NEDE}}/d (\ln a)^2 |_{*}$, respectively. We note that \eqref{eq:y} is a simple linear shift, and we, therefore, do not expect posteriors to depend on that choice.

The results from this analysis for the Taylor coefficients $w_*$, $dw/d\ln a|_*$, and  $d^2w/d(\ln a)^2|_*$ are shown in Fig.~\ref{fig:w_w1_w2}. While we performed runs with and without a prior on $H_0$, in this section, we always cite the values and show the contours for the analysis with the SH$_0$ES prior included where the new feature is maximally constrained. The red contours correspond to the \texttt{{CamSpec}}/NPIPE dataset and the blue ones to \texttt{Plik}/Planck 2018. Since the \texttt{HiLLiPoP}/NPIPE analysis was fully compatible with \texttt{CamSpec}/NPIPE for the base model, we refrained from doing further comparisons within the extended models.

In the $w$ vs $dw/d\ln a$ plane of Fig.~\ref{fig:w_w1_w2}, we observe an approximate degeneracy between these two parameters. A larger initial value of $w$ can be compensated by a more negative derivative. A similar degeneracy is observed in the other two panels that depict the $w$ vs $d^2 w / d (\ln a)^2$ and the $dw/d\ln a$ vs $d^2 w / d (\ln a)^2$ planes. There, however, one finds that the width of the degeneracy band increases towards large values of $w$. This feature reflects the fact that in that parameter region, the second derivative is less important for the parametrization of $w(a)$ and does not impact cosmological observables much. 
At the same time, there is a relatively rigid bound from below corresponding to $d^2 w / d (\ln a)^2 \gtrsim 0$ for both data set combinations. This, however, is due to the fact that in our simplified parametrization (\ref{eq_Taylor}), that parameter region would lead to negative $w(a)$ at late times which is excluded by construction. This is demonstrated in the right panel of Fig.~\ref{fig:w_w1_w2} for the fiducial value $w_{\mathrm{NEDE}}(a_*) = 5/6$. In the region below the black line, $w$ turns negative before today. We reiterate that this is an artifact of the truncation in \eqref{eq_Taylor}. In particular, including higher-order terms is expected to relax that bound. This, however, is not expected to change the results for $dw/d\ln a$, as data can only constrain $w(a)$ over a limited redshift range close to $z_*$.

\begin{figure}[tbp]
\centering
\includegraphics[width=\columnwidth*11/15]{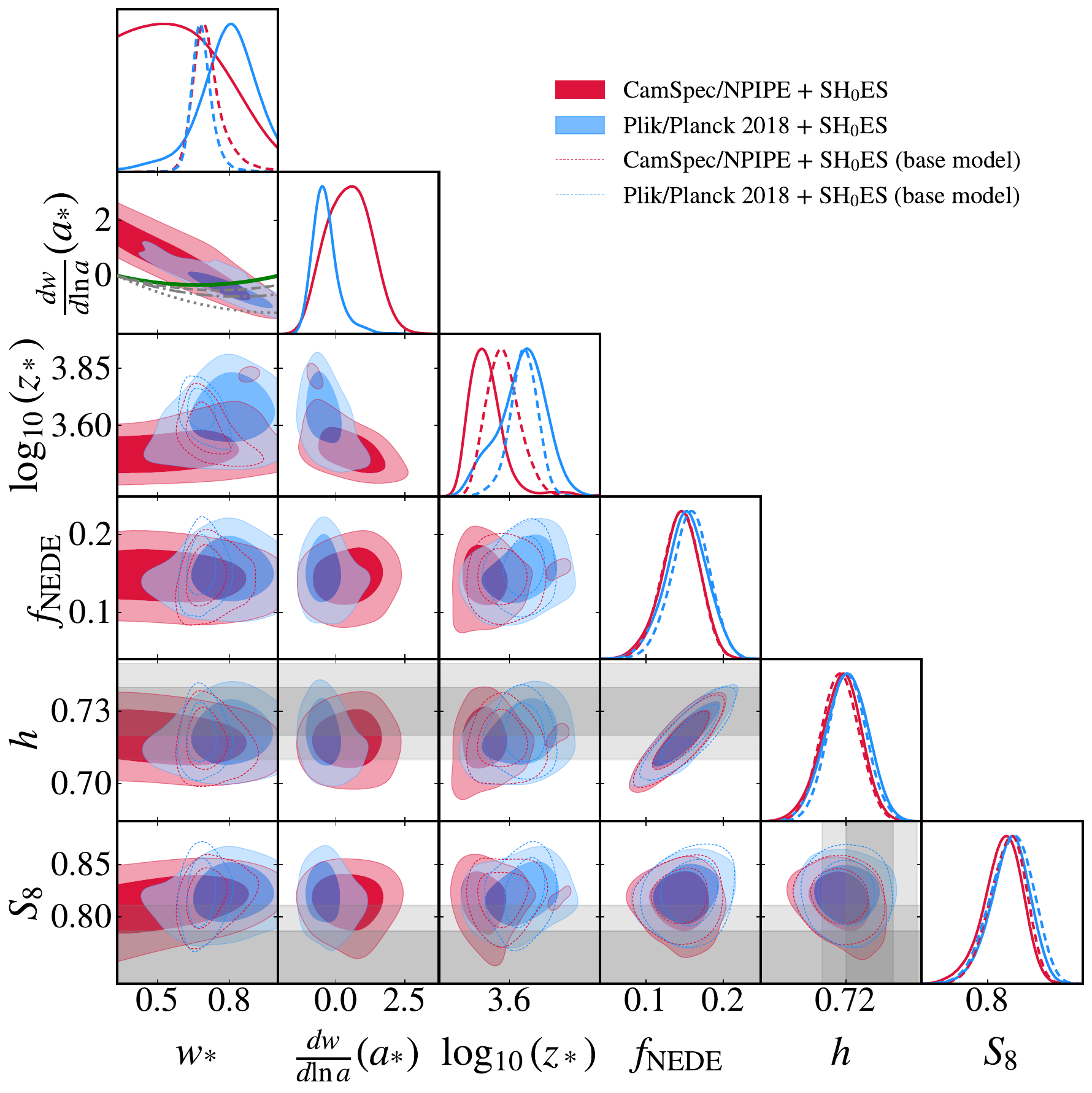}
\caption{Comparison of 2D distributions constructed using either the \texttt{CamSpec}/NPIPE (red)  or \texttt{Plik}/Planck 2018 (blue) datasets (alongside our baseline datasets and the SH$_0$ES prior on $M_b$) for the extension of the  NEDE model with a time-dependent equation of state (filled contours with solid lines) as well as for the base model (dashed lines). 
}
\label{fig:w_triangle}
\end{figure}

Having discussed the self-consistency of the Taylor ansatz~\eqref{eq_Taylor}, we move to discussing the entire data fit. To that end, we report the confidence intervals and best-fit values in Tab.~\ref{tab:extension_bf}, alongside the $\chi^2$ values in Tab.~\ref{tab:extension_chi2}. The corresponding marginalized posterior distributions are depicted in Fig.~\ref{fig:w_triangle} for the relevant NEDE parameters (together with $h$). The solid contours correspond to the extended model, and the dashed lines to the base model (with a constant equation of state parameter). 

Comparing \texttt{Camspec}/NPIPE and \texttt{Plik}/Planck 2018, we observe a preference towards smaller redshift $z_*$ in NPIPE for both the base model and the extension, moving the transition closer to matter-radiation equality, in agreement with the discussion in Sec.~\ref{sec:NEDE_NPIPE}. Moreover, we find that for \texttt{Camspec}/NPIPE, $w_*$ can take values in the whole range $1/3 < w_* < 1$, whereas the lower bound is tighter for Planck 2018 (excluding values $w_* \lesssim 0.5$). We attribute this to the fact that NPIPE prefers a later decay time. To be specific, for a given value of $w_*$,  the ratio $\rho / \rho_\mathrm{tot}$ falls off faster if the phase transition occurs closer to matter-radiation equality (since $\rho_\mathrm{tot}$ transitions from a $1/a^4$ to a $1/a^3$ scaling). As a result, for a later decay time, smaller values of $w_*$ (which correspond to a larger $dw/d(\ln a)|_*$) become accessible without enhancing the fraction of NEDE.

\begin{table}[tbp]
\centering

\begin{tabular}{|c|c|c|c|c|}
\hline
  & \multicolumn{2}{c|}{\texttt{CamSpec}/NPIPE}  &  \multicolumn{2}{c|}{\texttt{Plik}/ Planck 2018} \\
  \hline
SH$_0$ES prior? & no & yes & no & yes\\ 
\hline
\hline

$h$
& $0.6794(0.6826)^{+0.0048}_{-0.0095}$ 
& $0.7180(0.7233)^{+0.0089}_{-0.0077}$ 
& $0.6792(0.6786)^{+0.0047}_{-0.0079}$ 
& $0.7203(0.7225)\pm 0.0088$ 
\\

$f_{\mathrm{NEDE}}$
&  $ <0.0878(0.0316)$ 
& $0.145(0.156)^{+0.027}_{-0.023}$ 
&  $ <0.0776(0.0308)$ 
& $0.152(0.164)\pm 0.027$ 
\\
$\log_{10}(z_*)$
& unconstrained $(4.0)$ 
& $3.501(3.536)^{+0.046}_{-0.087}$ 
& $3.54(3.73)^{+0.27}_{-0.22}$ 
& $3.65(3.71)^{+0.12}_{-0.084}$ 
\\
$\Omega_{\phi}$
&  $ <0.00432(0.00051)$ 
& $ < 0.0077 (0.004)$ 
&  $ <0.00507(10^{-5})$ 
&  $ <0.00712(0.00174)$ 
\\
$3w_{*}$
&  unconstrained $(1.87)$ 
&  unconstrained $(1.61)$ 
&  unconstrained $(1.01)$ 
& $2.35(2.25)^{+0.39}_{-0.26}$ 
\\
$\frac{d w}{d\ln a}(a_{*})$
& $0.95(-0.56)^{+0.70}_{-1.7}$ 
& $0.46(0.57)\pm 0.81$ 
& $1.11(1.2)^{+0.81}_{-1.9}$ 
& $-0.38(-0.27)^{+0.31}_{-0.52}$ 
\\
$\frac{d^2 w}{d (\ln a)^2 }({a_{*}})$
& unconstrained $(0.65)$ 
& $0.69(-0.08)^{+0.51}_{-0.97}$ 
& unconstrained $(2.16)$ 
& $0.65(0.14)^{+0.24}_{-0.61}$ 
\\
$10^{2}\omega{}_{b }$
& $2.227(2.225)^{+0.016}_{-0.019}$ 
& $2.262(2.262)\pm 0.020$ 
& $2.250(2.255)\pm 0.019$ 
& $2.298(2.301)\pm 0.024$ 
\\
$\omega{}_{\rm cdm }$
& $0.1212(0.1225)^{+0.0011}_{-0.0026}$ 
& $0.1305(0.1313)\pm 0.0029$ 
& $0.1215(0.1222)^{+0.0013}_{-0.0025}$ 
& $0.132(0.1344)\pm 0.0033$ 
\\
$n_{s }$
& $0.9671(0.9733)^{+0.0043}_{-0.0063}$ 
& $0.9835(0.9881)^{+0.0060}_{-0.0053}$ 
& $0.9669(0.9693)^{+0.0044}_{-0.0052}$ 
& $0.9874(0.9919)\pm 0.0060$ 
\\
$10^9 A_s$
& $2.109(2.106)^{+0.027}_{-0.034}$ 
& $2.156(2.157)^{+0.029}_{-0.034}$ 
& $2.118(2.114)^{+0.028}_{-0.032}$ 
& $2.158(2.147)^{+0.028}_{-0.031}$ 
\\
$\tau{}_{\rm reio }$
& $0.0559(0.0532)^{+0.0061}_{-0.0072}$ 
& $0.0564(0.0561)^{+0.0067}_{-0.0079}$ 
& $0.0568(0.0543)^{+0.0066}_{-0.0078}$ 
& $0.0575(0.0555)^{+0.0064}_{-0.0073}$ 
\\
$\Omega_m$
& $0.3124(0.3111)\pm 0.0055$ 
& $0.3006(0.2983)\pm 0.0056$ 
& $0.3140(0.3144)\pm 0.0057$ 
& $0.3019(0.3032)^{+0.0049}_{-0.0056}$ 
\\
$S_8$
& $0.82(0.833)^{+0.013}_{-0.011}$ 
& $0.814(0.809)^{+0.021}_{-0.016}$ 
& $0.822(0.839)^{+0.015}_{-0.013}$ 
& $0.821(0.836)^{+0.020}_{-0.017}$ 
\\
\hline 
$\Delta \chi^2({\rm NEDE}-\Lambda$CDM) & $ -2.12 $ & $ -29 $ & $-1.41$ & $-35$ \\
\hline
$Q_{\rm DMAP}$ & \multicolumn{2}{c|}{3.5$\sigma$} & \multicolumn{2}{c|}{2.3$\sigma$} \\
\hline
\end{tabular}
\caption{Confidence intervals and $1\sigma$ error bars (best-fit in parenthesis) for the extended NEDE model with a time-dependent equation of state parameter, reconstructed from our baseline datasets together with either \texttt{CamSpec}/NPIPE or \texttt{Plik}/ Planck 2018  with and without the SH$_0$ES prior on $M_b$. 
}
\label{tab:extension_bf}
\end{table}

\begin{table}[!b]
\centering

\begin{tabular}{|l|c|c|c|c|}
\hline
 & \multicolumn{2}{c|}{\texttt{CamSpec}/NPIPE}  &  \multicolumn{2}{c|}{\texttt{Plik}/Planck 2018} \\
\hline
SH$_0$ES prior? & no & yes & no & yes\\ 
\hline
\hline
{\textit{Planck}}~high$-\ell$ TTTEEE & 11237.49 & 11245.84 & 2346.19 & 2348.64\\
{\textit{Planck}}~low$-\ell$ TT  & 22.17 &  21.25 & 21.34 & 20.83 \\
{\textit{Planck}}~low$-\ell$ EE & 395.89 & 396.34  & 396.04 & 396.12 \\
{\textit{Planck}}~lensing & 9.22 & 9.64 &  8.88 & 9.49 \\
BOSS BAO low$-z$ & 1.14 & 2.04 & 0.95 & 1.53 \\ 
BOSS BAO/$f\sigma_8$ DR12 & 6.81 & 5.99 & 7.95 & 6.44 \\
Pantheon+ & 1411.10 & 1413.39 & 1410.7 & 1412.39 \\
SH$_0$ES & $-$ & 1.9 & $-$ &  2 \\
\hline
total $\chi^2_{\rm min}$ & 13083.84 & 13096.39 &  4192.05  & 4197.44 \\
$\Delta \chi^2_{\rm min}({\rm NEDE}-\Lambda{\rm CDM})$ &  $ -2.12 $ & $ -29 $ & $-1.41$ & $-35$ \\
\hline
$Q_{\rm DMAP}$&\multicolumn{2}{c|}{3.5$\sigma$} & \multicolumn{2}{c|}{ 2.3$\sigma$} \\
\hline
AIC& 9.88&-17&10.59&-23\\
\hline
\end{tabular}
\caption{Best-fit $\chi^2$ per experiment (and total) for the extension of the NEDE model with a time-dependent equation of state, when fit to the our baseline datasets, with either the \texttt{CamSpec}/NPIPE or the \texttt{Plik}/ Planck 2018 likelihood. We compare the fits with and without the SH$_0$ES prior on $M_b$.}
\label{tab:extension_chi2}
\end{table}

We observe that Planck 2018 data favors the new feature more than NPIPE data. To be specific, for Planck 2018, we obtain $dw/d(\ln a) = {-0.38^{+0.31}_{-0.52}}$ ($68\%$ C.L.), corresponding to a slight preference for a negative derivative, whereas NPIPE with $dw/d(\ln a) = {0.95^{+0.70}_{-1.7}}$ ($68\%$ C.L.) does not show any preference for a non-vanishing value. Accordingly, only the Planck 2018 run improves the fit ( $\Delta \chi^2 = -4$) when the new feature is turned on. This is also reflected in the $Q_\mathrm{DMAP}$ tension, which for Planck 2018 data is lowered from $2.8 \sigma$ to  $2.3 \sigma$ (with the main improvement arising from the high-$\ell$ Planck data). On the other hand, the fit and residual tension remains unchanged for NPIPE data. Also, the constraints on $f_\mathrm{NEDE}$ are largely unchanged by including the new feature.

Let us emphasize that the purpose of the extended model is to get an understanding of what fluid features are compatible with data. Ultimately, the details of the fluid, including the time-dependence of its equation of state parameter, will be determined by the details of the underlying model (for example the decay channels of the tunneling field as discussed in Sec.~\ref{sec:micro}). After all, from a pure data fitting perspective, the mild improvement in $\chi_{\rm min}^2$ would not justify adding two more parameters. However, for the purpose of this analysis, this result simply shows that the constraining power current datasets have over the fluid details is still limited, and thus the assumption made in previous works that $w_\mathrm{NEDE}$ can be treated as a constant appears justified. We nevertheless evaluated the AIC criterion to provide a more quantitative measure of the overall fit quality in relation to the number of model parameters in Tab~\ref{tab:extension_bf}. While $\Delta {\rm AIC}$ is negative in all cases where the SH$_0$ES prior is included, it is indeed smaller for the base model.

The green curves show the two-fluid constraints (\ref{eq_mixture_dw}) and (\ref{eq_mixture_d2w}) for a mixture with $w_1=1$ and $w_2=1/3$. Each point on the curve corresponds to a certain choice of the initial ratio $r_f$ defined in \eqref{eq:r}. These results suggests that the mixture model is consistent with both dataset combinations when the two components initially have similar energy densities or the stiff component dominates. For example, for the case of the \texttt{Plik}/ Planck 2018 likelihood, the preferred values of $w_*$ translate into the fluid ratio $r_f = 0.68^{+0.20}_{-0.13}$ ($68\%$ C.L.), according to (\ref{eq:r}). Note that the case of $r_f \ll 1$ (the fluid consisting mostly of radiation at $a_*$) is disfavored by data as can be inferred from the $w$ vs $dw/d\ln a$ plane. This is despite the fact that a small second derivative with $w_*\approx 1/3$ can be consistent with data as can be seen from the $w$ vs $d^2w/d\ln a^2$ plane below. However this would require $dw/d\ln a$ to be positive, which is incompatible with the simple mixture model). On the other hand, values of $w_* \approx 0.8$ (or $r_f\approx0.75$) are consistent with data according to the $w$ vs $dw/d\ln a$ plane.\footnote{It might appear that this in slight tension with the contours in the $w$ vs $d^2w/d\ln a^2$ plane. This is however only a technical limitation due to the lower bound $w(a)>0$ that we imposed in our analysis. To explore this parameter space one would need to include higher derivative terms.}

The gray curves in the figure show the case when the stiff fluid decays into the radiation component with a decay rate controlled by $Q$, according to (\ref{eq_eom_Q}). The dashed, dash-dotted and dotted lines correspond to $Q_{*} = 1/2$, $Q_{*} = 1$ and  $Q_{*} = 2$, respectively. For these order unity values, a sizeable fraction of the stiff fluid decays within one Hubble time. In particular, we see that the decay scenarios are compatible with data as long as the decay rate satisfies $Q* \lesssim 2$. In particular, data appears to be compatible with a scenario where $Q_*=2$ and $r_f=1$, i.e., all of the latent heat is initially converted into a stiff fluid before it decays into radiation rather quickly. 
The gray curves are not shown in the $d^2w/dlna^2$ planes because they also depend on the time-dependence of the decay rate, i.e.~on the parameter $q_{*}$ according to~(\ref{eq_Q_dependence}) (making these contours less predictive for the decay scenario).

There are two takeaways from this analysis. First, the time-evolving equation of state is only (mildly) favored for the Planck 2018 data, while it is less constrained for NPIPE data. For both NPIPE and Planck 2018 data, there is a clear (and expected) degeneracy between  $dw/d(\ln a)$ and $w_*$. Moreover, both datasets are fully compatible with the assumption $dw/d(\ln a)=0$ made in the base model, and accordingly the data fit only gives a very weak evidence for the extension.   Second, in both cases, data appears to be compatible with a two-fluid system with $w_1=1$ and $w_2=1/3$, including the possibility of an energy transfer from the stiff to the radiation fluid. This motivates a more detailed study where both fluid components including their perturbations are evolved independently, without relying on the truncation.

\subsection{A First Test of the Thawing Assumption}
\label{sec:thawing}

\begin{figure}[tbp]
\centering
\includegraphics[width=0.5\textwidth]{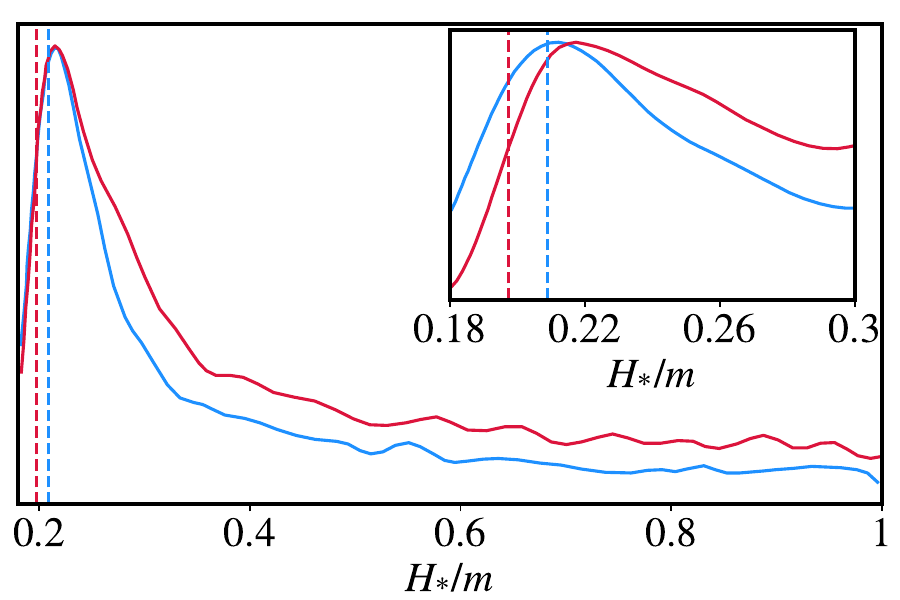}
\caption{Marginalized probability distribution for the trigger threshold parameter $H_*/m$ obtained by supplementing our baseline by the \texttt{CamSpec}/NPIPE (red) and \texttt{Plik}/Planck 2018 (blue) likelihoods alongside a prior on $H_0$. The vertical lines denote the corresponding best-fit values. The lower prior boundary corresponds to the first zero of the trigger field. The peak between $0.20$ and $0.22$ can be understood as a preliminary confirmation of the thawing assumption.} 
\label{fig:Hm}
\end{figure}

Allowing the equation of state to vary after the phase transition does not exhaust all possibilities for extending the phenomenological NEDE model. At the background level, another instructive example is provided by the trigger threshold parameter. It is given by the ratio $H_*/m$, where $H_{*} = H(z_{*})$ is the Hubble parameter at the transition time and $m$ is the mass of the trigger field. As it was already mentioned, the precise value of this parameter is determined by the underlying microscopic model. Typical thawing mechanisms trigger the transition when $H_* \lesssim m$ and, in particular, in the case of Cold NEDE~\cite{Niedermann:2020dwg} for $H_*\simeq 0.2 m$.

Remaining agnostic about the underlying model, here, we treat $H_{*}/m$ as a free parameter and fit it to both the \texttt{CamSpec}/NPIPE and \texttt{Plik}/Planck 2018 likelihoods (together with our baseline datasets and the SH$_0$ES prior on $M_b$). To be precise, we consider the model with parameters $\log_{10}(z_*)$, $f_\mathrm{NEDE}$, $w_*$ ($= \mathrm{const}$), and $H_*/m$ (while we set $\Omega_\phi \ll 1$ for simplicity). Note that on the background level, for a fixed $z_*$, this parameter controls the value of $m$, which enters the equations of motion (\ref{eq:KG}) for the trigger field. Since the trigger field is usually very subdominant, changing its mass has negligible effect on the cosmic evolution.  At the perturbation level, it however affects the matching equations~\eqref{eq:match} through their dependence on $\delta \phi$ and $\bar{\phi}^\prime$. In other words, the perturbations in the trigger field at the time of the phase transition control the initial amplitude of perturbations in the decaying NEDE fluid. We see that the matching equation diverges for $\bar{\phi}^\prime \to 0$, which during radiation domination imposes the lower bound $H_*/m > 0.12$. On the other hand, for $H_*/m \gg 1$, we enter the slow-roll regime, where the trigger is assumed to be inactive. We impose the flat prior $0.18 < H_*/m < 1$. The lower bound corresponds to the first zero of $\bar{\phi}$ after it has dropped out of slow-roll. In a specific trigger scenario such as Cold NEDE, we thus expect the phase transition to occur earlier, i.e., for larger values of $H_*/m$. The upper bound corresponds to a situation where the transition is triggered while $\phi$ is still marginally within the slow-roll regime (for even higher values the trigger perturbations are too suppressed for data to be sensitive to differences between different values of $H_*/m$).

The marginalized probability distribution for the $H_*/m$ parameter, obtained using the NPIPE (red) and Planck 2018 (blue) likelihood in the presence of the SH$_0$ES prior on $H_0$ is shown in Fig.~\ref{fig:Hm}. The vertical lines denote the corresponding best-fit values. Interestingly, the distribution is  peaked close to the value $0.2$ mentioned above. There is no obvious reason for data to prefer a value within the thawing regime $H_* \lesssim m$. Instead, the generic expectation would have been a lower bound on $H_*/m$ potentially tightening $H_*/m > 0.12$. In other words, data seems to be compatible with a thawing mechanism. Moreover, the distribution flattens for large values of $H_*/m$, which as mentioned above, corresponds to the regime where trigger perturbations are more and more suppressed and thus data cannot distinguish different values of $H_*/m$.We also did an additional run for the prior range $0.18 < H_*/m < 0.3$ in order to better resolve the peak (see the small panel in Fig.~\ref{fig:Hm}). In terms of the best-fit $\chi^2$ we do not observe a noticeable improvement compared to the base NEDE model where $H_*/m$ is fixed to $0.2$. 

We stress that these results should be only understood as a first indication. After all, the above results are obtained by combining with SH$_0$ES, which in the case of the \texttt{CamSpec} implementation of NPIPE is $3.5 \sigma$ discrepant. We leave a more systematic and general exploration, which requires a profile likelihood analysis to future work. This also includes other modifications such as a generalized sound speed.

\section{Conclusion}\label{sec:concl}

The Cold NEDE model introduces an early dark energy component that decays in a phase transition triggered by an ultralight scalar field. In the past, the model has been found to reduce both the $H_0$ and $S_8$ tensions below the $2 \sigma$ level. It admits a simple field theoretic realization in terms of a first-order vacuum phase transition that occurs in the dark sector before matter-radiation equality.
Cold NEDE features some important distinctions from the axiEDE model, which were highlighted in this work. AxiEDE involves only one scalar field which undergoes a smooth crossover transition and has a negligible relic abundance today. Instead, in Cold NEDE there is a first-order vacuum phase transition that is triggered by a scalar field when it thaws around matter-radiation equality. Unlike the scalar field in axiEDE, the trigger field in NEDE does not participate in the energy injection around the recombination time itself. It however seeds the perturbations of the fluid that is used to describe the post-phase transition state. At the same time, the oscillations of the trigger field can have a non-negligible energy today and act as a small fuzzy dark matter component. As a consequence, in contrast to axiEDE and other EDE type models, including the SH$_0$ES prior lowers the value of $S_8$ in Cold NEDE~\cite{Cruz:2023lmn}.

One of the benefits of the Cold NEDE model is its predictivity as it directly affects the physics of waves in the primordial plasma at scales that can be observed in the CMB and matter power spectra. This happens at the background level through a change of the sound horizon and at the perturbation level through the gravitational impact of dark sector perturbations associated with the decaying early dark energy fluid. Nevertheless, the model comes with freedom in how to realize the dissipation of the early dark energy component after the phase transition, which so far has been modeled in terms of an ideal fluid with a constant equation of state parameter $w_\mathrm{NEDE}$. At the same time, new CMB, LSS and supernovae observations offer the prospect of further constraining early dark energy models and, in the case of Cold NEDE, this should enable us to further constrain $w_\mathrm{NEDE}$ after the phase transition, providing a way to further narrow down viable microscopic scenarios.  

In line with this overarching goal, in this work, we have analyzed the cold NEDE model and some of its physically well motivated extensions in light of updated CMB and supernovae data. To be precise, the high-$\ell$ Planck 2018 polarization and temperature data and the Pantheon supernovae data were replaced by Planck NPIPE (using the \texttt{CamSpec} and \texttt{HiLLiPoP} likelihoods) and Pantheon+ data, respectively. While we find that the model's ability to resolve the Hubble tension is somewhat diminished with the inclusion of the new datasets, it still reduces the $Q_\mathrm{DMAP}$ tension considerably from $6.3 \sigma$ within $\Lambda$CDM to $3.5 \sigma$  when using the \texttt{CamSpec} likelihood code for the NPIPE data (and $3.7 \sigma$  for \texttt{HiLLiPoP}). This outcome is compatible with recent findings for the AxiEDE model, where a residual $3.7 \sigma$ tension was reported~\cite{Efstathiou:2023fbn}. It is important to note that the improvement over $\Lambda$CDM is almost unchanged compared to earlier works when it comes to relieving the Hubble tension, and the increased residual tension is mostly due to $\Lambda$CDM itself showing a stronger tension when using NPIPE and Pantheon+ data. Moreover, we performed a first test of the Cold NEDE model against the recently released DESI BAO data. We find that the tensions is significantly reduced to $2.6 \sigma$ when combining DESI with Pantheon+ and NPIPE data (also improving on the recent axiEDE result of $3.1\sigma$ in~\cite{Poulin:2024ken}). A complementary analysis relying on profile likelihoods confirms the above results.

In this work we find that the Cold NEDE model's ability to help with the $S_8$ tension is maintained in the presence of the new datasets. For concreteness, when using the weak lensing constraint from \cite{Joudaki:2019pmv} as reference value, our updated analyses yield residual Gaussian tensions between $1.3 \sigma$ and $2.1 \sigma$ (down from $\sim 2.5 \sigma$ within $\Lambda$CDM).   

In addition, for the first time, we have performed a systematic study of the post-phase transition fluid when its equation of state evolves with time. There are two takeaways from this: (i) The simplifying assumption made in previous work that $w_\mathrm{NEDE} = \mathrm{const}$ is still compatible with data. (ii) Nevertheless, data also allows for a significant time evolution that is compatible with a two-component fluid. In particular, such a fluid could be comprised of a stiff ($w = 1$) and radiation ($w = 1/3$) component with a possible decay channel between both components. The fit indicates that a fraction $r_f = 0.68^{+0.20}_{-0.13}$ ($68\%$ C.L.) of the latent heat is converted into a stiff fluid with the remaining energy going into a radiation component. While we have merely sketched a few field theoretic scenarios, it should be a priority of future work to come up with detailed and testable descriptions. In particular, each scenario is expected to affect the perturbation sector in unique ways, which offers the possibility to further decrease the tension. As another straightforward extension, we freed up the parameter combination $H(z_*)/m$, which determines by how much the trigger field has left the slow-roll regime when it triggers the phase transition. Interestingly, we find first evidence that data prefers a value that is close to the theoretical prediction of $H(z_*)/m \approx 0.2$ obtained in the tunneling scenario, which we reviewed in Sec.~\ref{sec:micro}.

Assuming future dataset updates do not revert the picture, one way of understanding our results is that the updated datasets indicate that the current form of the Cold NEDE model is not yet complete as a solution to the Hubble tension.
This would mean that data are finally able to refine our understanding of early dark energy models, enabling us to explore their unique features required to fully resolve the tension. For Cold NEDE this could involve relaxing assumptions made about the perturbation sector. In particular, it would be interesting to go beyond a tightly coupled barotropic fluid and for example study a more general $k$-dependent sound speed and viscosity parameter (following for example a general fluid approach as in~\cite{Hu:1998kj}). This phenomenological approach could be informed by one of the different microscopic dissipation scenarios mentioned in this work. Alternatively, it may turn out that there are yet unknown systematic errors affecting either Planck NPIPE data, or the SH$_0$ES measurement, that may go in the direction of improving or relaxing constraints to the model. In fact, the first results we obtained from including recent DESI BAO data relax the constrain significantly, and future BAO data will be thus be of utmost importance to test the model.
In any case, Cold NEDE remains so far one of the most promising and natural model for addressing the Hubble tension. Building upon it provides a constructive and promising way forward.

\bigskip

\begin{acknowledgments}
V.P. would like to thank Matthieu Tristram for providing him with a \texttt{MontePython} version of the \texttt{HiLLiPoP}/NPIPE likelihood.
The work of F.N. is supported by VR Starting Grant 2022-03160 of the Swedish Research Council. The computations were in parts enabled by resources provided by the National Academic Infrastructure for Supercomputing in Sweden (NAISS), partially funded by the Swedish Research Council through grant agreement no. 2022-06725. M.S.S. is supported by Independent Research Fund Denmark grant 0135-00378B.
The authors acknowledge the use of computational resources from the Excellence Initiative of Aix-Marseille University (A*MIDEX) of the “Investissements d’Avenir” programme. 
This project has received support from the European Union’s Horizon 2020 research and innovation program under the Marie Skodowska-Curie grant agreement No 860881-HIDDeN.  This project has received funding from the European Research Council (ERC) under the
European Union’s HORIZON-ERC-2022 (Grant agreement No. 101076865). 
\end{acknowledgments}

\appendix

\section{Profile likelihood for the Cold NEDE Model for NPIPE} 
\label{app:profile}

\begin{figure}[tbp]
\centering
\subfigure{
\adjustbox{valign=c}{\includegraphics[width=0.45\textwidth]{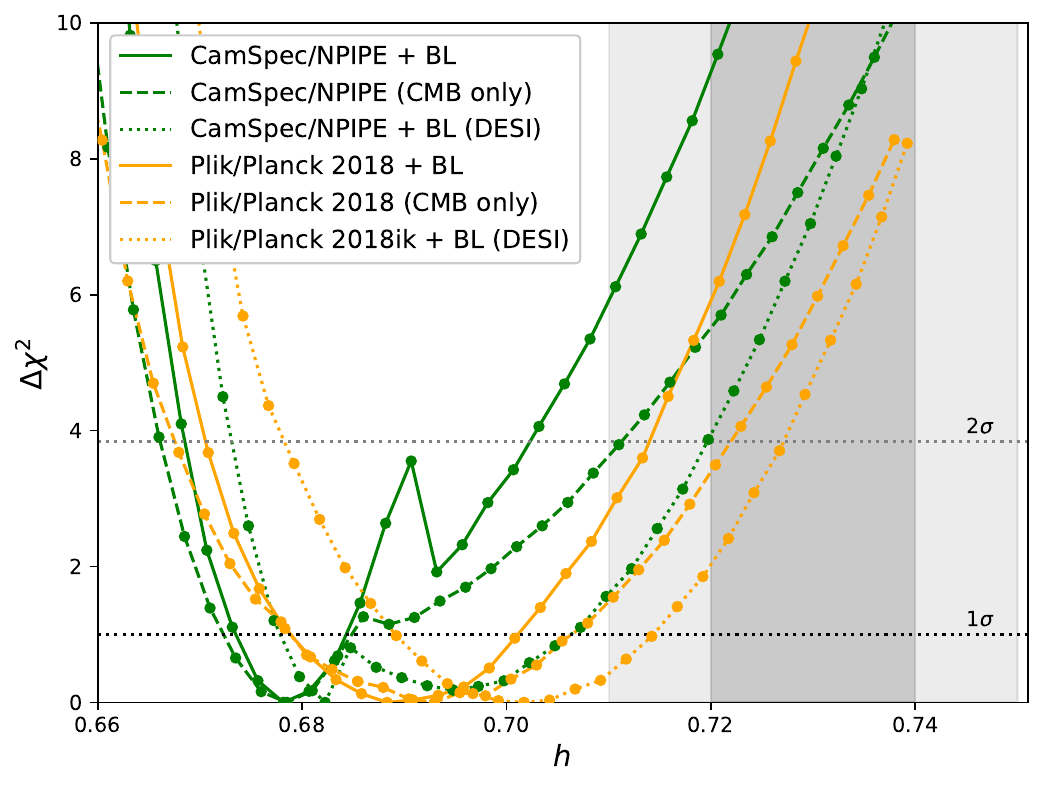}}
}
\hfill
\subfigure{
\adjustbox{valign=c}{\includegraphics[width=0.45\textwidth]{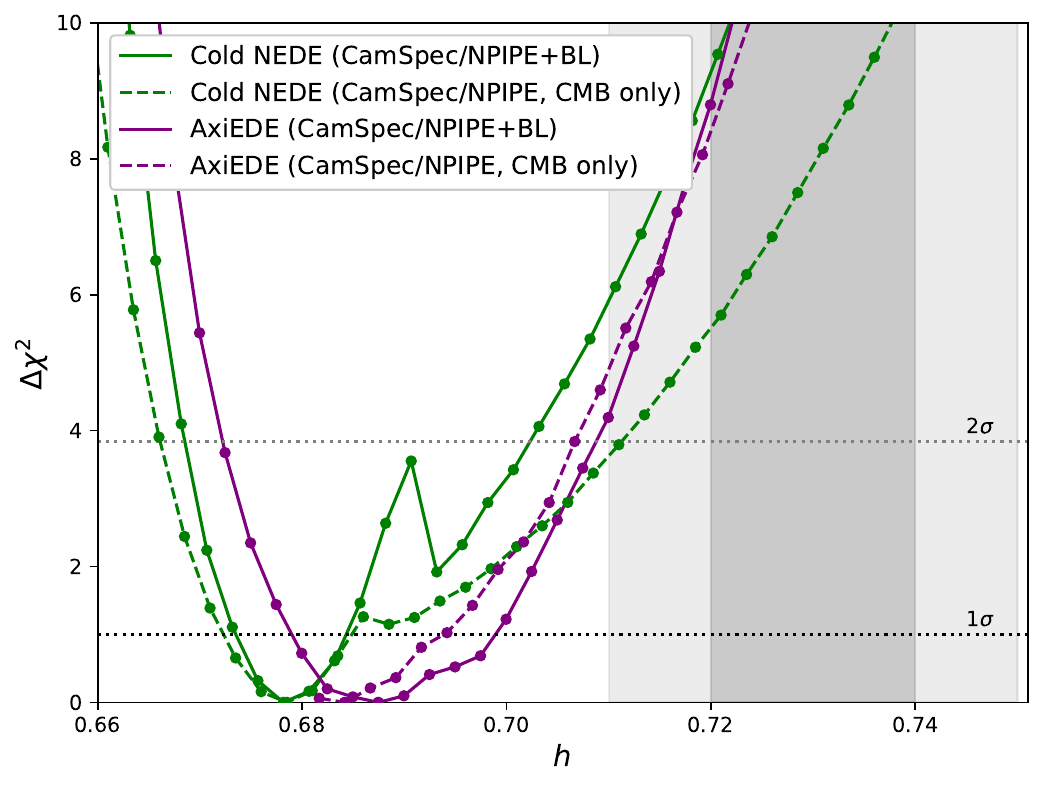}}
}
\caption{Profile likelihood curves for $h$ without including the SH$_0$ES prior on $H_0$ ($ \Delta \chi^2 = -2 \log (\mathcal{L}) - \chi_\mathrm{min}^2$). The gray bands show the $1\sigma$ and $2 \sigma$ constraints on $h$ from \cite{Riess:2021jrx}. The baseline (BL) datasets consist of  BAO and Pantheon+ data as described in the main text. {\bf Left:} Comparison between \texttt{CamSpec}/NPIPE (green) and \texttt{Plik}/Planck 2018 (orange) for the NEDE model. The \texttt{CamSpec} likelihood gives rise to a bi-modality not present for \texttt{Plik}. {\bf Right:} Comparison between AxiEDE (purple) and Cold NEDE (green)  using NPIPE CMB data. Both models show non-Gaussian features around the minimum.} 
\label{fig:profile}
\end{figure}

In order to complement the statistical analysis in the main part of the paper, we present here the profile likelihood constraints which we obtained by using the \texttt{Procoli} code~\cite{Karwal:2024qpt}. This approach computes the likelihood $\mathcal{L}$ of the data for a given choice of model parameters. It has the benefit of being insensitive to prior volume effects that otherwise lead to non-Gaussian features in the posteriors (for a detailed discussion within NEDE see also~\cite{Cruz:2023cxy}). In particular, it allows us to understand the observed differences between NPIPE and Planck 2018 in more detail. 

In the left panel of Fig.~\ref{fig:profile}, we compare the \texttt{CamSpec}/NPIPE (green) with the \texttt{Plik}/Planck 2018 (orange) profiles for different choices of supplemental datasets. The crucial observation, which also agrees with the findings in our main analysis, is that NPIPE tightens the upper limit on $H_0$ when comparing equivalent runs. For example,  the re-constructed $2\sigma$ intervals corresponding to the solid lines are   $h \in \left[ 0.669,0.698 \right]$ (\texttt{CamSpec}/NPIPE, green solid) and $h \in \left[ 0.671,0.714 \right]$ (\texttt{Plik}/Planck 2018, orange solid). These values are compatible with our results for $Q_\mathrm{DMAP}$ reported in Tab~\ref{tab:npipe} and Tab~\ref{tab:plik} when using a simple Gaussian tension measure. However, there is an additional difference, which the MCMC analysis did not reveal: While the \texttt{Plik} runs all correspond to an approximately Gaussian likelihood $\mathcal{L}$ (i.e.\ $ \chi^2 = -2 \log (\mathcal{L})$ is quadratic), the \texttt{CamSpec} runs deviate significantly from it. In fact, they all show signs of a slight bi-modality. Albeit less pronounced, the non-Gaussian features are also present for the axiEDE profiles in the right panel. Upon closer inspection we found that in the case of cold NEDE, this bi-modality manifests itself most prominently as a jump in the decay time, which transitions from $z_* \sim 10000 $ to $\sim 4000$ as $h$ increases. We did not observe this feature in the case of the \texttt{Plik} likelihood (as a side remark, in~\cite{Karwal:2024qpt}, the axiEDE model did exhibit a bi-modality also for \texttt{Plik}). At this stage, this is mainly an observation and it remains to be seen if this peculiar property can be reproduced with independent CMB data in the future or if it is a statistical fluke or the signs of a systematics within NPIPE. 

In addition, this analysis reproduces the observation that DESI data (dotted lines) allows for larger values of $H_0$, enabling Cold NEDE as a solution to the Hubble tension. To be specific, without including SH0ES we find  $h \in \left[ 0.673, 0.720 \right]$ (\texttt{CamSpec}/NPIPE, green dotted)  and $h \in \left[ 0.678, 0.727 \right]$ (\texttt{Plik}/Planck 2018, orange dotted), corresponding to $\sim 2\sigma$ Gaussian tensions. It is also noteworthy that the pure CMB runs (dashed) agree for large values of $H_0$ well with the combined DESI runs, whereas the runs including the previous BAO data (solid) give significantly less evidence for large values of $H_0$. This can be traced to the preference of DESI for lower $\Omega_m$/larger $H_0$ than Planck data, which is easier to accommodate in the (N)EDE cosmology.

Finally, we also compared Cold NEDE with AxiEDE in the right panel of Fig.~\ref{fig:profile}. While both models show overall very similar profiles, NEDE can accommodate higher values of $H_0$, which also explains the somewhat smaller $Q_\mathrm{DMAP}$ tension reported in the main part of the paper. This difference is most pronounced in the run with CMB/NPIPE data alone. 

\bibliography{ref}

\begin{thebibliography}{66}%
\makeatletter
\providecommand \@ifxundefined [1]{%
 \@ifx{#1\undefined}
}%
\providecommand \@ifnum [1]{%
 \ifnum #1\expandafter \@firstoftwo
 \else \expandafter \@secondoftwo
 \fi
}%
\providecommand \@ifx [1]{%
 \ifx #1\expandafter \@firstoftwo
 \else \expandafter \@secondoftwo
 \fi
}%
\providecommand \natexlab [1]{#1}%
\providecommand \enquote  [1]{``#1''}%
\providecommand \bibnamefont  [1]{#1}%
\providecommand \bibfnamefont [1]{#1}%
\providecommand \citenamefont [1]{#1}%
\providecommand \href@noop [0]{\@secondoftwo}%
\providecommand \href [0]{\begingroup \@sanitize@url \@href}%
\providecommand \@href[1]{\@@startlink{#1}\@@href}%
\providecommand \@@href[1]{\endgroup#1\@@endlink}%
\providecommand \@sanitize@url [0]{\catcode `\\12\catcode `\$12\catcode
  `\&12\catcode `\#12\catcode `\^12\catcode `\_12\catcode `\%12\relax}%
\providecommand \@@startlink[1]{}%
\providecommand \@@endlink[0]{}%
\providecommand \url  [0]{\begingroup\@sanitize@url \@url }%
\providecommand \@url [1]{\endgroup\@href {#1}{\urlprefix }}%
\providecommand \urlprefix  [0]{URL }%
\providecommand \Eprint [0]{\href }%
\providecommand \doibase [0]{https://doi.org/}%
\providecommand \selectlanguage [0]{\@gobble}%
\providecommand \bibinfo  [0]{\@secondoftwo}%
\providecommand \bibfield  [0]{\@secondoftwo}%
\providecommand \translation [1]{[#1]}%
\providecommand \BibitemOpen [0]{}%
\providecommand \bibitemStop [0]{}%
\providecommand \bibitemNoStop [0]{.\EOS\space}%
\providecommand \EOS [0]{\spacefactor3000\relax}%
\providecommand \BibitemShut  [1]{\csname bibitem#1\endcsname}%
\let\auto@bib@innerbib\@empty
\bibitem [{\citenamefont {Riess}\ \emph {et~al.}(2022)\citenamefont {Riess}
  \emph {et~al.}}]{Riess:2021jrx}%
  \BibitemOpen
  \bibfield  {author} {\bibinfo {author} {\bibfnamefont {A.~G.}\ \bibnamefont
  {Riess}} \emph {et~al.},\ }\bibfield  {title} {\bibinfo {title} {{A
  Comprehensive Measurement of the Local Value of the Hubble Constant with 1 km
  s$^{-1}$ Mpc$^{-1}$ Uncertainty from the Hubble Space Telescope and the SH0ES
  Team}},\ }\href {https://doi.org/10.3847/2041-8213/ac5c5b} {\bibfield
  {journal} {\bibinfo  {journal} {Astrophys. J. Lett.}\ }\textbf {\bibinfo
  {volume} {934}},\ \bibinfo {pages} {L7} (\bibinfo {year} {2022})},\ \Eprint
  {https://arxiv.org/abs/2112.04510} {arXiv:2112.04510 [astro-ph.CO]}
  \BibitemShut {NoStop}%
\bibitem [{\citenamefont {Aghanim}\ \emph
  {et~al.}(2020{\natexlab{a}})\citenamefont {Aghanim} \emph
  {et~al.}}]{Planck:2018vyg}%
  \BibitemOpen
  \bibfield  {author} {\bibinfo {author} {\bibfnamefont {N.}~\bibnamefont
  {Aghanim}} \emph {et~al.} (\bibinfo {collaboration} {Planck}),\ }\bibfield
  {title} {\bibinfo {title} {{Planck 2018 results. VI. Cosmological
  parameters}},\ }\href {https://doi.org/10.1051/0004-6361/201833910}
  {\bibfield  {journal} {\bibinfo  {journal} {Astron. Astrophys.}\ }\textbf
  {\bibinfo {volume} {641}},\ \bibinfo {pages} {A6} (\bibinfo {year}
  {2020}{\natexlab{a}})},\ \bibinfo {note} {[Erratum: Astron.Astrophys. 652, C4
  (2021)]},\ \Eprint {https://arxiv.org/abs/1807.06209} {arXiv:1807.06209
  [astro-ph.CO]} \BibitemShut {NoStop}%
\bibitem [{\citenamefont {Lee}\ \emph {et~al.}(2024)\citenamefont {Lee},
  \citenamefont {Freedman}, \citenamefont {Madore}, \citenamefont {Jang},
  \citenamefont {Owens},\ and\ \citenamefont {Hoyt}}]{Lee:2024qzr}%
  \BibitemOpen
  \bibfield  {author} {\bibinfo {author} {\bibfnamefont {A.~J.}\ \bibnamefont
  {Lee}}, \bibinfo {author} {\bibfnamefont {W.~L.}\ \bibnamefont {Freedman}},
  \bibinfo {author} {\bibfnamefont {B.~F.}\ \bibnamefont {Madore}}, \bibinfo
  {author} {\bibfnamefont {I.~S.}\ \bibnamefont {Jang}}, \bibinfo {author}
  {\bibfnamefont {K.~A.}\ \bibnamefont {Owens}},\ and\ \bibinfo {author}
  {\bibfnamefont {T.~J.}\ \bibnamefont {Hoyt}},\ }\bibfield  {title} {\bibinfo
  {title} {{The Chicago-Carnegie Hubble Program: The JWST J-region Asymptotic
  Giant Branch (JAGB) Extragalactic Distance Scale}},\ }\href@noop {} {\
  (\bibinfo {year} {2024})},\ \Eprint {https://arxiv.org/abs/2408.03474}
  {arXiv:2408.03474 [astro-ph.GA]} \BibitemShut {NoStop}%
\bibitem [{\citenamefont {Freedman}\ \emph {et~al.}(2024)\citenamefont
  {Freedman}, \citenamefont {Madore}, \citenamefont {Jang}, \citenamefont
  {Hoyt}, \citenamefont {Lee},\ and\ \citenamefont {Owens}}]{Freedman:2024eph}%
  \BibitemOpen
  \bibfield  {author} {\bibinfo {author} {\bibfnamefont {W.~L.}\ \bibnamefont
  {Freedman}}, \bibinfo {author} {\bibfnamefont {B.~F.}\ \bibnamefont
  {Madore}}, \bibinfo {author} {\bibfnamefont {I.~S.}\ \bibnamefont {Jang}},
  \bibinfo {author} {\bibfnamefont {T.~J.}\ \bibnamefont {Hoyt}}, \bibinfo
  {author} {\bibfnamefont {A.~J.}\ \bibnamefont {Lee}},\ and\ \bibinfo {author}
  {\bibfnamefont {K.~A.}\ \bibnamefont {Owens}},\ }\bibfield  {title} {\bibinfo
  {title} {{Status Report on the Chicago-Carnegie Hubble Program (CCHP): Three
  Independent Astrophysical Determinations of the Hubble Constant Using the
  James Webb Space Telescope}},\ }\href@noop {} {\  (\bibinfo {year} {2024})},\
  \Eprint {https://arxiv.org/abs/2408.06153} {arXiv:2408.06153 [astro-ph.CO]}
  \BibitemShut {NoStop}%
\bibitem [{\citenamefont {Anand}\ \emph {et~al.}(2024)\citenamefont {Anand}
  \emph {et~al.}}]{Anand:2024nim}%
  \BibitemOpen
  \bibfield  {author} {\bibinfo {author} {\bibfnamefont {G.~S.}\ \bibnamefont
  {Anand}} \emph {et~al.},\ }\bibfield  {title} {\bibinfo {title} {{Tip of the
  Red Giant Branch Distances with JWST: An Absolute Calibration in NGC 4258 and
  First Applications to Type Ia Supernova Hosts}},\ }\href
  {https://doi.org/10.3847/1538-4357/ad2e0a} {\bibfield  {journal} {\bibinfo
  {journal} {Astrophys. J.}\ }\textbf {\bibinfo {volume} {966}},\ \bibinfo
  {pages} {89} (\bibinfo {year} {2024})},\ \Eprint
  {https://arxiv.org/abs/2401.04776} {arXiv:2401.04776 [astro-ph.CO]}
  \BibitemShut {NoStop}%
\bibitem [{\citenamefont {Li}\ \emph {et~al.}(2024{\natexlab{a}})\citenamefont
  {Li}, \citenamefont {Riess}, \citenamefont {Casertano}, \citenamefont
  {Anand}, \citenamefont {Scolnic}, \citenamefont {Yuan}, \citenamefont
  {Breuval},\ and\ \citenamefont {Huang}}]{Li:2024yoe}%
  \BibitemOpen
  \bibfield  {author} {\bibinfo {author} {\bibfnamefont {S.}~\bibnamefont
  {Li}}, \bibinfo {author} {\bibfnamefont {A.~G.}\ \bibnamefont {Riess}},
  \bibinfo {author} {\bibfnamefont {S.}~\bibnamefont {Casertano}}, \bibinfo
  {author} {\bibfnamefont {G.~S.}\ \bibnamefont {Anand}}, \bibinfo {author}
  {\bibfnamefont {D.~M.}\ \bibnamefont {Scolnic}}, \bibinfo {author}
  {\bibfnamefont {W.}~\bibnamefont {Yuan}}, \bibinfo {author} {\bibfnamefont
  {L.}~\bibnamefont {Breuval}},\ and\ \bibinfo {author} {\bibfnamefont {C.~D.}\
  \bibnamefont {Huang}},\ }\bibfield  {title} {\bibinfo {title}
  {{Reconnaissance with JWST of the J-region Asymptotic Giant Branch in
  Distance Ladder Galaxies: From Irregular Luminosity Functions to
  Approximation of the Hubble Constant}},\ }\href
  {https://doi.org/10.3847/1538-4357/ad2f2b} {\bibfield  {journal} {\bibinfo
  {journal} {Astrophys. J.}\ }\textbf {\bibinfo {volume} {966}},\ \bibinfo
  {pages} {20} (\bibinfo {year} {2024}{\natexlab{a}})},\ \Eprint
  {https://arxiv.org/abs/2401.04777} {arXiv:2401.04777 [astro-ph.CO]}
  \BibitemShut {NoStop}%
\bibitem [{\citenamefont {Li}\ \emph {et~al.}(2024{\natexlab{b}})\citenamefont
  {Li}, \citenamefont {Anand}, \citenamefont {Riess}, \citenamefont
  {Casertano}, \citenamefont {Yuan}, \citenamefont {Breuval}, \citenamefont
  {Macri}, \citenamefont {Scolnic}, \citenamefont {Beaton},\ and\ \citenamefont
  {Anderson}}]{Li:2024pjo}%
  \BibitemOpen
  \bibfield  {author} {\bibinfo {author} {\bibfnamefont {S.}~\bibnamefont
  {Li}}, \bibinfo {author} {\bibfnamefont {G.~S.}\ \bibnamefont {Anand}},
  \bibinfo {author} {\bibfnamefont {A.~G.}\ \bibnamefont {Riess}}, \bibinfo
  {author} {\bibfnamefont {S.}~\bibnamefont {Casertano}}, \bibinfo {author}
  {\bibfnamefont {W.}~\bibnamefont {Yuan}}, \bibinfo {author} {\bibfnamefont
  {L.}~\bibnamefont {Breuval}}, \bibinfo {author} {\bibfnamefont {L.~M.}\
  \bibnamefont {Macri}}, \bibinfo {author} {\bibfnamefont {D.}~\bibnamefont
  {Scolnic}}, \bibinfo {author} {\bibfnamefont {R.}~\bibnamefont {Beaton}},\
  and\ \bibinfo {author} {\bibfnamefont {R.~I.}\ \bibnamefont {Anderson}},\
  }\bibfield  {title} {\bibinfo {title} {{Tip of the Red Giant Branch Distances
  with JWST. II. I-band Measurements in a Sample of Hosts of 10 SN Ia Match HST
  Cepheids}},\ }\href@noop {} {\  (\bibinfo {year} {2024}{\natexlab{b}})},\
  \Eprint {https://arxiv.org/abs/2408.00065} {arXiv:2408.00065 [astro-ph.CO]}
  \BibitemShut {NoStop}%
\bibitem [{\citenamefont {Riess}\ \emph {et~al.}(2024)\citenamefont {Riess}
  \emph {et~al.}}]{Riess:2024vfa}%
  \BibitemOpen
  \bibfield  {author} {\bibinfo {author} {\bibfnamefont {A.~G.}\ \bibnamefont
  {Riess}} \emph {et~al.},\ }\bibfield  {title} {\bibinfo {title} {{JWST
  Validates HST Distance Measurements: Selection of Supernova Subsample
  Explains Differences in JWST Estimates of Local H0}},\ }\href@noop {} {\
  (\bibinfo {year} {2024})},\ \Eprint {https://arxiv.org/abs/2408.11770}
  {arXiv:2408.11770 [astro-ph.CO]} \BibitemShut {NoStop}%
\bibitem [{\citenamefont {Scolnic}\ \emph {et~al.}(2023)\citenamefont
  {Scolnic}, \citenamefont {Riess}, \citenamefont {Wu}, \citenamefont {Li},
  \citenamefont {Anand}, \citenamefont {Beaton}, \citenamefont {Casertano},
  \citenamefont {Anderson}, \citenamefont {Dhawan},\ and\ \citenamefont
  {Ke}}]{Scolnic:2023mrv}%
  \BibitemOpen
  \bibfield  {author} {\bibinfo {author} {\bibfnamefont {D.}~\bibnamefont
  {Scolnic}}, \bibinfo {author} {\bibfnamefont {A.~G.}\ \bibnamefont {Riess}},
  \bibinfo {author} {\bibfnamefont {J.}~\bibnamefont {Wu}}, \bibinfo {author}
  {\bibfnamefont {S.}~\bibnamefont {Li}}, \bibinfo {author} {\bibfnamefont
  {G.~S.}\ \bibnamefont {Anand}}, \bibinfo {author} {\bibfnamefont
  {R.}~\bibnamefont {Beaton}}, \bibinfo {author} {\bibfnamefont
  {S.}~\bibnamefont {Casertano}}, \bibinfo {author} {\bibfnamefont {R.~I.}\
  \bibnamefont {Anderson}}, \bibinfo {author} {\bibfnamefont {S.}~\bibnamefont
  {Dhawan}},\ and\ \bibinfo {author} {\bibfnamefont {X.}~\bibnamefont {Ke}},\
  }\bibfield  {title} {\bibinfo {title} {{CATS: The Hubble Constant from
  Standardized TRGB and Type Ia Supernova Measurements}},\ }\href
  {https://doi.org/10.3847/2041-8213/ace978} {\bibfield  {journal} {\bibinfo
  {journal} {Astrophys. J. Lett.}\ }\textbf {\bibinfo {volume} {954}},\
  \bibinfo {pages} {L31} (\bibinfo {year} {2023})},\ \Eprint
  {https://arxiv.org/abs/2304.06693} {arXiv:2304.06693 [astro-ph.CO]}
  \BibitemShut {NoStop}%
\bibitem [{\citenamefont {Kamionkowski}\ and\ \citenamefont
  {Riess}(2023)}]{Kamionkowski:2022pkx}%
  \BibitemOpen
  \bibfield  {author} {\bibinfo {author} {\bibfnamefont {M.}~\bibnamefont
  {Kamionkowski}}\ and\ \bibinfo {author} {\bibfnamefont {A.~G.}\ \bibnamefont
  {Riess}},\ }\bibfield  {title} {\bibinfo {title} {{The Hubble Tension and
  Early Dark Energy}},\ }\href
  {https://doi.org/10.1146/annurev-nucl-111422-024107} {\bibfield  {journal}
  {\bibinfo  {journal} {Ann. Rev. Nucl. Part. Sci.}\ }\textbf {\bibinfo
  {volume} {73}},\ \bibinfo {pages} {153} (\bibinfo {year} {2023})},\ \Eprint
  {https://arxiv.org/abs/2211.04492} {arXiv:2211.04492 [astro-ph.CO]}
  \BibitemShut {NoStop}%
\bibitem [{\citenamefont {Poulin}\ \emph {et~al.}(2023)\citenamefont {Poulin},
  \citenamefont {Smith},\ and\ \citenamefont {Karwal}}]{Poulin:2023lkg}%
  \BibitemOpen
  \bibfield  {author} {\bibinfo {author} {\bibfnamefont {V.}~\bibnamefont
  {Poulin}}, \bibinfo {author} {\bibfnamefont {T.~L.}\ \bibnamefont {Smith}},\
  and\ \bibinfo {author} {\bibfnamefont {T.}~\bibnamefont {Karwal}},\
  }\bibfield  {title} {\bibinfo {title} {{The Ups and Downs of Early Dark
  Energy solutions to the Hubble tension: A review of models, hints and
  constraints circa 2023}},\ }\href
  {https://doi.org/10.1016/j.dark.2023.101348} {\bibfield  {journal} {\bibinfo
  {journal} {Phys. Dark Univ.}\ }\textbf {\bibinfo {volume} {42}},\ \bibinfo
  {pages} {101348} (\bibinfo {year} {2023})},\ \Eprint
  {https://arxiv.org/abs/2302.09032} {arXiv:2302.09032 [astro-ph.CO]}
  \BibitemShut {NoStop}%
\bibitem [{\citenamefont {McDonough}\ \emph {et~al.}(2023)\citenamefont
  {McDonough}, \citenamefont {Hill}, \citenamefont {Ivanov}, \citenamefont
  {La~Posta},\ and\ \citenamefont {Toomey}}]{McDonough:2023qcu}%
  \BibitemOpen
  \bibfield  {author} {\bibinfo {author} {\bibfnamefont {E.}~\bibnamefont
  {McDonough}}, \bibinfo {author} {\bibfnamefont {J.~C.}\ \bibnamefont {Hill}},
  \bibinfo {author} {\bibfnamefont {M.~M.}\ \bibnamefont {Ivanov}}, \bibinfo
  {author} {\bibfnamefont {A.}~\bibnamefont {La~Posta}},\ and\ \bibinfo
  {author} {\bibfnamefont {M.~W.}\ \bibnamefont {Toomey}},\ }\bibfield  {title}
  {\bibinfo {title} {{Observational constraints on early dark energy}},\
  }\href@noop {} {\  (\bibinfo {year} {2023})},\ \Eprint
  {https://arxiv.org/abs/2310.19899} {arXiv:2310.19899 [astro-ph.CO]}
  \BibitemShut {NoStop}%
\bibitem [{\citenamefont {Karwal}\ and\ \citenamefont
  {Kamionkowski}(2016)}]{Karwal:2016vyq}%
  \BibitemOpen
  \bibfield  {author} {\bibinfo {author} {\bibfnamefont {T.}~\bibnamefont
  {Karwal}}\ and\ \bibinfo {author} {\bibfnamefont {M.}~\bibnamefont
  {Kamionkowski}},\ }\bibfield  {title} {\bibinfo {title} {{Dark energy at
  early times, the Hubble parameter, and the string axiverse}},\ }\href
  {https://doi.org/10.1103/PhysRevD.94.103523} {\bibfield  {journal} {\bibinfo
  {journal} {Phys. Rev. D}\ }\textbf {\bibinfo {volume} {94}},\ \bibinfo
  {pages} {103523} (\bibinfo {year} {2016})},\ \Eprint
  {https://arxiv.org/abs/1608.01309} {arXiv:1608.01309 [astro-ph.CO]}
  \BibitemShut {NoStop}%
\bibitem [{\citenamefont {Poulin}\ \emph {et~al.}(2019)\citenamefont {Poulin},
  \citenamefont {Smith}, \citenamefont {Karwal},\ and\ \citenamefont
  {Kamionkowski}}]{Poulin:2018cxd}%
  \BibitemOpen
  \bibfield  {author} {\bibinfo {author} {\bibfnamefont {V.}~\bibnamefont
  {Poulin}}, \bibinfo {author} {\bibfnamefont {T.~L.}\ \bibnamefont {Smith}},
  \bibinfo {author} {\bibfnamefont {T.}~\bibnamefont {Karwal}},\ and\ \bibinfo
  {author} {\bibfnamefont {M.}~\bibnamefont {Kamionkowski}},\ }\bibfield
  {title} {\bibinfo {title} {{Early Dark Energy Can Resolve The Hubble
  Tension}},\ }\href {https://doi.org/10.1103/PhysRevLett.122.221301}
  {\bibfield  {journal} {\bibinfo  {journal} {Phys. Rev. Lett.}\ }\textbf
  {\bibinfo {volume} {122}},\ \bibinfo {pages} {221301} (\bibinfo {year}
  {2019})},\ \Eprint {https://arxiv.org/abs/1811.04083} {arXiv:1811.04083
  [astro-ph.CO]} \BibitemShut {NoStop}%
\bibitem [{\citenamefont {Smith}\ \emph {et~al.}(2020)\citenamefont {Smith},
  \citenamefont {Poulin},\ and\ \citenamefont {Amin}}]{Smith:2019ihp}%
  \BibitemOpen
  \bibfield  {author} {\bibinfo {author} {\bibfnamefont {T.~L.}\ \bibnamefont
  {Smith}}, \bibinfo {author} {\bibfnamefont {V.}~\bibnamefont {Poulin}},\ and\
  \bibinfo {author} {\bibfnamefont {M.~A.}\ \bibnamefont {Amin}},\ }\bibfield
  {title} {\bibinfo {title} {{Oscillating scalar fields and the Hubble tension:
  a resolution with novel signatures}},\ }\href
  {https://doi.org/10.1103/PhysRevD.101.063523} {\bibfield  {journal} {\bibinfo
   {journal} {Phys. Rev. D}\ }\textbf {\bibinfo {volume} {101}},\ \bibinfo
  {pages} {063523} (\bibinfo {year} {2020})},\ \Eprint
  {https://arxiv.org/abs/1908.06995} {arXiv:1908.06995 [astro-ph.CO]}
  \BibitemShut {NoStop}%
\bibitem [{\citenamefont {Niedermann}\ and\ \citenamefont
  {Sloth}(2024)}]{Niedermann:2023ssr}%
  \BibitemOpen
  \bibfield  {author} {\bibinfo {author} {\bibfnamefont {F.}~\bibnamefont
  {Niedermann}}\ and\ \bibinfo {author} {\bibfnamefont {M.~S.}\ \bibnamefont
  {Sloth}},\ }\bibinfo {title} {New early dark energy as a solution to the
  h$_0$ and s$_8$ tensions},\ in\ \href
  {https://doi.org/10.1007/978-981-99-0177-7_23} {\emph {\bibinfo {booktitle}
  {The Hubble Constant Tension}}},\ \bibinfo {editor} {edited by\ \bibinfo
  {editor} {\bibfnamefont {E.}~\bibnamefont {Di~Valentino}}\ and\ \bibinfo
  {editor} {\bibfnamefont {D.}~\bibnamefont {Brout}}}\ (\bibinfo  {publisher}
  {Springer Nature Singapore},\ \bibinfo {address} {Singapore},\ \bibinfo
  {year} {2024})\ pp.\ \bibinfo {pages} {431--456}\BibitemShut {NoStop}%
\bibitem [{\citenamefont {Niedermann}\ and\ \citenamefont
  {Sloth}(2021)}]{Niedermann:2019olb}%
  \BibitemOpen
  \bibfield  {author} {\bibinfo {author} {\bibfnamefont {F.}~\bibnamefont
  {Niedermann}}\ and\ \bibinfo {author} {\bibfnamefont {M.~S.}\ \bibnamefont
  {Sloth}},\ }\bibfield  {title} {\bibinfo {title} {{New early dark energy}},\
  }\href {https://doi.org/10.1103/PhysRevD.103.L041303} {\bibfield  {journal}
  {\bibinfo  {journal} {Phys. Rev. D}\ }\textbf {\bibinfo {volume} {103}},\
  \bibinfo {pages} {L041303} (\bibinfo {year} {2021})},\ \Eprint
  {https://arxiv.org/abs/1910.10739} {arXiv:1910.10739 [astro-ph.CO]}
  \BibitemShut {NoStop}%
\bibitem [{\citenamefont {Niedermann}\ and\ \citenamefont
  {Sloth}(2020)}]{Niedermann:2020dwg}%
  \BibitemOpen
  \bibfield  {author} {\bibinfo {author} {\bibfnamefont {F.}~\bibnamefont
  {Niedermann}}\ and\ \bibinfo {author} {\bibfnamefont {M.~S.}\ \bibnamefont
  {Sloth}},\ }\bibfield  {title} {\bibinfo {title} {{Resolving the Hubble
  tension with new early dark energy}},\ }\href
  {https://doi.org/10.1103/PhysRevD.102.063527} {\bibfield  {journal} {\bibinfo
   {journal} {Phys. Rev. D}\ }\textbf {\bibinfo {volume} {102}},\ \bibinfo
  {pages} {063527} (\bibinfo {year} {2020})},\ \Eprint
  {https://arxiv.org/abs/2006.06686} {arXiv:2006.06686 [astro-ph.CO]}
  \BibitemShut {NoStop}%
\bibitem [{\citenamefont {Cruz}\ \emph
  {et~al.}(2023{\natexlab{a}})\citenamefont {Cruz}, \citenamefont
  {Niedermann},\ and\ \citenamefont {Sloth}}]{Cruz:2023lmn}%
  \BibitemOpen
  \bibfield  {author} {\bibinfo {author} {\bibfnamefont {J.~S.}\ \bibnamefont
  {Cruz}}, \bibinfo {author} {\bibfnamefont {F.}~\bibnamefont {Niedermann}},\
  and\ \bibinfo {author} {\bibfnamefont {M.~S.}\ \bibnamefont {Sloth}},\
  }\bibfield  {title} {\bibinfo {title} {{Cold New Early Dark Energy pulls the
  trigger on the H$_{0}$ and S$_{8}$ tensions: a simultaneous solution to both
  tensions without new ingredients}},\ }\href
  {https://doi.org/10.1088/1475-7516/2023/11/033} {\bibfield  {journal}
  {\bibinfo  {journal} {JCAP}\ }\textbf {\bibinfo {volume} {11}},\ \bibinfo
  {pages} {033}},\ \Eprint {https://arxiv.org/abs/2305.08895} {arXiv:2305.08895
  [astro-ph.CO]} \BibitemShut {NoStop}%
\bibitem [{\citenamefont {Niedermann}\ and\ \citenamefont
  {Sloth}(2022{\natexlab{a}})}]{Niedermann:2021vgd}%
  \BibitemOpen
  \bibfield  {author} {\bibinfo {author} {\bibfnamefont {F.}~\bibnamefont
  {Niedermann}}\ and\ \bibinfo {author} {\bibfnamefont {M.~S.}\ \bibnamefont
  {Sloth}},\ }\bibfield  {title} {\bibinfo {title} {{Hot new early dark
  energy}},\ }\href {https://doi.org/10.1103/PhysRevD.105.063509} {\bibfield
  {journal} {\bibinfo  {journal} {Phys. Rev. D}\ }\textbf {\bibinfo {volume}
  {105}},\ \bibinfo {pages} {063509} (\bibinfo {year} {2022}{\natexlab{a}})},\
  \Eprint {https://arxiv.org/abs/2112.00770} {arXiv:2112.00770 [hep-ph]}
  \BibitemShut {NoStop}%
\bibitem [{\citenamefont {Niedermann}\ and\ \citenamefont
  {Sloth}(2022{\natexlab{b}})}]{Niedermann:2021ijp}%
  \BibitemOpen
  \bibfield  {author} {\bibinfo {author} {\bibfnamefont {F.}~\bibnamefont
  {Niedermann}}\ and\ \bibinfo {author} {\bibfnamefont {M.~S.}\ \bibnamefont
  {Sloth}},\ }\bibfield  {title} {\bibinfo {title} {{Hot new early dark energy:
  Towards a unified dark sector of neutrinos, dark energy and dark matter}},\
  }\href {https://doi.org/10.1016/j.physletb.2022.137555} {\bibfield  {journal}
  {\bibinfo  {journal} {Phys. Lett. B}\ }\textbf {\bibinfo {volume} {835}},\
  \bibinfo {pages} {137555} (\bibinfo {year} {2022}{\natexlab{b}})},\ \Eprint
  {https://arxiv.org/abs/2112.00759} {arXiv:2112.00759 [hep-ph]} \BibitemShut
  {NoStop}%
\bibitem [{\citenamefont {Garny}\ \emph {et~al.}(2024)\citenamefont {Garny},
  \citenamefont {Niedermann}, \citenamefont {Rubira},\ and\ \citenamefont
  {Sloth}}]{Garny:2024ums}%
  \BibitemOpen
  \bibfield  {author} {\bibinfo {author} {\bibfnamefont {M.}~\bibnamefont
  {Garny}}, \bibinfo {author} {\bibfnamefont {F.}~\bibnamefont {Niedermann}},
  \bibinfo {author} {\bibfnamefont {H.}~\bibnamefont {Rubira}},\ and\ \bibinfo
  {author} {\bibfnamefont {M.~S.}\ \bibnamefont {Sloth}},\ }\bibfield  {title}
  {\bibinfo {title} {{Hot new early dark energy bridging cosmic gaps:
  Supercooled phase transition reconciles stepped dark radiation solutions to
  the Hubble tension with BBN}},\ }\href
  {https://doi.org/10.1103/PhysRevD.110.023531} {\bibfield  {journal} {\bibinfo
   {journal} {Phys. Rev. D}\ }\textbf {\bibinfo {volume} {110}},\ \bibinfo
  {pages} {023531} (\bibinfo {year} {2024})},\ \Eprint
  {https://arxiv.org/abs/2404.07256} {arXiv:2404.07256 [astro-ph.CO]}
  \BibitemShut {NoStop}%
\bibitem [{\citenamefont {Cruz}\ \emph
  {et~al.}(2023{\natexlab{b}})\citenamefont {Cruz}, \citenamefont
  {Niedermann},\ and\ \citenamefont {Sloth}}]{Cruz:2022oqk}%
  \BibitemOpen
  \bibfield  {author} {\bibinfo {author} {\bibfnamefont {J.~S.}\ \bibnamefont
  {Cruz}}, \bibinfo {author} {\bibfnamefont {F.}~\bibnamefont {Niedermann}},\
  and\ \bibinfo {author} {\bibfnamefont {M.~S.}\ \bibnamefont {Sloth}},\
  }\bibfield  {title} {\bibinfo {title} {{A grounded perspective on new early
  dark energy using ACT, SPT, and BICEP/Keck}},\ }\href
  {https://doi.org/10.1088/1475-7516/2023/02/041} {\bibfield  {journal}
  {\bibinfo  {journal} {JCAP}\ }\textbf {\bibinfo {volume} {02}},\ \bibinfo
  {pages} {041}},\ \Eprint {https://arxiv.org/abs/2209.02708} {arXiv:2209.02708
  [astro-ph.CO]} \BibitemShut {NoStop}%
\bibitem [{\citenamefont {Cruz}\ \emph
  {et~al.}(2023{\natexlab{c}})\citenamefont {Cruz}, \citenamefont {Hannestad},
  \citenamefont {Holm}, \citenamefont {Niedermann}, \citenamefont {Sloth},\
  and\ \citenamefont {Tram}}]{Cruz:2023cxy}%
  \BibitemOpen
  \bibfield  {author} {\bibinfo {author} {\bibfnamefont {J.~S.}\ \bibnamefont
  {Cruz}}, \bibinfo {author} {\bibfnamefont {S.}~\bibnamefont {Hannestad}},
  \bibinfo {author} {\bibfnamefont {E.~B.}\ \bibnamefont {Holm}}, \bibinfo
  {author} {\bibfnamefont {F.}~\bibnamefont {Niedermann}}, \bibinfo {author}
  {\bibfnamefont {M.~S.}\ \bibnamefont {Sloth}},\ and\ \bibinfo {author}
  {\bibfnamefont {T.}~\bibnamefont {Tram}},\ }\bibfield  {title} {\bibinfo
  {title} {{Profiling cold new early dark energy}},\ }\href
  {https://doi.org/10.1103/PhysRevD.108.023518} {\bibfield  {journal} {\bibinfo
   {journal} {Phys. Rev. D}\ }\textbf {\bibinfo {volume} {108}},\ \bibinfo
  {pages} {023518} (\bibinfo {year} {2023}{\natexlab{c}})},\ \Eprint
  {https://arxiv.org/abs/2302.07934} {arXiv:2302.07934 [astro-ph.CO]}
  \BibitemShut {NoStop}%
\bibitem [{\citenamefont {Akrami}\ \emph {et~al.}(2020)\citenamefont {Akrami}
  \emph {et~al.}}]{Planck:2020olo}%
  \BibitemOpen
  \bibfield  {author} {\bibinfo {author} {\bibfnamefont {Y.}~\bibnamefont
  {Akrami}} \emph {et~al.} (\bibinfo {collaboration} {Planck}),\ }\bibfield
  {title} {\bibinfo {title} {{$Planck$ intermediate results. LVII. Joint Planck
  LFI and HFI data processing}},\ }\href
  {https://doi.org/10.1051/0004-6361/202038073} {\bibfield  {journal} {\bibinfo
   {journal} {Astron. Astrophys.}\ }\textbf {\bibinfo {volume} {643}},\
  \bibinfo {pages} {A42} (\bibinfo {year} {2020})},\ \Eprint
  {https://arxiv.org/abs/2007.04997} {arXiv:2007.04997 [astro-ph.CO]}
  \BibitemShut {NoStop}%
\bibitem [{\citenamefont {Efstathiou}\ \emph {et~al.}(2024)\citenamefont
  {Efstathiou}, \citenamefont {Rosenberg},\ and\ \citenamefont
  {Poulin}}]{Efstathiou:2023fbn}%
  \BibitemOpen
  \bibfield  {author} {\bibinfo {author} {\bibfnamefont {G.}~\bibnamefont
  {Efstathiou}}, \bibinfo {author} {\bibfnamefont {E.}~\bibnamefont
  {Rosenberg}},\ and\ \bibinfo {author} {\bibfnamefont {V.}~\bibnamefont
  {Poulin}},\ }\bibfield  {title} {\bibinfo {title} {{Improved Planck
  Constraints on Axionlike Early Dark Energy as a Resolution of the Hubble
  Tension}},\ }\href {https://doi.org/10.1103/PhysRevLett.132.221002}
  {\bibfield  {journal} {\bibinfo  {journal} {Phys. Rev. Lett.}\ }\textbf
  {\bibinfo {volume} {132}},\ \bibinfo {pages} {221002} (\bibinfo {year}
  {2024})},\ \Eprint {https://arxiv.org/abs/2311.00524} {arXiv:2311.00524
  [astro-ph.CO]} \BibitemShut {NoStop}%
\bibitem [{\citenamefont {Rosenberg}\ \emph {et~al.}(2022)\citenamefont
  {Rosenberg}, \citenamefont {Gratton},\ and\ \citenamefont
  {Efstathiou}}]{Rosenberg:2022sdy}%
  \BibitemOpen
  \bibfield  {author} {\bibinfo {author} {\bibfnamefont {E.}~\bibnamefont
  {Rosenberg}}, \bibinfo {author} {\bibfnamefont {S.}~\bibnamefont {Gratton}},\
  and\ \bibinfo {author} {\bibfnamefont {G.}~\bibnamefont {Efstathiou}},\
  }\bibfield  {title} {\bibinfo {title} {{CMB power spectra and cosmological
  parameters from Planck PR4 with CamSpec}},\ }\href
  {https://doi.org/10.1093/mnras/stac2744} {\bibfield  {journal} {\bibinfo
  {journal} {Mon. Not. Roy. Astron. Soc.}\ }\textbf {\bibinfo {volume} {517}},\
  \bibinfo {pages} {4620} (\bibinfo {year} {2022})},\ \Eprint
  {https://arxiv.org/abs/2205.10869} {arXiv:2205.10869 [astro-ph.CO]}
  \BibitemShut {NoStop}%
\bibitem [{\citenamefont {Tristram}\ \emph {et~al.}(2024)\citenamefont
  {Tristram} \emph {et~al.}}]{Tristram:2023haj}%
  \BibitemOpen
  \bibfield  {author} {\bibinfo {author} {\bibfnamefont {M.}~\bibnamefont
  {Tristram}} \emph {et~al.},\ }\bibfield  {title} {\bibinfo {title}
  {{Cosmological parameters derived from the final Planck data release
  (PR4)}},\ }\href {https://doi.org/10.1051/0004-6361/202348015} {\bibfield
  {journal} {\bibinfo  {journal} {Astron. Astrophys.}\ }\textbf {\bibinfo
  {volume} {682}},\ \bibinfo {pages} {A37} (\bibinfo {year} {2024})},\ \Eprint
  {https://arxiv.org/abs/2309.10034} {arXiv:2309.10034 [astro-ph.CO]}
  \BibitemShut {NoStop}%
\bibitem [{\citenamefont {Brout}\ \emph {et~al.}(2022)\citenamefont {Brout}
  \emph {et~al.}}]{Brout:2022vxf}%
  \BibitemOpen
  \bibfield  {author} {\bibinfo {author} {\bibfnamefont {D.}~\bibnamefont
  {Brout}} \emph {et~al.},\ }\bibfield  {title} {\bibinfo {title} {{The
  Pantheon+ Analysis: Cosmological Constraints}},\ }\href
  {https://doi.org/10.3847/1538-4357/ac8e04} {\bibfield  {journal} {\bibinfo
  {journal} {Astrophys. J.}\ }\textbf {\bibinfo {volume} {938}},\ \bibinfo
  {pages} {110} (\bibinfo {year} {2022})},\ \Eprint
  {https://arxiv.org/abs/2202.04077} {arXiv:2202.04077 [astro-ph.CO]}
  \BibitemShut {NoStop}%
\bibitem [{\citenamefont {Adame}\ \emph {et~al.}(2024)\citenamefont {Adame}
  \emph {et~al.}}]{DESI:2024mwx}%
  \BibitemOpen
  \bibfield  {author} {\bibinfo {author} {\bibfnamefont {A.~G.}\ \bibnamefont
  {Adame}} \emph {et~al.} (\bibinfo {collaboration} {DESI}),\ }\bibfield
  {title} {\bibinfo {title} {{DESI 2024 VI: Cosmological Constraints from the
  Measurements of Baryon Acoustic Oscillations}},\ }\href@noop {} {\  (\bibinfo
  {year} {2024})},\ \Eprint {https://arxiv.org/abs/2404.03002}
  {arXiv:2404.03002 [astro-ph.CO]} \BibitemShut {NoStop}%
\bibitem [{\citenamefont {Qu}\ \emph {et~al.}(2024)\citenamefont {Qu},
  \citenamefont {Surrao}, \citenamefont {Bolliet}, \citenamefont {Hill},
  \citenamefont {Sherwin},\ and\ \citenamefont {Jense}}]{Qu:2024lpx}%
  \BibitemOpen
  \bibfield  {author} {\bibinfo {author} {\bibfnamefont {F.~J.}\ \bibnamefont
  {Qu}}, \bibinfo {author} {\bibfnamefont {K.~M.}\ \bibnamefont {Surrao}},
  \bibinfo {author} {\bibfnamefont {B.}~\bibnamefont {Bolliet}}, \bibinfo
  {author} {\bibfnamefont {J.~C.}\ \bibnamefont {Hill}}, \bibinfo {author}
  {\bibfnamefont {B.~D.}\ \bibnamefont {Sherwin}},\ and\ \bibinfo {author}
  {\bibfnamefont {H.~T.}\ \bibnamefont {Jense}},\ }\bibfield  {title} {\bibinfo
  {title} {{Accelerated inference on accelerated cosmic expansion: New
  constraints on axion-like early dark energy with DESI BAO and ACT DR6 CMB
  lensing}},\ }\href@noop {} {\  (\bibinfo {year} {2024})},\ \Eprint
  {https://arxiv.org/abs/2404.16805} {arXiv:2404.16805 [astro-ph.CO]}
  \BibitemShut {NoStop}%
\bibitem [{\citenamefont {Poulin}\ \emph {et~al.}(2024)\citenamefont {Poulin},
  \citenamefont {Smith}, \citenamefont {Calder\'on},\ and\ \citenamefont
  {Simon}}]{Poulin:2024ken}%
  \BibitemOpen
  \bibfield  {author} {\bibinfo {author} {\bibfnamefont {V.}~\bibnamefont
  {Poulin}}, \bibinfo {author} {\bibfnamefont {T.~L.}\ \bibnamefont {Smith}},
  \bibinfo {author} {\bibfnamefont {R.}~\bibnamefont {Calder\'on}},\ and\
  \bibinfo {author} {\bibfnamefont {T.}~\bibnamefont {Simon}},\ }\bibfield
  {title} {\bibinfo {title} {{On the implications of the `cosmic calibration
  tension' beyond $H_0$ and the synergy between early- and late-time new
  physics}},\ }\href@noop {} {\  (\bibinfo {year} {2024})},\ \Eprint
  {https://arxiv.org/abs/2407.18292} {arXiv:2407.18292 [astro-ph.CO]}
  \BibitemShut {NoStop}%
\bibitem [{\citenamefont {Preskill}\ \emph {et~al.}(1983)\citenamefont
  {Preskill}, \citenamefont {Wise},\ and\ \citenamefont
  {Wilczek}}]{Preskill:1982cy}%
  \BibitemOpen
  \bibfield  {author} {\bibinfo {author} {\bibfnamefont {J.}~\bibnamefont
  {Preskill}}, \bibinfo {author} {\bibfnamefont {M.~B.}\ \bibnamefont {Wise}},\
  and\ \bibinfo {author} {\bibfnamefont {F.}~\bibnamefont {Wilczek}},\
  }\bibfield  {title} {\bibinfo {title} {{Cosmology of the Invisible Axion}},\
  }\href {https://doi.org/10.1016/0370-2693(83)90637-8} {\bibfield  {journal}
  {\bibinfo  {journal} {Phys. Lett. B}\ }\textbf {\bibinfo {volume} {120}},\
  \bibinfo {pages} {127} (\bibinfo {year} {1983})}\BibitemShut {NoStop}%
\bibitem [{\citenamefont {Abbott}\ and\ \citenamefont
  {Sikivie}(1983)}]{Abbott:1982af}%
  \BibitemOpen
  \bibfield  {author} {\bibinfo {author} {\bibfnamefont {L.~F.}\ \bibnamefont
  {Abbott}}\ and\ \bibinfo {author} {\bibfnamefont {P.}~\bibnamefont
  {Sikivie}},\ }\bibfield  {title} {\bibinfo {title} {{A Cosmological Bound on
  the Invisible Axion}},\ }\href {https://doi.org/10.1016/0370-2693(83)90638-X}
  {\bibfield  {journal} {\bibinfo  {journal} {Phys. Lett. B}\ }\textbf
  {\bibinfo {volume} {120}},\ \bibinfo {pages} {133} (\bibinfo {year}
  {1983})}\BibitemShut {NoStop}%
\bibitem [{\citenamefont {Dine}\ and\ \citenamefont
  {Fischler}(1983)}]{Dine:1982ah}%
  \BibitemOpen
  \bibfield  {author} {\bibinfo {author} {\bibfnamefont {M.}~\bibnamefont
  {Dine}}\ and\ \bibinfo {author} {\bibfnamefont {W.}~\bibnamefont
  {Fischler}},\ }\bibfield  {title} {\bibinfo {title} {{The Not So Harmless
  Axion}},\ }\href {https://doi.org/10.1016/0370-2693(83)90639-1} {\bibfield
  {journal} {\bibinfo  {journal} {Phys. Lett. B}\ }\textbf {\bibinfo {volume}
  {120}},\ \bibinfo {pages} {137} (\bibinfo {year} {1983})}\BibitemShut
  {NoStop}%
\bibitem [{\citenamefont {Turner}(1983)}]{Turner:1983he}%
  \BibitemOpen
  \bibfield  {author} {\bibinfo {author} {\bibfnamefont {M.~S.}\ \bibnamefont
  {Turner}},\ }\bibfield  {title} {\bibinfo {title} {{Coherent Scalar Field
  Oscillations in an Expanding Universe}},\ }\href
  {https://doi.org/10.1103/PhysRevD.28.1243} {\bibfield  {journal} {\bibinfo
  {journal} {Phys. Rev. D}\ }\textbf {\bibinfo {volume} {28}},\ \bibinfo
  {pages} {1243} (\bibinfo {year} {1983})}\BibitemShut {NoStop}%
\bibitem [{\citenamefont {Brandenberger}(1985)}]{Brandenberger:1984jq}%
  \BibitemOpen
  \bibfield  {author} {\bibinfo {author} {\bibfnamefont {R.~H.}\ \bibnamefont
  {Brandenberger}},\ }\bibfield  {title} {\bibinfo {title} {{Cosmological
  Perturbations in a Universe Dominated by a Coherent Scalar Field}},\ }\href
  {https://doi.org/10.1103/PhysRevD.32.501} {\bibfield  {journal} {\bibinfo
  {journal} {Phys. Rev. D}\ }\textbf {\bibinfo {volume} {32}},\ \bibinfo
  {pages} {501} (\bibinfo {year} {1985})}\BibitemShut {NoStop}%
\bibitem [{\citenamefont {Hu}\ \emph {et~al.}(2000)\citenamefont {Hu},
  \citenamefont {Barkana},\ and\ \citenamefont {Gruzinov}}]{Hu:2000ke}%
  \BibitemOpen
  \bibfield  {author} {\bibinfo {author} {\bibfnamefont {W.}~\bibnamefont
  {Hu}}, \bibinfo {author} {\bibfnamefont {R.}~\bibnamefont {Barkana}},\ and\
  \bibinfo {author} {\bibfnamefont {A.}~\bibnamefont {Gruzinov}},\ }\bibfield
  {title} {\bibinfo {title} {{Cold and fuzzy dark matter}},\ }\href
  {https://doi.org/10.1103/PhysRevLett.85.1158} {\bibfield  {journal} {\bibinfo
   {journal} {Phys. Rev. Lett.}\ }\textbf {\bibinfo {volume} {85}},\ \bibinfo
  {pages} {1158} (\bibinfo {year} {2000})},\ \Eprint
  {https://arxiv.org/abs/astro-ph/0003365} {arXiv:astro-ph/0003365}
  \BibitemShut {NoStop}%
\bibitem [{\citenamefont {Hwang}\ and\ \citenamefont
  {Noh}(2009)}]{Hwang:2009js}%
  \BibitemOpen
  \bibfield  {author} {\bibinfo {author} {\bibfnamefont {J.-c.}\ \bibnamefont
  {Hwang}}\ and\ \bibinfo {author} {\bibfnamefont {H.}~\bibnamefont {Noh}},\
  }\bibfield  {title} {\bibinfo {title} {{Axion as a Cold Dark Matter
  candidate}},\ }\href {https://doi.org/10.1016/j.physletb.2009.08.031}
  {\bibfield  {journal} {\bibinfo  {journal} {Phys. Lett. B}\ }\textbf
  {\bibinfo {volume} {680}},\ \bibinfo {pages} {1} (\bibinfo {year} {2009})},\
  \Eprint {https://arxiv.org/abs/0902.4738} {arXiv:0902.4738 [astro-ph.CO]}
  \BibitemShut {NoStop}%
\bibitem [{\citenamefont {Israel}(1966)}]{Israel:1966rt}%
  \BibitemOpen
  \bibfield  {author} {\bibinfo {author} {\bibfnamefont {W.}~\bibnamefont
  {Israel}},\ }\bibfield  {title} {\bibinfo {title} {{Singular hypersurfaces
  and thin shells in general relativity}},\ }\href
  {https://doi.org/10.1007/BF02710419} {\bibfield  {journal} {\bibinfo
  {journal} {Nuovo Cim. B}\ }\textbf {\bibinfo {volume} {44S10}},\ \bibinfo
  {pages} {1} (\bibinfo {year} {1966})},\ \bibinfo {note} {[Erratum: Nuovo
  Cim.B 48, 463 (1967)]}\BibitemShut {NoStop}%
\bibitem [{\citenamefont {Deruelle}\ and\ \citenamefont
  {Mukhanov}(1995)}]{Deruelle:1995kd}%
  \BibitemOpen
  \bibfield  {author} {\bibinfo {author} {\bibfnamefont {N.}~\bibnamefont
  {Deruelle}}\ and\ \bibinfo {author} {\bibfnamefont {V.~F.}\ \bibnamefont
  {Mukhanov}},\ }\bibfield  {title} {\bibinfo {title} {{On matching conditions
  for cosmological perturbations}},\ }\href
  {https://doi.org/10.1103/PhysRevD.52.5549} {\bibfield  {journal} {\bibinfo
  {journal} {Phys. Rev. D}\ }\textbf {\bibinfo {volume} {52}},\ \bibinfo
  {pages} {5549} (\bibinfo {year} {1995})},\ \Eprint
  {https://arxiv.org/abs/gr-qc/9503050} {arXiv:gr-qc/9503050} \BibitemShut
  {NoStop}%
\bibitem [{\citenamefont {Elor}\ \emph {et~al.}(2023)\citenamefont {Elor},
  \citenamefont {Jinno}, \citenamefont {Kumar}, \citenamefont {McGehee},\ and\
  \citenamefont {Tsai}}]{Elor:2023xbz}%
  \BibitemOpen
  \bibfield  {author} {\bibinfo {author} {\bibfnamefont {G.}~\bibnamefont
  {Elor}}, \bibinfo {author} {\bibfnamefont {R.}~\bibnamefont {Jinno}},
  \bibinfo {author} {\bibfnamefont {S.}~\bibnamefont {Kumar}}, \bibinfo
  {author} {\bibfnamefont {R.}~\bibnamefont {McGehee}},\ and\ \bibinfo {author}
  {\bibfnamefont {Y.}~\bibnamefont {Tsai}},\ }\bibfield  {title} {\bibinfo
  {title} {{Finite Bubble Statistics Constrain Late Cosmological Phase
  Transitions}},\ }\href@noop {} {\  (\bibinfo {year} {2023})},\ \Eprint
  {https://arxiv.org/abs/2311.16222} {arXiv:2311.16222 [hep-ph]} \BibitemShut
  {NoStop}%
\bibitem [{\citenamefont {Poulin}\ \emph {et~al.}(2018)\citenamefont {Poulin},
  \citenamefont {Smith}, \citenamefont {Grin}, \citenamefont {Karwal},\ and\
  \citenamefont {Kamionkowski}}]{Poulin:2018dzj}%
  \BibitemOpen
  \bibfield  {author} {\bibinfo {author} {\bibfnamefont {V.}~\bibnamefont
  {Poulin}}, \bibinfo {author} {\bibfnamefont {T.~L.}\ \bibnamefont {Smith}},
  \bibinfo {author} {\bibfnamefont {D.}~\bibnamefont {Grin}}, \bibinfo {author}
  {\bibfnamefont {T.}~\bibnamefont {Karwal}},\ and\ \bibinfo {author}
  {\bibfnamefont {M.}~\bibnamefont {Kamionkowski}},\ }\bibfield  {title}
  {\bibinfo {title} {{Cosmological implications of ultralight axionlike
  fields}},\ }\href {https://doi.org/10.1103/PhysRevD.98.083525} {\bibfield
  {journal} {\bibinfo  {journal} {Phys. Rev. D}\ }\textbf {\bibinfo {volume}
  {98}},\ \bibinfo {pages} {083525} (\bibinfo {year} {2018})},\ \Eprint
  {https://arxiv.org/abs/1806.10608} {arXiv:1806.10608 [astro-ph.CO]}
  \BibitemShut {NoStop}%
\bibitem [{\citenamefont {Bringmann}\ \emph {et~al.}(2018)\citenamefont
  {Bringmann}, \citenamefont {Kahlhoefer}, \citenamefont {Schmidt-Hoberg},\
  and\ \citenamefont {Walia}}]{Bringmann:2018jpr}%
  \BibitemOpen
  \bibfield  {author} {\bibinfo {author} {\bibfnamefont {T.}~\bibnamefont
  {Bringmann}}, \bibinfo {author} {\bibfnamefont {F.}~\bibnamefont
  {Kahlhoefer}}, \bibinfo {author} {\bibfnamefont {K.}~\bibnamefont
  {Schmidt-Hoberg}},\ and\ \bibinfo {author} {\bibfnamefont {P.}~\bibnamefont
  {Walia}},\ }\bibfield  {title} {\bibinfo {title} {{Converting nonrelativistic
  dark matter to radiation}},\ }\href
  {https://doi.org/10.1103/PhysRevD.98.023543} {\bibfield  {journal} {\bibinfo
  {journal} {Phys. Rev. D}\ }\textbf {\bibinfo {volume} {98}},\ \bibinfo
  {pages} {023543} (\bibinfo {year} {2018})},\ \Eprint
  {https://arxiv.org/abs/1803.03644} {arXiv:1803.03644 [astro-ph.CO]}
  \BibitemShut {NoStop}%
\bibitem [{\citenamefont {Adams}\ and\ \citenamefont
  {Freese}(1991)}]{Adams:1990ds}%
  \BibitemOpen
  \bibfield  {author} {\bibinfo {author} {\bibfnamefont {F.~C.}\ \bibnamefont
  {Adams}}\ and\ \bibinfo {author} {\bibfnamefont {K.}~\bibnamefont {Freese}},\
  }\bibfield  {title} {\bibinfo {title} {{Double field inflation}},\ }\href
  {https://doi.org/10.1103/PhysRevD.43.353} {\bibfield  {journal} {\bibinfo
  {journal} {Phys. Rev. D}\ }\textbf {\bibinfo {volume} {43}},\ \bibinfo
  {pages} {353} (\bibinfo {year} {1991})},\ \Eprint
  {https://arxiv.org/abs/hep-ph/0504135} {arXiv:hep-ph/0504135} \BibitemShut
  {NoStop}%
\bibitem [{\citenamefont {Linde}(1990)}]{Linde:1990gz}%
  \BibitemOpen
  \bibfield  {author} {\bibinfo {author} {\bibfnamefont {A.~D.}\ \bibnamefont
  {Linde}},\ }\bibfield  {title} {\bibinfo {title} {{Eternal extended inflation
  and graceful exit from old inflation without Jordan-Brans-Dicke}},\ }\href
  {https://doi.org/10.1016/0370-2693(90)90521-7} {\bibfield  {journal}
  {\bibinfo  {journal} {Phys. Lett. B}\ }\textbf {\bibinfo {volume} {249}},\
  \bibinfo {pages} {18} (\bibinfo {year} {1990})}\BibitemShut {NoStop}%
\bibitem [{\citenamefont {Xue}\ and\ \citenamefont
  {Steinhardt}(2011)}]{Xue:2011nw}%
  \BibitemOpen
  \bibfield  {author} {\bibinfo {author} {\bibfnamefont {B.}~\bibnamefont
  {Xue}}\ and\ \bibinfo {author} {\bibfnamefont {P.~J.}\ \bibnamefont
  {Steinhardt}},\ }\bibfield  {title} {\bibinfo {title} {{Evolution of
  curvature and anisotropy near a nonsingular bounce}},\ }\href
  {https://doi.org/10.1103/PhysRevD.84.083520} {\bibfield  {journal} {\bibinfo
  {journal} {Phys. Rev. D}\ }\textbf {\bibinfo {volume} {84}},\ \bibinfo
  {pages} {083520} (\bibinfo {year} {2011})},\ \Eprint
  {https://arxiv.org/abs/1106.1416} {arXiv:1106.1416 [hep-th]} \BibitemShut
  {NoStop}%
\bibitem [{\citenamefont {Zel'dovich}(1961)}]{Zeldovich:1961sbr}%
  \BibitemOpen
  \bibfield  {author} {\bibinfo {author} {\bibfnamefont {Y.~B.}\ \bibnamefont
  {Zel'dovich}},\ }\bibfield  {title} {\bibinfo {title} {{The equation of state
  at ultrahigh densities and its relativistic limitations}},\ }\href@noop {}
  {\bibfield  {journal} {\bibinfo  {journal} {Zh. Eksp. Teor. Fiz.}\ }\textbf
  {\bibinfo {volume} {41}},\ \bibinfo {pages} {1609} (\bibinfo {year}
  {1961})}\BibitemShut {NoStop}%
\bibitem [{\citenamefont {Niedermann}\ and\ \citenamefont
  {Sloth}()}]{NiedermannandSlothtoappear}%
  \BibitemOpen
  \bibfield  {author} {\bibinfo {author} {\bibfnamefont {F.}~\bibnamefont
  {Niedermann}}\ and\ \bibinfo {author} {\bibfnamefont {M.~S.}\ \bibnamefont
  {Sloth}},\ }\bibfield  {title} {\bibinfo {title} {{to appear}},\ }\href@noop
  {} {\ }\BibitemShut {NoStop}%
\bibitem [{\citenamefont {Blas}\ \emph {et~al.}(2011)\citenamefont {Blas},
  \citenamefont {Lesgourgues},\ and\ \citenamefont {Tram}}]{Blas:2011rf}%
  \BibitemOpen
  \bibfield  {author} {\bibinfo {author} {\bibfnamefont {D.}~\bibnamefont
  {Blas}}, \bibinfo {author} {\bibfnamefont {J.}~\bibnamefont {Lesgourgues}},\
  and\ \bibinfo {author} {\bibfnamefont {T.}~\bibnamefont {Tram}},\ }\bibfield
  {title} {\bibinfo {title} {{The Cosmic Linear Anisotropy Solving System
  (CLASS) II: Approximation schemes}},\ }\href
  {https://doi.org/10.1088/1475-7516/2011/07/034} {\bibfield  {journal}
  {\bibinfo  {journal} {JCAP}\ }\textbf {\bibinfo {volume} {07}},\ \bibinfo
  {pages} {034}},\ \Eprint {https://arxiv.org/abs/1104.2933} {arXiv:1104.2933
  [astro-ph.CO]} \BibitemShut {NoStop}%
\bibitem [{\citenamefont {Audren}\ \emph {et~al.}(2013)\citenamefont {Audren},
  \citenamefont {Lesgourgues}, \citenamefont {Benabed},\ and\ \citenamefont
  {Prunet}}]{Audren:2012wb}%
  \BibitemOpen
  \bibfield  {author} {\bibinfo {author} {\bibfnamefont {B.}~\bibnamefont
  {Audren}}, \bibinfo {author} {\bibfnamefont {J.}~\bibnamefont {Lesgourgues}},
  \bibinfo {author} {\bibfnamefont {K.}~\bibnamefont {Benabed}},\ and\ \bibinfo
  {author} {\bibfnamefont {S.}~\bibnamefont {Prunet}},\ }\bibfield  {title}
  {\bibinfo {title} {{Conservative Constraints on Early Cosmology: an
  illustration of the Monte Python cosmological parameter inference code}},\
  }\href {https://doi.org/10.1088/1475-7516/2013/02/001} {\bibfield  {journal}
  {\bibinfo  {journal} {JCAP}\ }\textbf {\bibinfo {volume} {02}},\ \bibinfo
  {pages} {001}},\ \Eprint {https://arxiv.org/abs/1210.7183} {arXiv:1210.7183
  [astro-ph.CO]} \BibitemShut {NoStop}%
\bibitem [{\citenamefont {Brinckmann}\ and\ \citenamefont
  {Lesgourgues}(2019)}]{Brinckmann:2018cvx}%
  \BibitemOpen
  \bibfield  {author} {\bibinfo {author} {\bibfnamefont {T.}~\bibnamefont
  {Brinckmann}}\ and\ \bibinfo {author} {\bibfnamefont {J.}~\bibnamefont
  {Lesgourgues}},\ }\bibfield  {title} {\bibinfo {title} {{MontePython 3:
  boosted MCMC sampler and other features}},\ }\href
  {https://doi.org/10.1016/j.dark.2018.100260} {\bibfield  {journal} {\bibinfo
  {journal} {Phys. Dark Univ.}\ }\textbf {\bibinfo {volume} {24}},\ \bibinfo
  {pages} {100260} (\bibinfo {year} {2019})},\ \Eprint
  {https://arxiv.org/abs/1804.07261} {arXiv:1804.07261 [astro-ph.CO]}
  \BibitemShut {NoStop}%
\bibitem [{\citenamefont {Efstathiou}\ and\ \citenamefont
  {Gratton}(2019)}]{Efstathiou:2019mdh}%
  \BibitemOpen
  \bibfield  {author} {\bibinfo {author} {\bibfnamefont {G.}~\bibnamefont
  {Efstathiou}}\ and\ \bibinfo {author} {\bibfnamefont {S.}~\bibnamefont
  {Gratton}},\ }\bibfield  {title} {\bibinfo {title} {{A Detailed Description
  of the CamSpec Likelihood Pipeline and a Reanalysis of the Planck High
  Frequency Maps}}\ }\href {https://doi.org/10.21105/astro.1910.00483}
  {10.21105/astro.1910.00483} (\bibinfo {year} {2019}),\ \Eprint
  {https://arxiv.org/abs/1910.00483} {arXiv:1910.00483 [astro-ph.CO]}
  \BibitemShut {NoStop}%
\bibitem [{\citenamefont {Aghanim}\ \emph
  {et~al.}(2020{\natexlab{b}})\citenamefont {Aghanim} \emph
  {et~al.}}]{Planck:2018lbu}%
  \BibitemOpen
  \bibfield  {author} {\bibinfo {author} {\bibfnamefont {N.}~\bibnamefont
  {Aghanim}} \emph {et~al.} (\bibinfo {collaboration} {Planck}),\ }\bibfield
  {title} {\bibinfo {title} {{Planck 2018 results. VIII. Gravitational
  lensing}},\ }\href {https://doi.org/10.1051/0004-6361/201833886} {\bibfield
  {journal} {\bibinfo  {journal} {Astron. Astrophys.}\ }\textbf {\bibinfo
  {volume} {641}},\ \bibinfo {pages} {A8} (\bibinfo {year}
  {2020}{\natexlab{b}})},\ \Eprint {https://arxiv.org/abs/1807.06210}
  {arXiv:1807.06210 [astro-ph.CO]} \BibitemShut {NoStop}%
\bibitem [{\citenamefont {Beutler}\ \emph {et~al.}(2011)\citenamefont
  {Beutler}, \citenamefont {Blake}, \citenamefont {Colless}, \citenamefont
  {Jones}, \citenamefont {Staveley-Smith}, \citenamefont {Campbell},
  \citenamefont {Parker}, \citenamefont {Saunders},\ and\ \citenamefont
  {Watson}}]{Beutler:2011hx}%
  \BibitemOpen
  \bibfield  {author} {\bibinfo {author} {\bibfnamefont {F.}~\bibnamefont
  {Beutler}}, \bibinfo {author} {\bibfnamefont {C.}~\bibnamefont {Blake}},
  \bibinfo {author} {\bibfnamefont {M.}~\bibnamefont {Colless}}, \bibinfo
  {author} {\bibfnamefont {D.~H.}\ \bibnamefont {Jones}}, \bibinfo {author}
  {\bibfnamefont {L.}~\bibnamefont {Staveley-Smith}}, \bibinfo {author}
  {\bibfnamefont {L.}~\bibnamefont {Campbell}}, \bibinfo {author}
  {\bibfnamefont {Q.}~\bibnamefont {Parker}}, \bibinfo {author} {\bibfnamefont
  {W.}~\bibnamefont {Saunders}},\ and\ \bibinfo {author} {\bibfnamefont
  {F.}~\bibnamefont {Watson}},\ }\bibfield  {title} {\bibinfo {title} {{The 6dF
  Galaxy Survey: Baryon Acoustic Oscillations and the Local Hubble Constant}},\
  }\href {https://doi.org/10.1111/j.1365-2966.2011.19250.x} {\bibfield
  {journal} {\bibinfo  {journal} {Mon. Not. Roy. Astron. Soc.}\ }\textbf
  {\bibinfo {volume} {416}},\ \bibinfo {pages} {3017} (\bibinfo {year}
  {2011})},\ \Eprint {https://arxiv.org/abs/1106.3366} {arXiv:1106.3366
  [astro-ph.CO]} \BibitemShut {NoStop}%
\bibitem [{\citenamefont {Ross}\ \emph {et~al.}(2015)\citenamefont {Ross},
  \citenamefont {Samushia}, \citenamefont {Howlett}, \citenamefont {Percival},
  \citenamefont {Burden},\ and\ \citenamefont {Manera}}]{Ross:2014qpa}%
  \BibitemOpen
  \bibfield  {author} {\bibinfo {author} {\bibfnamefont {A.~J.}\ \bibnamefont
  {Ross}}, \bibinfo {author} {\bibfnamefont {L.}~\bibnamefont {Samushia}},
  \bibinfo {author} {\bibfnamefont {C.}~\bibnamefont {Howlett}}, \bibinfo
  {author} {\bibfnamefont {W.~J.}\ \bibnamefont {Percival}}, \bibinfo {author}
  {\bibfnamefont {A.}~\bibnamefont {Burden}},\ and\ \bibinfo {author}
  {\bibfnamefont {M.}~\bibnamefont {Manera}},\ }\bibfield  {title} {\bibinfo
  {title} {{The clustering of the SDSS DR7 main Galaxy sample \textendash{} I.
  A 4 per cent distance measure at $z = 0.15$}},\ }\href
  {https://doi.org/10.1093/mnras/stv154} {\bibfield  {journal} {\bibinfo
  {journal} {Mon. Not. Roy. Astron. Soc.}\ }\textbf {\bibinfo {volume} {449}},\
  \bibinfo {pages} {835} (\bibinfo {year} {2015})},\ \Eprint
  {https://arxiv.org/abs/1409.3242} {arXiv:1409.3242 [astro-ph.CO]}
  \BibitemShut {NoStop}%
\bibitem [{\citenamefont {Alam}\ \emph {et~al.}(2017)\citenamefont {Alam} \emph
  {et~al.}}]{BOSS:2016wmc}%
  \BibitemOpen
  \bibfield  {author} {\bibinfo {author} {\bibfnamefont {S.}~\bibnamefont
  {Alam}} \emph {et~al.} (\bibinfo {collaboration} {BOSS}),\ }\bibfield
  {title} {\bibinfo {title} {{The clustering of galaxies in the completed
  SDSS-III Baryon Oscillation Spectroscopic Survey: cosmological analysis of
  the DR12 galaxy sample}},\ }\href {https://doi.org/10.1093/mnras/stx721}
  {\bibfield  {journal} {\bibinfo  {journal} {Mon. Not. Roy. Astron. Soc.}\
  }\textbf {\bibinfo {volume} {470}},\ \bibinfo {pages} {2617} (\bibinfo {year}
  {2017})},\ \Eprint {https://arxiv.org/abs/1607.03155} {arXiv:1607.03155
  [astro-ph.CO]} \BibitemShut {NoStop}%
\bibitem [{\citenamefont {Joudaki}\ \emph {et~al.}(2020)\citenamefont {Joudaki}
  \emph {et~al.}}]{Joudaki:2019pmv}%
  \BibitemOpen
  \bibfield  {author} {\bibinfo {author} {\bibfnamefont {S.}~\bibnamefont
  {Joudaki}} \emph {et~al.},\ }\bibfield  {title} {\bibinfo {title}
  {{KiDS+VIKING-450 and DES-Y1 combined: Cosmology with cosmic shear}},\ }\href
  {https://doi.org/10.1051/0004-6361/201936154} {\bibfield  {journal} {\bibinfo
   {journal} {Astron. Astrophys.}\ }\textbf {\bibinfo {volume} {638}},\
  \bibinfo {pages} {L1} (\bibinfo {year} {2020})},\ \Eprint
  {https://arxiv.org/abs/1906.09262} {arXiv:1906.09262 [astro-ph.CO]}
  \BibitemShut {NoStop}%
\bibitem [{\citenamefont {Raveri}\ and\ \citenamefont
  {Hu}(2019)}]{Raveri:2018wln}%
  \BibitemOpen
  \bibfield  {author} {\bibinfo {author} {\bibfnamefont {M.}~\bibnamefont
  {Raveri}}\ and\ \bibinfo {author} {\bibfnamefont {W.}~\bibnamefont {Hu}},\
  }\bibfield  {title} {\bibinfo {title} {{Concordance and Discordance in
  Cosmology}},\ }\href {https://doi.org/10.1103/PhysRevD.99.043506} {\bibfield
  {journal} {\bibinfo  {journal} {Phys. Rev. D}\ }\textbf {\bibinfo {volume}
  {99}},\ \bibinfo {pages} {043506} (\bibinfo {year} {2019})},\ \Eprint
  {https://arxiv.org/abs/1806.04649} {arXiv:1806.04649 [astro-ph.CO]}
  \BibitemShut {NoStop}%
\bibitem [{\citenamefont {Sch\"oneberg}\ \emph {et~al.}(2022)\citenamefont
  {Sch\"oneberg}, \citenamefont {Franco~Abell\'an}, \citenamefont
  {P\'erez~S\'anchez}, \citenamefont {Witte}, \citenamefont {Poulin},\ and\
  \citenamefont {Lesgourgues}}]{Schoneberg:2021qvd}%
  \BibitemOpen
  \bibfield  {author} {\bibinfo {author} {\bibfnamefont {N.}~\bibnamefont
  {Sch\"oneberg}}, \bibinfo {author} {\bibfnamefont {G.}~\bibnamefont
  {Franco~Abell\'an}}, \bibinfo {author} {\bibfnamefont {A.}~\bibnamefont
  {P\'erez~S\'anchez}}, \bibinfo {author} {\bibfnamefont {S.~J.}\ \bibnamefont
  {Witte}}, \bibinfo {author} {\bibfnamefont {V.}~\bibnamefont {Poulin}},\ and\
  \bibinfo {author} {\bibfnamefont {J.}~\bibnamefont {Lesgourgues}},\
  }\bibfield  {title} {\bibinfo {title} {{The H0 Olympics: A fair ranking of
  proposed models}},\ }\href {https://doi.org/10.1016/j.physrep.2022.07.001}
  {\bibfield  {journal} {\bibinfo  {journal} {Phys. Rept.}\ }\textbf {\bibinfo
  {volume} {984}},\ \bibinfo {pages} {1} (\bibinfo {year} {2022})},\ \Eprint
  {https://arxiv.org/abs/2107.10291} {arXiv:2107.10291 [astro-ph.CO]}
  \BibitemShut {NoStop}%
\bibitem [{\citenamefont {Herold}\ and\ \citenamefont
  {Ferreira}(2023)}]{Herold:2022iib}%
  \BibitemOpen
  \bibfield  {author} {\bibinfo {author} {\bibfnamefont {L.}~\bibnamefont
  {Herold}}\ and\ \bibinfo {author} {\bibfnamefont {E.~G.~M.}\ \bibnamefont
  {Ferreira}},\ }\bibfield  {title} {\bibinfo {title} {{Resolving the Hubble
  tension with early dark energy}},\ }\href
  {https://doi.org/10.1103/PhysRevD.108.043513} {\bibfield  {journal} {\bibinfo
   {journal} {Phys. Rev. D}\ }\textbf {\bibinfo {volume} {108}},\ \bibinfo
  {pages} {043513} (\bibinfo {year} {2023})},\ \Eprint
  {https://arxiv.org/abs/2210.16296} {arXiv:2210.16296 [astro-ph.CO]}
  \BibitemShut {NoStop}%
\bibitem [{\citenamefont {Karwal}\ \emph {et~al.}(2024)\citenamefont {Karwal},
  \citenamefont {Patel}, \citenamefont {Bartlett}, \citenamefont {Poulin},
  \citenamefont {Smith},\ and\ \citenamefont {Pfeffer}}]{Karwal:2024qpt}%
  \BibitemOpen
  \bibfield  {author} {\bibinfo {author} {\bibfnamefont {T.}~\bibnamefont
  {Karwal}}, \bibinfo {author} {\bibfnamefont {Y.}~\bibnamefont {Patel}},
  \bibinfo {author} {\bibfnamefont {A.}~\bibnamefont {Bartlett}}, \bibinfo
  {author} {\bibfnamefont {V.}~\bibnamefont {Poulin}}, \bibinfo {author}
  {\bibfnamefont {T.~L.}\ \bibnamefont {Smith}},\ and\ \bibinfo {author}
  {\bibfnamefont {D.~N.}\ \bibnamefont {Pfeffer}},\ }\bibfield  {title}
  {\bibinfo {title} {{Procoli: Profiles of cosmological likelihoods}},\
  }\href@noop {} {\  (\bibinfo {year} {2024})},\ \Eprint
  {https://arxiv.org/abs/2401.14225} {arXiv:2401.14225 [astro-ph.CO]}
  \BibitemShut {NoStop}%
\bibitem [{\citenamefont {Gelman}\ and\ \citenamefont
  {Rubin}(1992)}]{Gelman:1992zz}%
  \BibitemOpen
  \bibfield  {author} {\bibinfo {author} {\bibfnamefont {A.}~\bibnamefont
  {Gelman}}\ and\ \bibinfo {author} {\bibfnamefont {D.~B.}\ \bibnamefont
  {Rubin}},\ }\bibfield  {title} {\bibinfo {title} {{Inference from Iterative
  Simulation Using Multiple Sequences}},\ }\href
  {https://doi.org/10.1214/ss/1177011136} {\bibfield  {journal} {\bibinfo
  {journal} {Statist. Sci.}\ }\textbf {\bibinfo {volume} {7}},\ \bibinfo
  {pages} {457} (\bibinfo {year} {1992})}\BibitemShut {NoStop}%
\bibitem [{\citenamefont {Murgia}\ \emph {et~al.}(2021)\citenamefont {Murgia},
  \citenamefont {Abell\'an},\ and\ \citenamefont {Poulin}}]{Murgia:2020ryi}%
  \BibitemOpen
  \bibfield  {author} {\bibinfo {author} {\bibfnamefont {R.}~\bibnamefont
  {Murgia}}, \bibinfo {author} {\bibfnamefont {G.~F.}\ \bibnamefont
  {Abell\'an}},\ and\ \bibinfo {author} {\bibfnamefont {V.}~\bibnamefont
  {Poulin}},\ }\bibfield  {title} {\bibinfo {title} {{Early dark energy
  resolution to the Hubble tension in light of weak lensing surveys and lensing
  anomalies}},\ }\href {https://doi.org/10.1103/PhysRevD.103.063502} {\bibfield
   {journal} {\bibinfo  {journal} {Phys. Rev. D}\ }\textbf {\bibinfo {volume}
  {103}},\ \bibinfo {pages} {063502} (\bibinfo {year} {2021})},\ \Eprint
  {https://arxiv.org/abs/2009.10733} {arXiv:2009.10733 [astro-ph.CO]}
  \BibitemShut {NoStop}%
\bibitem [{\citenamefont {Poulin}\ \emph {et~al.}(2021)\citenamefont {Poulin},
  \citenamefont {Smith},\ and\ \citenamefont {Bartlett}}]{Poulin:2021bjr}%
  \BibitemOpen
  \bibfield  {author} {\bibinfo {author} {\bibfnamefont {V.}~\bibnamefont
  {Poulin}}, \bibinfo {author} {\bibfnamefont {T.~L.}\ \bibnamefont {Smith}},\
  and\ \bibinfo {author} {\bibfnamefont {A.}~\bibnamefont {Bartlett}},\
  }\bibfield  {title} {\bibinfo {title} {{Dark energy at early times and ACT
  data: A larger Hubble constant without late-time priors}},\ }\href
  {https://doi.org/10.1103/PhysRevD.104.123550} {\bibfield  {journal} {\bibinfo
   {journal} {Phys. Rev. D}\ }\textbf {\bibinfo {volume} {104}},\ \bibinfo
  {pages} {123550} (\bibinfo {year} {2021})},\ \Eprint
  {https://arxiv.org/abs/2109.06229} {arXiv:2109.06229 [astro-ph.CO]}
  \BibitemShut {NoStop}%
\bibitem [{\citenamefont {Hu}(1998)}]{Hu:1998kj}%
  \BibitemOpen
  \bibfield  {author} {\bibinfo {author} {\bibfnamefont {W.}~\bibnamefont
  {Hu}},\ }\bibfield  {title} {\bibinfo {title} {{Structure formation with
  generalized dark matter}},\ }\href {https://doi.org/10.1086/306274}
  {\bibfield  {journal} {\bibinfo  {journal} {Astrophys. J.}\ }\textbf
  {\bibinfo {volume} {506}},\ \bibinfo {pages} {485} (\bibinfo {year}
  {1998})},\ \Eprint {https://arxiv.org/abs/astro-ph/9801234}
  {arXiv:astro-ph/9801234} \BibitemShut {NoStop}%
\end{thebibliography}%
\end{document}